\documentclass[onecolumn,longauth]{aa}  

\usepackage{color}
\definecolor{black}{rgb}{0,0,0}
\definecolor{red}{rgb}{1,0,0}
\definecolor{darkblue}{rgb}{0,0,0.7}
\definecolor{blue}{rgb}{0,0,1} 
\definecolor{green}{rgb}{0,0.5,0} 
\definecolor{orange}{rgb}{0.8,0.6,0} 
\definecolor{purple}{rgb}{1,0,1}

\usepackage{longtable,array}
\usepackage{graphicx}
\usepackage{txfonts}
\usepackage[utf8]{inputenc} 

\begin{document} 

   \title{Constraints on the structure and seasonal variations of Triton's atmosphere from the 5 October 2017 stellar occultation and previous observations}
      \subtitle{}

\author{
J. Marques Oliveira\inst{1} 
\and
B. Sicardy\inst{1} 
\and
A. R. Gomes-Júnior\inst{2,3} 
\and
J. L. Ortiz\inst{4} 
\and
D. F. Strobel\inst{5} %
\and
T. Bertrand\inst{1,6} %
\and
F. Forget\inst{7} %
\and
E.~Lellouch\inst{1} %
\and
J. Desmars\inst{8,9} 
\and
D. Bérard\inst{1} %
% ...
}

\institute{LESIA, Observatoire de Paris, Université PSL, CNRS, Sorbonne Université, 5 place Jules Janssen, 92190 Meudon, France\\
\email{Joana.Oliveira@obspm.fr}
\and
UNESP - São Paulo State University, Grupo de Dinâmica Orbital e Planetologia, Guaratinguetá, SP, 12516-410, Brazil
\and
Laboratório Interinstitucional de e-Astronomia - LIneA, Rua Gal. José Cristino 77, Rio de Janeiro, RJ, 20921-400, Brazil
\and
Instituto de Astrofísica de Andalucía (IAA-CSIC). Glorieta de la Astronomía s/n. 18008-Granada, Spain
\and
Departments of Earth \& Planetary Sciences and Physics \& Astronomy, Johns Hopkins University, 3400 N. Charles Street, Baltimore, MD 21218, United States
\and
National Aeronautics and Space Administration (NASA), Ames Research Center, Space Science Division, Moffett Field, CA 94035, USA
\and
Laboratoire de Météorologie Dynamique, IPSL, Université PSL, CNRS, Sorbonne Université, 4 place Jussieu, 75005 Paris, France
\and
Institut Polytechnique des Sciences Avancées IPSA, 63 boulevard de Brandebourg, F-94200 Ivry-sur-Seine, France
\and
IMCCE/Observatoire de Paris, Université PSL, CNRS, Sorbonne Université, Univ. Lille - 77 Avenue Denfert Rochereau, 75014 Paris, France
% ...
}
 
 \titlerunning{Triton’s atmosphere from the 5 October 2017 stellar occultation}
\date{Received June 01, 2021; accepted Nov. 17, 2021}

\clearpage \newpage

  \abstract
   {A stellar occultation by Neptune's main satellite, Triton, was observed on 5 October 2017 from Europe, 
   North Africa, and the USA. We derived 90 light curves from this event, 
   42 of which yielded a central flash detection.}
   {We aimed at constraining Triton's atmospheric structure and the seasonal variations of its atmospheric pressure since the Voyager 2 epoch (1989). We also derived the shape of the lower atmosphere from central flash analysis.}
   {We used Abel inversions and direct ray-tracing code to provide the density, pressure, and temperature profiles 
   in the altitude range $\sim 8$~km to $\sim 190$~km, 
   corresponding to pressure levels from 9~$\mu$bar down to a few nanobars.}
   {$(i)$ A pressure of 1.18 $\pm$ 0.03 $\mu$bar is found at a reference radius of 1400~km  (47 km altitude).
   $(ii)$ A new analysis of the Voyager 2 radio science occultation shows that this is consistent with an extrapolation of pressure down to the surface pressure obtained in 1989.
   $(iii)$ A survey of occultations obtained between 1989 and 2017
   suggests that an enhancement in surface pressure as reported during the 1990s might be real, but debatable, due to very few high S/N light curves and data accessible for reanalysis. 
   The volatile transport model analysed supports a moderate increase in surface pressure, with a maximum value around 2005-2015 no higher than $23~\mu$bar. The pressures observed in 1995-1997 and 2017 appear mutually inconsistent with the volatile transport model presented here.
   $(iv)$ The central flash structure does not show evidence of an atmospheric distortion. 
   We find an upper limit of 0.0011 for the apparent oblateness of the atmosphere near the 8~km altitude.}
   {}
   

   \maketitle

\section{Introduction}

The large satellite Triton was discovered in 1846, only 17 days after the discovery 
of its planet, Neptune. An atmosphere was first speculated by \cite{crsi79}, 
with the claimed detection of the gaseous CH$_4$ spectral signature, although 
in retrospect the features were due to methane ice on the surface.
In any case, the presence of this volatile ice did suggest the existence of 
an atmosphere. 
This was to be confirmed ten years later, when the NASA Voyager 2 (V2) spacecraft flew 
by the Neptunian system in August 1989. During this flyby, Triton's tenuous atmosphere
(mainly nitrogen N$_2$) was detected during the Radio Science Subsystem (RSS)
occultation, providing its surface density, pressure, and temperature \citep{tyl89}.
These results were later improved by \cite{gur95} (G95 hereafter), who derived a surface
pressure of $p_{\rm surf}= 14 \pm 2$~$\mu$bar. The studies that used RSS data did not, however,
provide a thermal profile.
Dynamical aspects, such as wind regimes, were studied using V2 images of plumes near Triton's surface \citep{yelle91}, as well as vertical profiles of methane and hazes using V2's UV images \citep{str90,her91,kra93,kras93,krcr95,str95}. 

\begin{figure}[h!]
\centerline{
\includegraphics[totalheight=8cm,trim=0 0 0 0,angle=0]{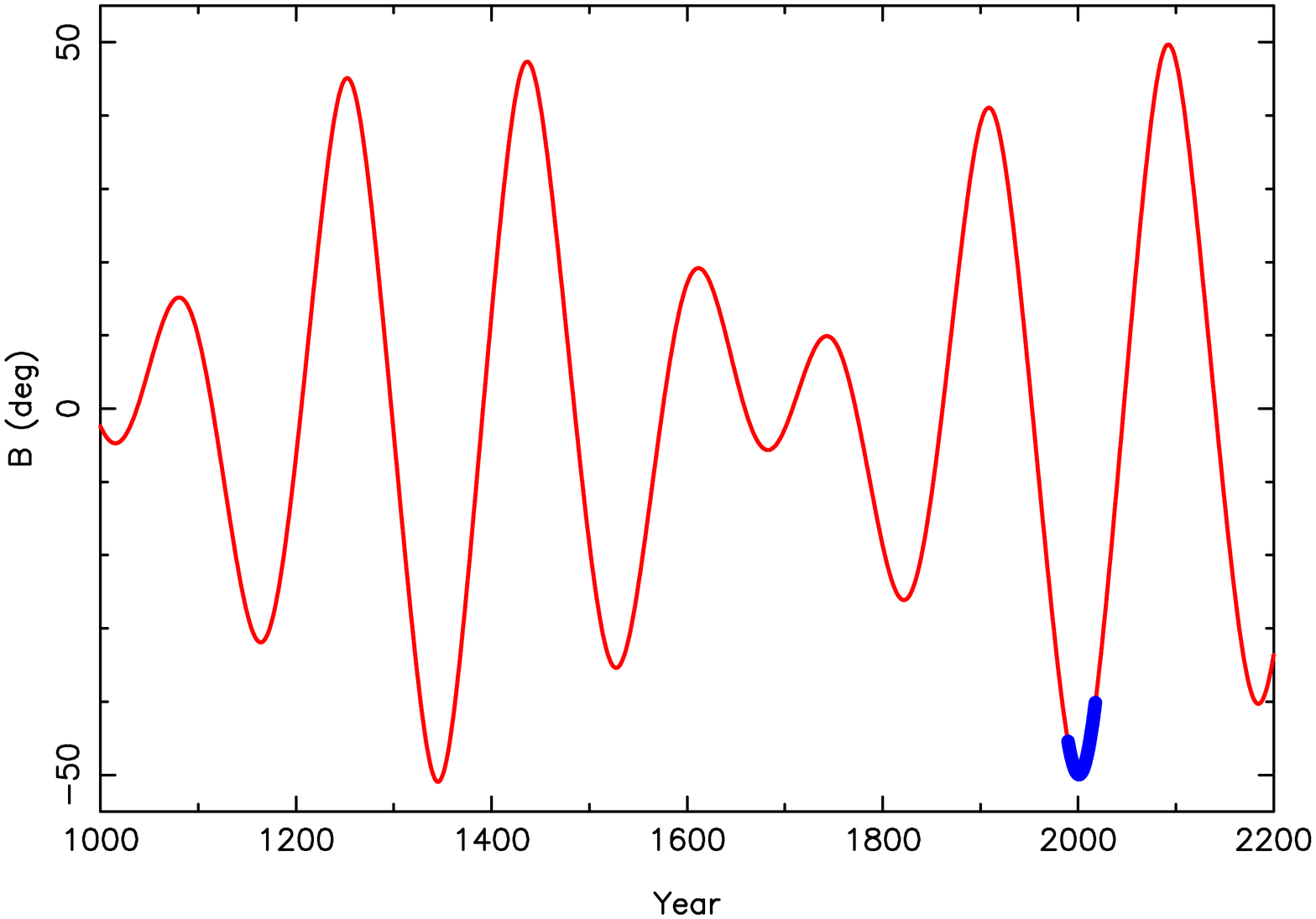}
}
\centerline{
\includegraphics[totalheight=8cm,trim=0 0 0 0]{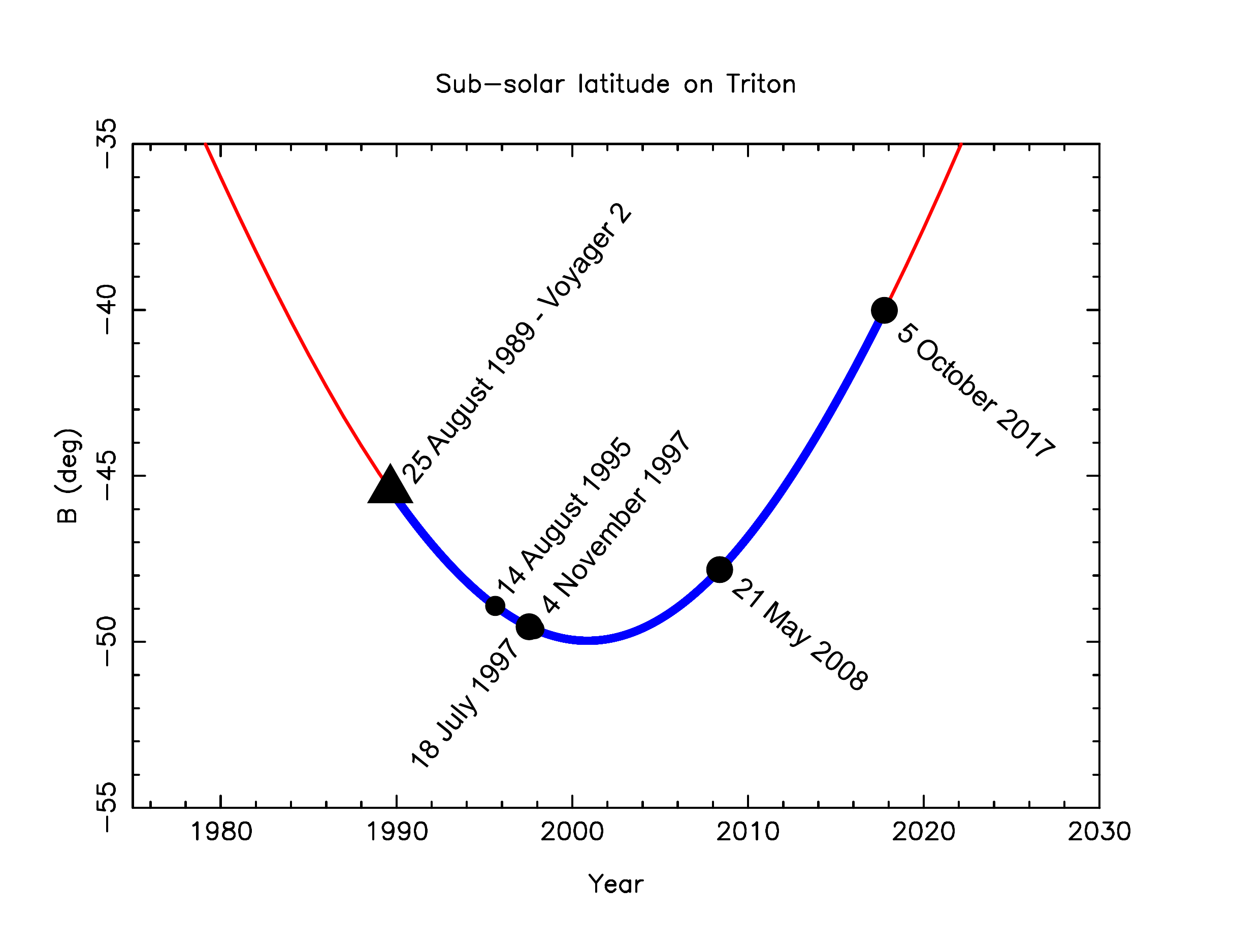}
}
\caption{
Sub-solar latitude on Triton over time.
\textit{Upper panel:} 
Overview of the sub-solar latitude on Triton versus time over the last millennium. 
The blue part corresponds to the period from the V2 encounter (August 1989)
to the 5 October 2017 stellar occultation. It shows that during this interval, 
Triton experienced an extreme summer solstice in its southern hemisphere, 
with a minimum sub-solar latitude of 50$^\circ$ S in 2000.
\textit{Lower panel:}
Close-up view of the upper panel around the year 2000.
The black points correspond to occultations observed, distinguishing from the black triangle 
(V2 RSS experiment). The larger symbols are from the data that we use in this paper.
}
\label{fig_year_B}
\end{figure}

Triton is currently experiencing a rare `extreme southern solstice', 
a configuration that occurs every $\thicksim$~650 years (Fig.~\ref{fig_year_B}). 
In particular, the sub-solar latitude on the satellite reached about 
50$^\circ$ S in 2000.
The various measurements of Triton's atmospheric pressure using occultations bracket
that epoch, from the RSS results in 1989 to the ground-based stellar occultation 
of 2017 discussed here.
In this context, it is interesting to look for ongoing seasonal effects (if any)
occurring in Triton's atmosphere, especially large pressure variations in the last
three decades. Such seasonal variations (or its absence) can then constrain
global climate models (GCMs) and volatile transport models (VTMs) that account for volatile transport induced by insolation changes.

Since 1989, only a handful of Earth-based stellar occultations have been observed, and there have been a few spectroscopic studies that detected CO and CH$_4$ in the near-IR
\citep{lel10} and CO and HCN with the Atacama Large Millimeter/submillimeter Array (ALMA) in the millimetre \citep{gur19}.
As discussed later, the deepest layers accessible during Triton Earth-based occultations (when a central flash\footnote{A central flash is a sharp increase in the intensity of the stellar light observed during a stellar occultation. It is observed near the central path of the occultation shadow and is produced by the refraction of light in the atmosphere of the occulting object.} is observed) typically lie at 8~km altitude 
($\sim 9$~$\mu$bar pressure level). Conversely, the best light curves can provide 
information up to an altitude of about 190~km, corresponding to a few nanobars.

Here, we report on results obtained from the 5 October 2017 ground-based stellar
occultation. 
Attempts to view this rare event were made from more than 100 sites in Europe, northern Africa, 
and the eastern USA.
We extracted 90 occultation light curves from this campaign, 
among which 42 show a central flash. 
This is by far the most observed stellar occultation by Triton ever monitored 
(and among the most observed event of its kind, all Solar System objects combined)
both in terms of latitudinal coverage of the satellite and central flash sampling.

Our goals are to
(1)~provide Triton's atmospheric profiles (density, pressure, temperature) derived from this event, 
(2)~compare the results with those obtained from previous occultations, including the V2 RSS experiment, 
(3)~constrain the seasonal variations of Triton's atmosphere,
(4)~compare the results with current GCMs,
and (5)~and derive the shape of the central flash layer.

The 5 October 2017 event is discussed in Sect.~\ref{sec_data_analysis}. 
The methods used are given in Sect.~\ref{sec_retrieving_atmo}, and in Sect.~\ref{sec_results} we present the results. 
We reanalyse previous events, including a new approach to retrieve new information from the V2 experiment in Sect.~\ref{sec_previous_occs}.
In Sect.~\ref{sec_pres_evolution} we discuss the atmospheric seasonal variations, using pressure values from our work and from other works.
Section~\ref{sec_lower_atmo} focuses on the analysis of the central flashes.
We mention some issues that are to be addressed at a future date in Sect.~\ref{sec_discuss}.
Concluding remarks are provided in Sect.~\ref{sec_conclusion}.

\section{The 5 October 2017 stellar occultation}
\label{sec_data_analysis}

\subsection{Prediction}

In the past two decades, Neptune has been crossing regions with low surface-density of stars.
In the early 2010s, faint star surveys up to R = 19 were made using the Wide Field Imager attached to
the 2.2 m Max-Planck telescope at the European Southern Observatory (ESO) \citep{ass10,ass12,cam14}. They provided many candidates for
occultations by Pluto, large Trans-Neptunian Objects and Centaurs, but no suitable Triton events 
for 2008-2015.

Predicting occultations by Triton is problematic for two reasons:
(1)~Neptune's orbit may have systematic errors, causing a systematic shift of Triton's position 
with respect to the stars; 
(2)~Triton's neptunocentric orbit may have systematic errors due to the large brightness and colour differences
between Neptune and Triton, and also from changes in Neptune's magnitude, making their relative colour
variable \citep{sch16}.
Both points affect differently the ground-based measurements of Triton and Neptune, 
resulting in a distorted neptunocentric orbit for the satellite.
A way to overcome these problems is 
to use R or I filters to minimise differential refraction during observations and 
distribute Triton observations evenly along its orbit around Neptune. 
Triton's path is then set by the average ephemeris offsets found for 
right ascension ($\alpha$) and declination ($\delta$).

In this context, we gathered more than 4700 charge-coupled-device (CCD) images of Triton in R between 1992 and 2016 with the 
0.6-m B\&C and 1.6-m P\&E telescopes at Pico dos Dias Observatory (OPD) in Brazil (IAU code 874). 
The observations were reduced with the best available astrometric catalogue at that time, the UCAC4 \citep{zac13},
following the same procedures described in \cite{gom15}. 
From the average ephemeris offsets of many nights, and after sigma-clipping, we found an overall offset 
($\Delta \alpha \cos \delta, \Delta \delta) = (+1 \pm 45, -16 \pm 45$) milliarcsecond (mas), 
with error bars at 1$\sigma$ level, with respect to the DE435/NEP081 ephemeris from the Jet Propulsion Laboratory (JPL).
Applying this offset and searching for post-2015 events, we uncovered the promising occultation by Triton of 
a relatively bright star with V=12.7, G=12.2 (UCAC4 410-143659; \textit{Gaia} DR2 2610107911326516992) 
that was to occur on 5 October 2017, crossing all Europe and  northern Africa 
and reaching the eastern USA.

Closer in time to the event, a dedicated 8-night run on the OPD 1.6 m telescope was conducted between 15 and 23 September 2017 to further improve the accuracy of the prediction, in particular to pin down the path of the central flash. 
The field of view of the images was $6.1 \arcmin \times 6.1 \arcmin$ with a pixel scale of 180 mas/pixel. 
At that point, digital coronagraphy \citep{ass09,cam15}, which mitigates Neptune's scattered light, proved 
to be unnecessary.
Chromatic refraction corrections to Triton's position were carried out, but proved to be negligible too, 
as the observations were made in the I band. Observations from two nights were discarded due to bad weather, 
but we were able to cover a complete orbit of Triton around Neptune (about 6 days) with more than 1000 images. 

\begin{figure}[!t]
\begin{center}
\includegraphics[totalheight=6cm,trim=0 0 0 0]{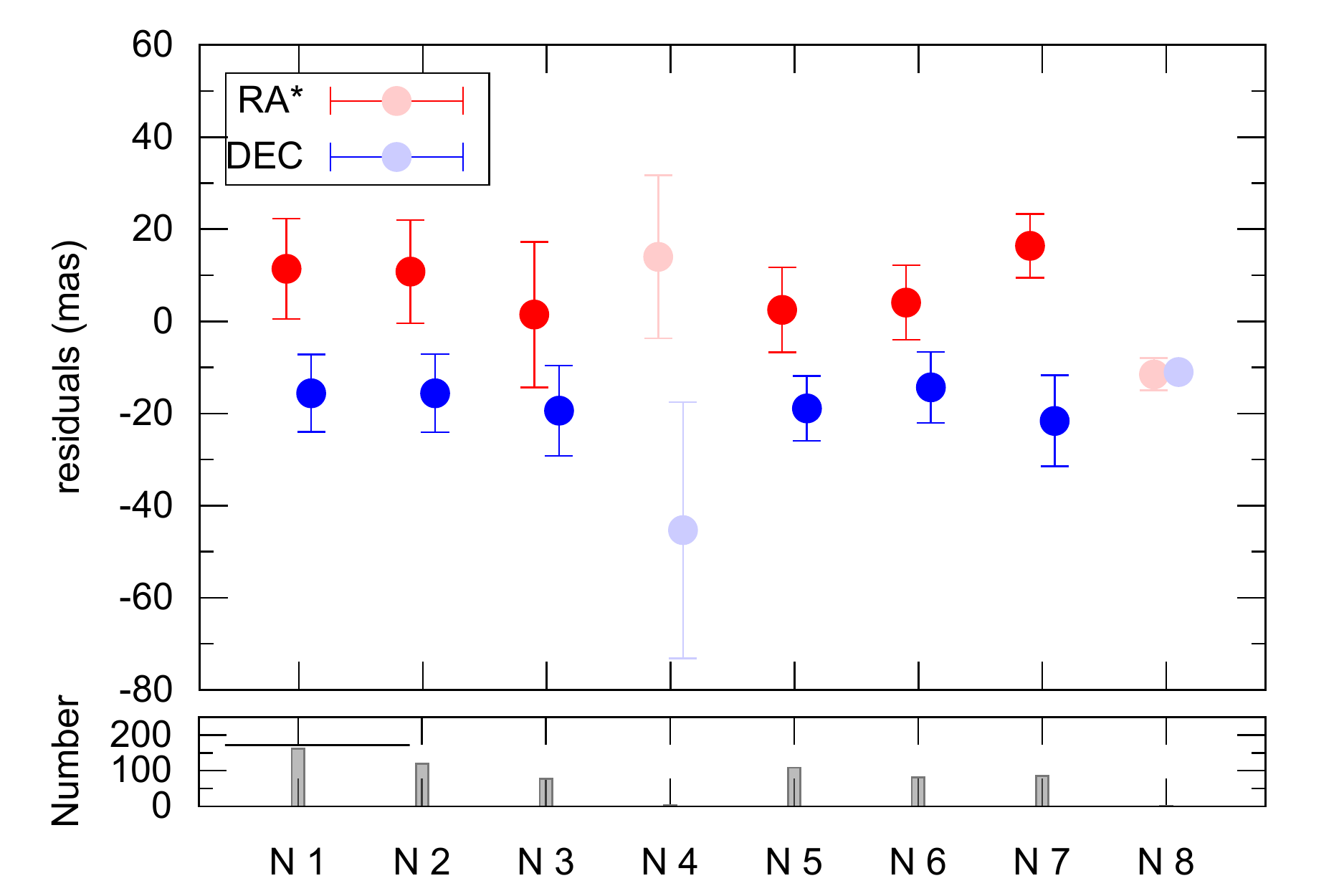}
\end{center}
\caption[]{
Average offsets in ($\alpha$, $\delta$) mas of the eight nights of OPD observations using 
the preliminary \textit{Gaia} DR2 catalogue for the astrometric reduction. 
Positions in lighter colours were not accounted for to compute the average offset,
due to lower quality associated with poor weather conditions.
The lower panel refers to the number of images taken for each night.
}
\label{fig:redGDR}
\end{figure}

Prior to \textit{Gaia} Data Release 2 (DR2), published in April 2018 \citep{gaia16b,gaia18b},
the \textit{Gaia} team\footnote{%
https://www.cosmos.esa.int/web/gaia/news\_20170930
}
released a preliminary \textit{Gaia} subset DR2 with 431 stars (R = 12-17) surrounding Triton's path 
in the sky plane during the eight nights of our run. 
The main improvement with respect to DR1 is the inclusion of the stars' proper motions, 
leading to mas-level accuracy of stellar positions at epoch. 
Using these reference stars and the PRAIA package \citep{ass11}, we obtained a mean offset 
($\Delta \alpha \cos \delta, \Delta\delta) = (+7.8 \pm 5.4,-17.6 \pm 2.6)$ mas 
with respect to the JPL DE435/NEP081 ephemeris.
The corresponding prediction uncertainty was about 60~km cross-track and 8~seconds in time. 
Figure~\ref{fig:redGDR} displays the mean offset ($\Delta \alpha \cos \delta$, $\Delta \delta$) 
for the eight nights, using the \textit{Gaia} DR2 catalogue. 
This is a large improvement from a prediction using only the \textit{Gaia} DR1 catalogue, 
shifting the shadow path about 370~km to the south in the sky plane, or some 500-700~km 
when projected on Earth, depending on the station considered, 
as well as reducing the offsets uncertainties by factors of about 1.5 and 2.2, respectively.
Since the run covers a full synchronous revolution, and rotational, period of Triton, the average offset reflects an error 
in Neptune's heliocentric position rather than a neptunocentric error in Triton's ephemeris.

The $\pm 60$~km cross-track 
uncertainty on the prediction was essential for better planning observations of the central flash, keeping in mind 
that the width of the region where that central flash is significant (e.g. more than 20\% of the unocculted stellar 
flux; see Figs.~\ref{fig_fit_flash_1}-\ref{fig_fit_flash_2}) typically spans $\pm 100$~km in the sky plane.

\begin{figure}[!t]
\centerline{
\includegraphics[width=9cm,trim=0 0 0 0,angle=0]{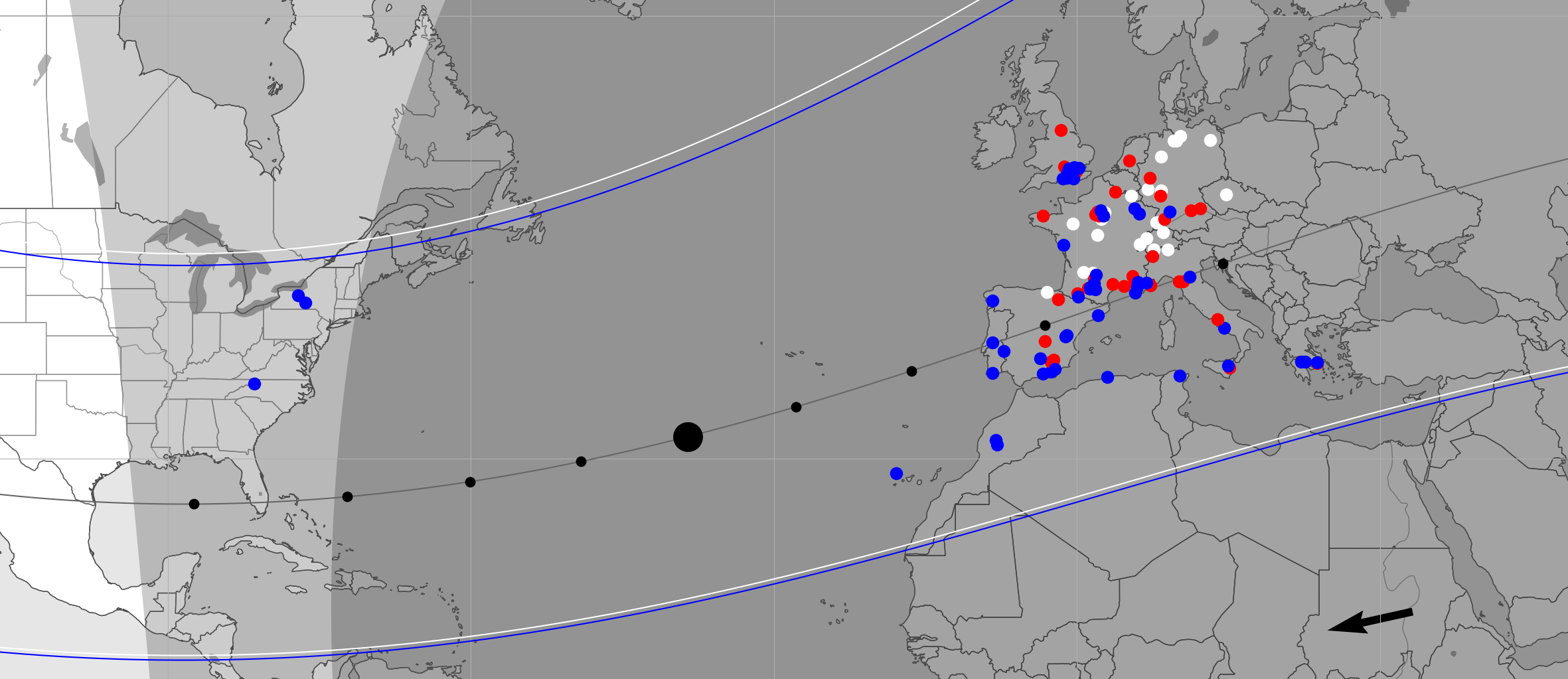}
}
\centerline{
\includegraphics[width=9cm,trim=0 0 0 0]{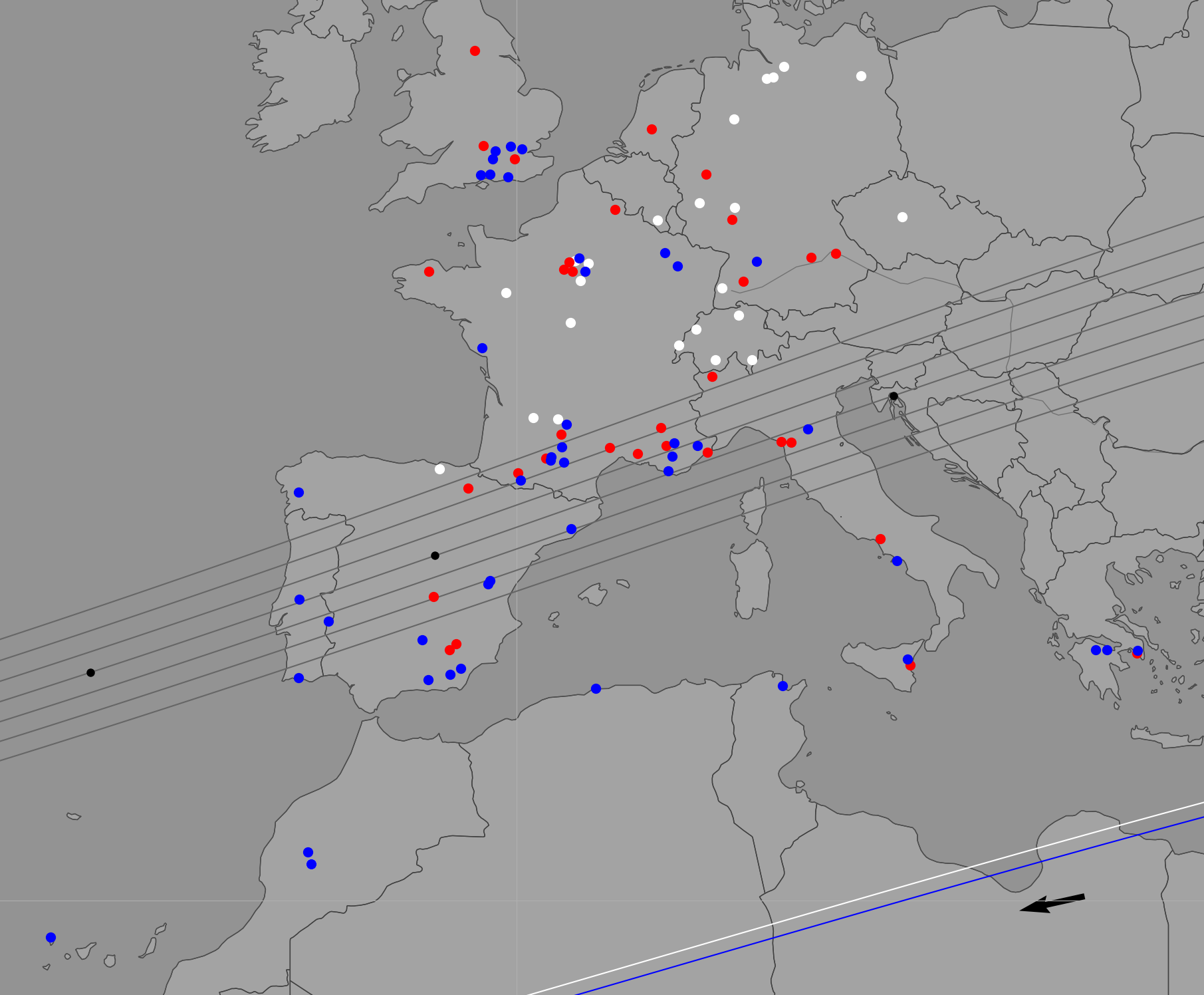}
}
\caption{
Triton's shadow path on 5 October 2017.
The black dots are spaced by one minute, the arrow indicating the direction of motion of the shadow. The northern and southern limits of the solid body assuming a radius of 1353~km are also represented, with the predicted path as the white lines and the effective path as the blue lines. 
The grey area represents night on Earth (dark grey for full astronomical night and light grey for twilight).
Stations with a successful observation that have been used in our fit are represented by blue dots, while the ones that were not used are shown as red dots. The white dots are the stations that attempted observation but were clouded out or had technical difficulties.
\textit{Upper panel:} 
Overview of all observing stations.
The larger black dot along the black line corresponds to the closest approach of the shadow centre to the geocentre, at around 23h52 UTC (see Table \ref{tab:OccultationCircumstances}).
\textit{Lower panel:}
Closer look at the central flash path across Europe. 
The grey lines around the centremost line correspond to a spacing of 50 km (corresponding to about 2.5 mas) 
once projected in the sky plane.
}
\label{fig:prediction}
\end{figure}

Combining the pre-event Triton's ephemeris offset with the geocentric astrometric 
\textit{Gaia} DR2 position of the occulted
star\footnote{https://www.cosmos.esa.int/web/gaia/news\_20170523}, 
we obtained the parameters describing the nominal prediction listed in
Table~\ref{tab:OccultationCircumstances}.
This updated prediction was promptly released to the scientific community before the
event\footnote{https://www.cosmos.esa.int/web/gaia/iow\_20171005} and the corresponding map 
of the shadow track on Earth is displayed in Fig.~\ref{fig:prediction}. 

\begin{table*}[h]
\centering
\caption{
Occultation prediction using the \textit{Gaia} DR2 position for the star and JPL DE435/NEP081 for Neptune's and Triton's ephemerides and the additional offset deduced from the OPD observations (see text for details).
\label{tab:OccultationCircumstances}
}
\begin{tabular}{ll}
\hline
\hline
  \multicolumn{2}{c}{\textbf{Occultation circumstances (5 October 2017)}}\\
\hline
Predicted time of geocentric closest approach               & 23h51m38.0s $\pm$ 8~s UTC      \\
Predicted geocentric closest approach Triton-star           & 196 mas                       \\
Position angle between Triton and star at closest approach  & 347.51 deg                    \\
Geocentric shadow velocity at closest approach              & 16.80 km s$^{-1}$             \\
\hline
Retrieved time of geocentric closest approach               & 23h51m28.92s $\pm$ 0.02~s UTC  \\
Retrieved geocentric Triton-star closest approach           & 196.58 $\pm$ 0.05 mas         \\
\hline
  \multicolumn{2}{c}{\textbf{Occulted Star (from \textit{Gaia} DR2)}}\\
\hline
Star source ID (stellar catalogue)                          & 2610107911326516992                       \\  
Geocentric star position (ICRF) at epoch & $\alpha=$ 22h54m18.4364s $\pm$ 0.2~mas, $\delta=-08^\circ 00'08.318" \pm 0.2$~mas\\
G-mag / RP mag                                              &  12.5 / 12.0                              \\
Stellar diameter projected at Triton's distance\tablefootmark{1}    & 0.65~km                      \\ 
\hline  \hline
\end{tabular}
\tablefoot{
\tablefoottext{1}{%
Using \cite{van99}'s formulae and magnitudes 
B= 13.305, V= 12.655 (AAVSO photometric survey) and 
K= 11.080 (2mass) for the star;
see http://vizier.u-strasbg.fr/viz-bin/VizieR
}%
}
\end{table*} 

\subsection{Observations of the occultation}

\begin{figure}[!t]
\centerline{
\includegraphics[width=7cm,trim=0 0 0 0]{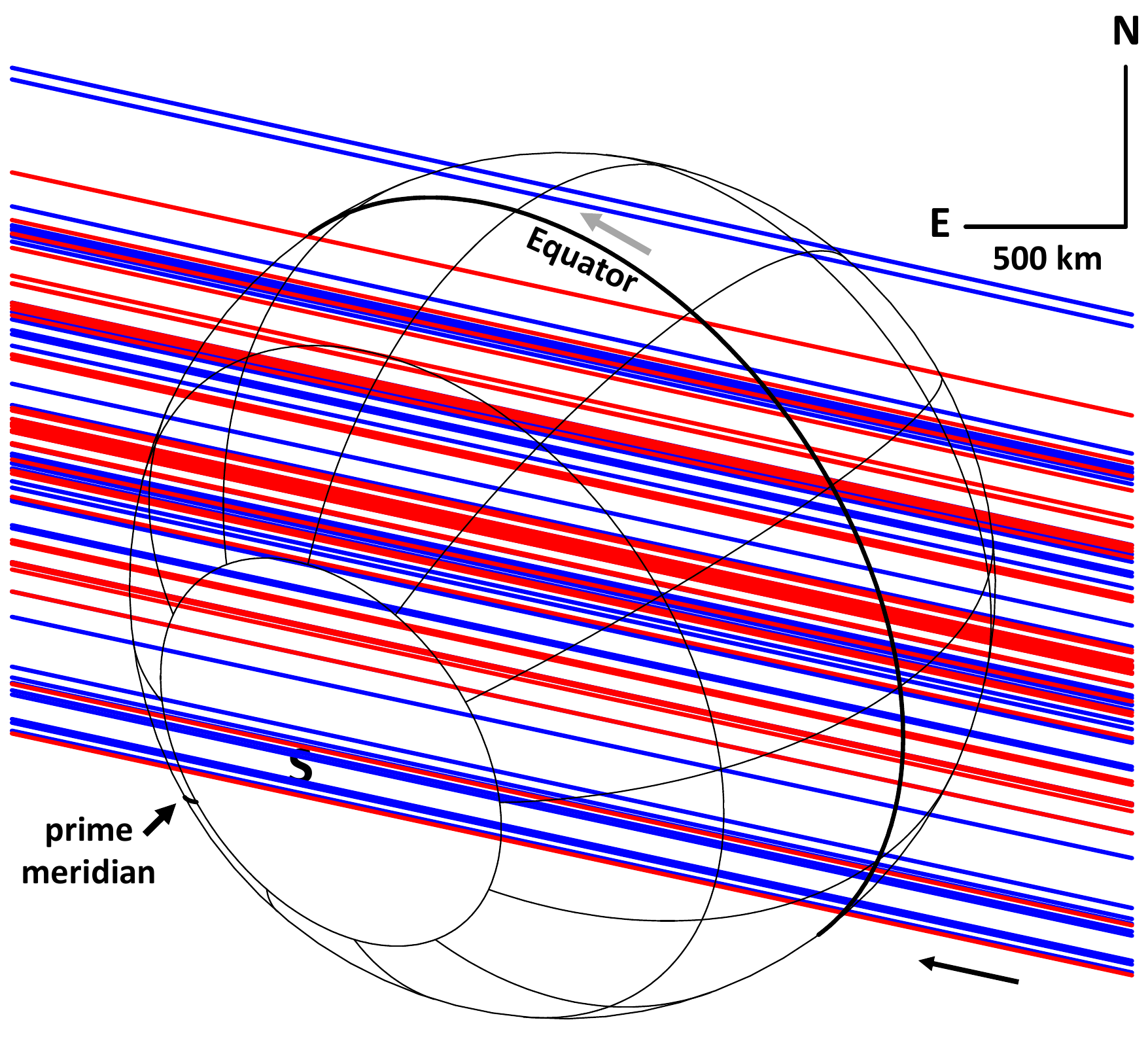}
}
\caption{
Geometry of the 5 October 2017 stellar occultation by Triton, as seen in the sky plane.
The J2000 celestial north (N) and east (E) directions and the scale are indicated 
in the upper right corner.
Triton's radius is fixed at $R_T= 1353$~km, and the grey arrow near the equator shows 
the  direction of rotation of the satellite.
The (Neptune-facing) prime meridian is drawn as a thicker line compared to the
other meridians, and the south pole is marked by the label S.
The inclined lines are the trajectories of the star relative to Triton
(or `occultation chords') as observed from various stations,
with the black arrow indicating the direction of motion. 
We gathered a total of 90 occultation light curves, 52 of which 
(corresponding to the blue colour, as in Fig.~\ref{fig:prediction}) had sufficient S/N to be included in a global atmospheric fit; the remaining 38 (red colour) with lower S/N were not included in the fit.
} 
\label{fig_chords}
\end{figure}

The event was attempted from over one hundred stations in Europe, northern Africa, and the eastern USA,
resulting in a total of 90 occultation light curves. 
Figure~\ref{fig_chords} displays the corresponding occultation chords and
Table~\ref{tab_sites} lists the circumstances of observations for the respective sites.

The first part of this table lists the stations that were eventually used in a simultaneous 
fit to a Triton's atmospheric template model.
The second part of the table lists other stations that were not used because the light curves had
insufficient signal-to-noise ratios (S/N) to provide relevant contribution to that fit. 
The last part of the table provides information on the sites that were involved in the campaign,
but could not gather data due to bad weather or technical problems.

The rule we used for deciding to include or not a light curve in the global fit is as follows. The S/N was first estimated by calculating the standard deviation $\sigma$ of the flux outside the occultation. The resulting S/N 1/$\sigma$ per data point was then re-calculated for a fixed time interval of one second. Assuming a gaussian noise, this implies a multiplicative factor of $\sqrt{1/{\rm exposure~time}}$ to be applied to the S/N obtained above. Finally, the noise level must be compared to the `useful' information, that is, the actual drop of signal caused by the occultation, not the total signal. This implies a new multiplicative factor of $\Delta \phi$ applied to the S/N per second obtained above, where $\Delta \phi$ is the flux drop observed during the occultation. Consequently, some light curves were eliminated due to a low contrast caused by light contamination from Neptune (see for instance the Abington, Caserta, Agerola~50~cm, or Catania~28~cm light curves in Figs.~\ref{fig_no_fit_data_1}-\ref{fig_no_fit_data_4}).

The normalisation described above then allows us to compared consistently the various datasets. We used a normalised S/N cut-off of ten prior to the global fit. There are a few exceptions to that rule for light curves containing a strong central flash (see for instance the Le Beausset and Felsina data in Fig.~\ref{fig_fit_data_3}). They have a poor overall S/N, but we have kept them in the global fit (without flash; Sect.~\ref{sec_ray_tracing_results}) to test the effect of the central flash inclusion (see Fig.~\ref{fig_fit_flash_1} and Sect.~\ref{sec_spherical_fit}). This approach allowed us to ensure that the central flashes provide a shadow centre witch is consistent with, but more accurate than the one given by the global fit (see Sect.~\ref{sec_spherical_fit}).

This said, the cut-off of ten remains somehow arbitrary. We have tested other cut-off values, and we did not obtain any significant changes in the results of the fits presented in the rest of the paper.

The analysis of the light curves (as described in the next section) allowed us to reconstruct the geometry 
of the event by providing Triton's position with respect to the occulted star (see Table~\ref{tab:OccultationCircumstances}). 
The reconstructed geometry of the occultation implies a shift of 
($\Delta\alpha \cos\delta, \Delta\delta) = (-7.2, +0.6$) mas of Triton's position with respect to 
the latest prediction described in the previous section. 
This means that the occultation occurred about 9 s earlier than expected, 
and that the shadow centre was 12 km north of the predicted path in the sky plane.

This mismatch between observation and prediction is at a $\sim$1.3$\sigma$-level, and is thus insignificant at our accuracy level. 
It shows in particular that the \textit{Gaia} DR2 astrometry was crucial in getting an accurate 
prediction that allowed the detection of the central flash at various stations.

\subsection{Occultation light curves}

\begin{figure*}[!t]
\centerline{
\includegraphics[width=9cm]{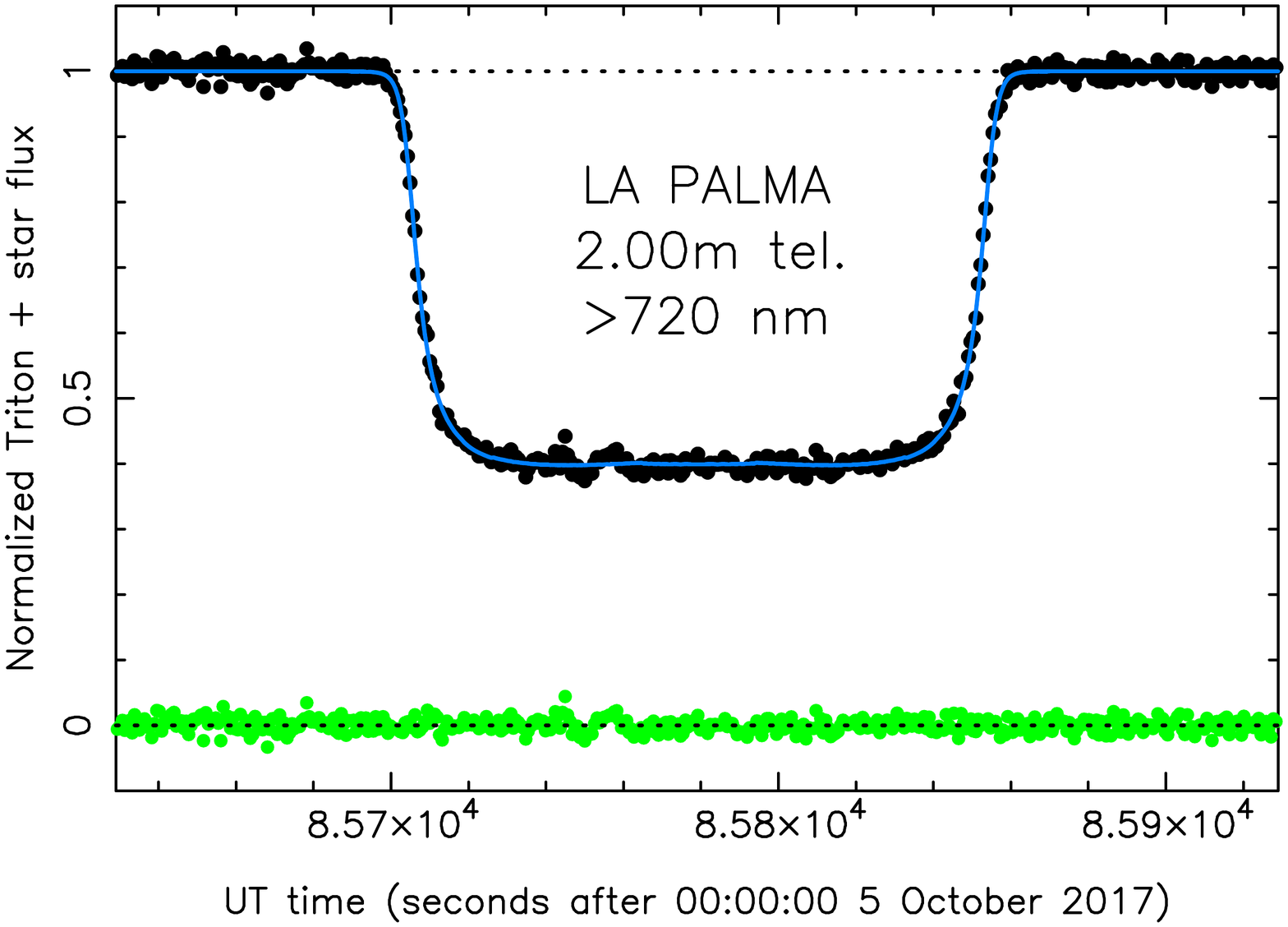}
\includegraphics[width=9cm]{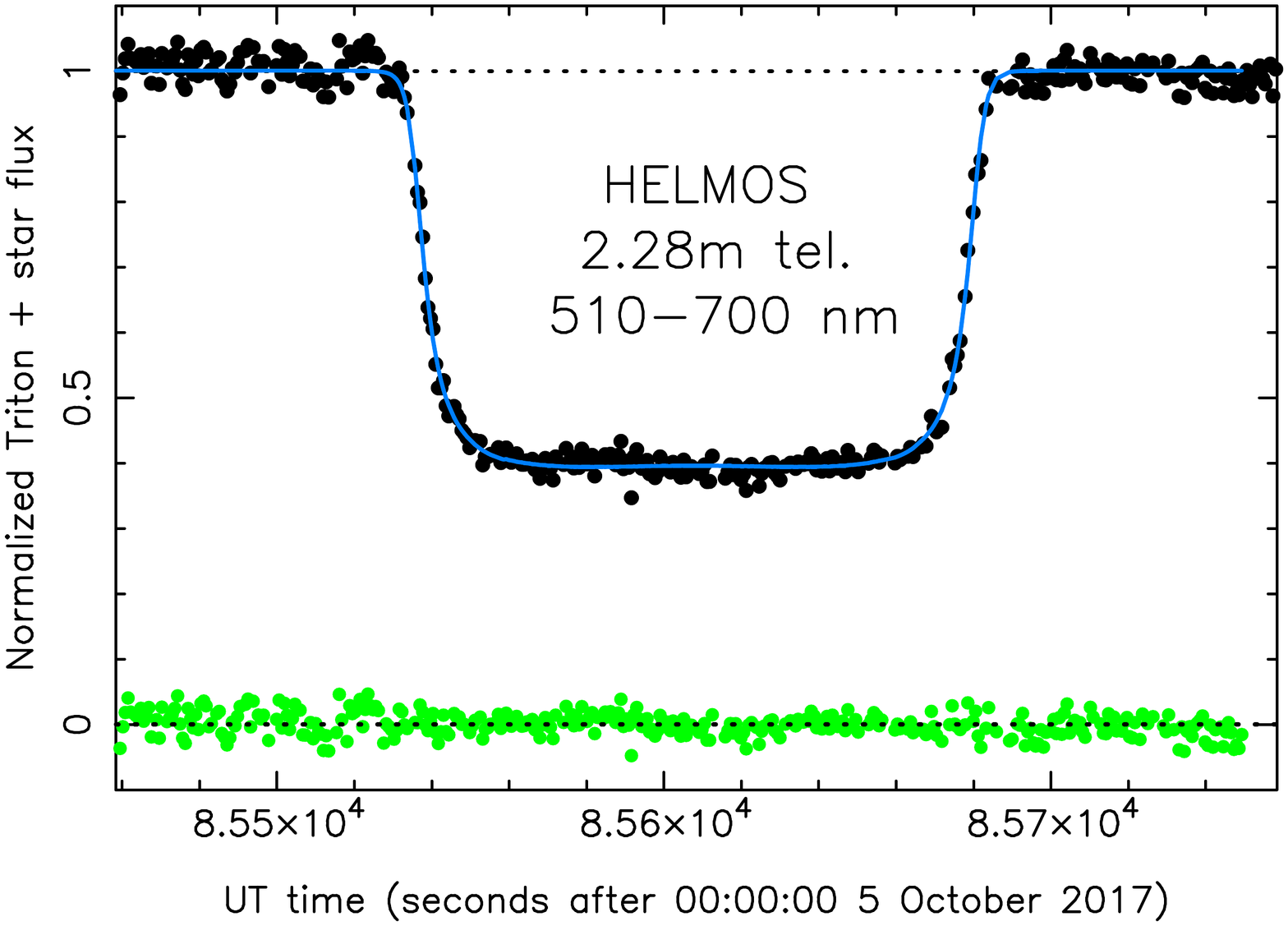}
}
\caption{
Best two light curves obtained during the Triton occultation of 5 October 2017,
at La Palma and Helmos stations (see details in Table~\ref{tab:OccultationCircumstances}).
Both telescopes were equipped with the same E2V CCD 47-20 detector with
quantum efficiency peaking at 600 nm and reaching zero near 300 and 1000 nm, respectively.
\textit{Left panel}:
Full resolution light curve (cycle time 0.635~s) for the La Palma station.
\textit{Right panel}:
Same for the Helmos station (cycle time 0.674~s).
The spectral ranges used for each instrument are indicated in the figures
(I+z at La Palma and V+R at Helmos).
The blue lines are the best simultaneous fits obtained with our ray-tracing approach (see Sect.~\ref{sec_results}).
} 
\label{fig_liv_ari}
\end{figure*}

All our occultation light curves are displayed in 
Figs.~\ref{fig_fit_data_1}-\ref{fig_fit_data_5} and
Figs.~\ref{fig_no_fit_data_1}-\ref{fig_no_fit_data_4}.
As was done in Table~\ref{tab_sites}, 
the first group of figures corresponds to light curves that had 
sufficient S/N to be used in Triton's atmospheric fit, 
while the second group is for light curves with lower S/N that were not used in the fit.
It should be noted, however, that the best synthetic models expected for those light curves
(plotted in grey in Figs.~\ref{fig_no_fit_data_1}-\ref{fig_no_fit_data_4}) 
are fully consistent with the observations, in the limit of the noise level.

Among those light curves, three were used for obtaining atmospheric profiles
from an Abel inversion procedure (see Sect.~\ref{sec_results}).
Two of them (La Palma and Helmos) are displayed in Fig.~\ref{fig_liv_ari}, 
and the third one (Calern) is shown in the upper panel of Fig.~\ref{fig_fit_flash_cal_con}.

\section{Retrieving Triton's atmospheric structure}
\label{sec_retrieving_atmo}

\subsection{Methodology}

We first adopt a bootstrap method to retrieve the molecular density $n(r)$, 
pressure $p(r)$, and temperature $T(r)$ of Triton's atmosphere as a function of 
the distance to Triton's centre, $r$.
To do so, one approach is the Abel inversion of our refractive occultation 
light curves that have the highest S/N. Its primary result is the density profiles $n(r)$, 
from which the $p(r)$ and $T(r)$ profiles are derived by using the hydrostatic 
and ideal gas equations.

The other approach is a direct one. 
It is used once the inversion procedures have provided the density profiles $n(r)$.
These profiles are smoothed and parameterised according to physical arguments (discussed later).
Then, a ray-tracing scheme generates synthetic light curves that are fitted to the occultation light curves, 
thus describing the global structure of Triton's atmosphere, as illustrated in Fig.~\ref{fig_chords}. 
One product of this fit is the location of Triton's centre, 
which is used iteratively with the Abel inversion, thus improving the accuracy on the altitude scale.
The other product of our approach is the value of the pressure at a prescribed radius 
for comparison with previous results, aimed at detecting possible long-term seasonal effects.

The final approach is to fit the central flashes observed in some light curves and 
to measure the departure (if any) from the spherical shape of Triton's deep atmosphere, 
which is eventually used to constrain its wind regime.
The other goal of this fit is to reveal possible absorbing material 
(by comparing the height of the central flash in stations that provided observations 
in different wavelengths) along the line of sight, such as hazes just above Triton's surface.

Technical details on the inversion technique are given in \cite{vap73}, 
ray-tracing schemes and central flash fitting are described in \cite{sic99}, and 
applications to Pluto's atmosphere are presented in 
\cite{dia15} (DO15 hereafter), and \cite{mez19}.
Numerical values used in both our inversion and ray-tracing codes are summarised in Table~\ref{tab_param}.

\begin{table}[!b]
\caption{Adopted physical parameters for Triton and its atmosphere.}
\label{tab_param}
\centering
\begin{tabular}{ll}
\hline \hline
\multicolumn{2}{c}{Triton's body} \\
\hline
Mass\tablefootmark{1}                   & $GM_T=  1.428 \times 10^{12}$ m$^3$ s$^{-2}$ \\
Radius\tablefootmark{1}                 & $R_T= 1353$~km \\
\hline
\multicolumn{2}{c}{Triton's geometry on 5 October 2017} \\
\hline
Triton pole position\tablefootmark{2}   & $\alpha_{\rm p}$= 20h 09m 29.40s \\
(J2000)                                 & $\delta_{\rm p}$=  20$^\circ$ 25' 34.2" \\
Sub-solar latitude                      & 40.0$^\circ$ S \\
Sub-observer latitude                   & 40.5$^\circ$ S \\
Sub-observer longitude                  & 169.9$^\circ$ E \\
N. pole position angle\tablefootmark{3} & 305.7$^\circ$ \\
Geocentric distance                     & $D= 4.3506 \times 10^9$~km \\ 
\hline
\multicolumn{2}{c}{Triton's atmosphere parameters} \\
\hline
N$_2$ molecular mass                    & $\mu= 4.652 \times 10^{-26}$ kg                       \\
N$_2$ specific heat                     & $c_p = 1.04 \times 10^3$ J K$^{-1}$ kg$^{-1}$         \\
at constant pressure                    &                                                       \\
N$_2$ molecular                         & $K = 1.091 \times 10^{-23}$                           \\ 
refractivity                            & $+ (6.282 \times 10^{-26}/\lambda_{\rm \mu m}^2)$     \\
(visible bands\tablefootmark{4})        & cm$^3$ molecule$^{-1}$                                \\
Refractivity at 3.6 cm\tablefootmark{5} & $K = 1.0945 \times 10^{-23}$ cm$^3$ molecule$^{-1}$   \\
Boltzmann constant                      & $k_B= 1.380626 \times 10^{-23}$ J K$^{-1}$            \\
\hline
\end{tabular}
\tablefoot{
\tablefoottext{1}{\cite{mck95}, where $G$ is the constant of gravitation.}
\tablefoottext{2}{On 5 October 2017, using \cite{dav96}, with corrections available at
ftp://ftp.imcce.fr/pub/iauwg/poles.pdf.}
\tablefoottext{3}{Position angle of Triton's north pole projected in the sky plane. 
Counted positively from celestial north to celestial east.}
\tablefoottext{4}{\cite{was30}.}
\tablefoottext{5}{G95.}
}
\end{table}

\subsection{Assumptions}

Our inversion and ray-tracing schemes assume that: \\
(1) The atmosphere is composed of pure N$_2$. 
The next most abundant species (CH$_4$) has a volume mixing ratio (hereafter referred to as mixing ratio) [CH$_4$/N$_2$] of less than $10^{-3}$
\citep{str95,lel10}. 
Our ray-tracing code shows that such an abundance causes a fractional change of the synthetic flux
of about 10$^{-5}$ near the half-light level, these effects are negligible considering the noise level
of the data. \\
(2) The atmosphere is transparent.
The deepest layers reached during Earth-based occultations are those that cause the central flash, 
at an altitude of about 8~km (see Appendix~\ref{sec_appen_validity}).
The validity of this assumption will be discussed in Sect.~\ref{sec_discuss}. \\
(3) The upper atmosphere is globally spherical. 
This hypothesis is supported by the fact that the observed central flashes are consistent 
with a spherical shape (Sect.~\ref{sec_lower_atmo}).
Small departures from the spherical model, however, are observed in 
some central flash shapes, and are discussed in Sect.~\ref{sec_lower_atmo}.

The limitations of our approach described above are presented in
Appendix~\ref{sec_appen_validity}.

\section{Results}
\label{sec_results}

\subsection{Inverted profiles}

\begin{figure*}
\centering
\includegraphics[width=12cm,trim=300 0 300 0]{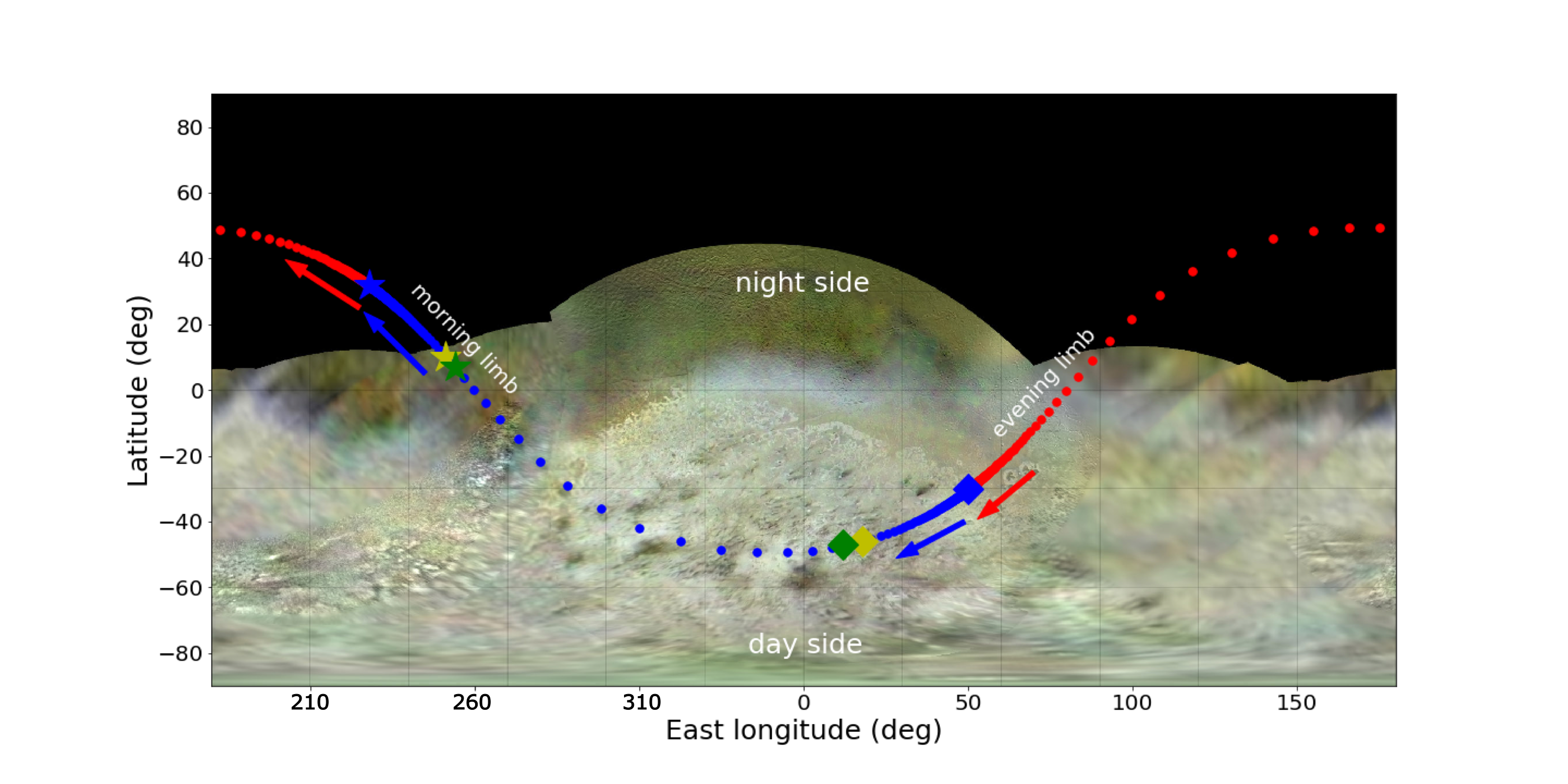}
\caption{
Tracks of the primary (red dots) and secondary (blue dots) stellar images above Triton's surface,
as seen from Constância, plotted every 0.1~s.
The junctions between the red and blue paths correspond to ingress (left)
and egress (right) points for the Constância station. The arrows show the direction of the stellar images' path.
The regions probed by the central flash are those where the dots are more spaced.
All the other stations probed essentially the same path (or part of it), 
with the primary and secondary images being swapped (as well as their directions of motion),
depending on whether the station probed north or south of the shadow centre. 
Since the Earth and the Sun are angularly close ($\sim 1^\circ$) to each other as seen from Triton,
the stellar paths essentially mark Triton's terminator, 
the night side extending above the terminator in this figure.
The two yellow symbols are for La Palma station, 
with ingress plotted as a star and egress plotted as a diamond.
The two green symbols are the same for Helmos station and the two white symbols for Calern station (see Table~\ref{tab_ingress_egress_liv_ari_cal}
for the corresponding values of the latitudes and longitudes).
The background image is a global colour map of Triton, produced using V2 data and orange, green, and blue filter images
in order to obtain an approximation of Triton's natural colours.
Background image credits: Image selection, radiometric calibration, geographic registration and photometric correction, and final mosaic assembly were performed by Dr. Paul Schenk at the Lunar and Planetary Institute, Houston, Texas. Image data are from V2 (NASA, JPL).
}
\label{fig_long_lat_constancia}
\end{figure*}

We used the three datasets with highest S/N to perform our inversion method, more precisely the light curves from
La Palma (2-m Liverpool telescope, Spain),  
Helmos (2.28-m Aristarchos telescope, Greece) and
Calern (1.04-m C2PU telescope, France) (see Table~\ref{tab_sites}).
At half-light times (where the star flux has been reduced by 50\%)
and for ingress and egress, each of these stations probe 
different locations in Triton's atmosphere.
The corresponding latitudes, longitudes, and local solar times of the 
sub-occultation points are provided in Table~\ref{tab_ingress_egress_liv_ari_cal}.
%
\begin{table*}
\caption{
Local circumstances at the three stations (ingress and egress) used 
for the Abel inversion analysis.
\label{tab_ingress_egress_liv_ari_cal}
}
\centering
\begin{tabular}{lccc}
\hline
Site & Time (UT)\tablefootmark{1} & Location on surface & Local solar time\tablefootmark{2} \\
\hline
La Palma, ingress           &  23:48:27  & 251$^\circ$E, 10$^\circ$N & 06:36 (sunrise) \\
\hline
La Palma, egress            &  23:50:52  & 18$^\circ$E, 46$^\circ$S  & 22:08 (sunset)  \\
\hline
Helmos, ingress             &  23:45:38  & 254$^\circ$E,  7$^\circ$N & 06:24 (sunrise) \\
\hline
Helmos, egress              &  23:47:58  & 12$^\circ$E, 47$^\circ$S  & 22:32 (sunset)  \\
\hline
Calern, ingress             &  23:46:28  & 228$^\circ$E, 32$^\circ$N & 08:08 (sunrise) \\
\hline
Calern, egress              &  23:49:15  & 50$^\circ$E, 30$^\circ$S  & 19:00 (sunset)  \\
\hline
\end{tabular}
\tablefoot{
\tablefoottext{1}{UTC time at half-light level, 5 October 2017}
\tablefoottext{2}{One `hour' corresponds to a 15$^\circ$ rotation of Triton.
A local time before (resp. after) 12.0~h means morning (resp. evening) limb.}
}
\end{table*} 
%
The paths of the stellar images over Triton's surface as seen from these
stations are plotted in Fig.~\ref{fig_long_lat_constancia}.

\begin{figure}[!t]
\centerline{
\includegraphics[totalheight=7cm]{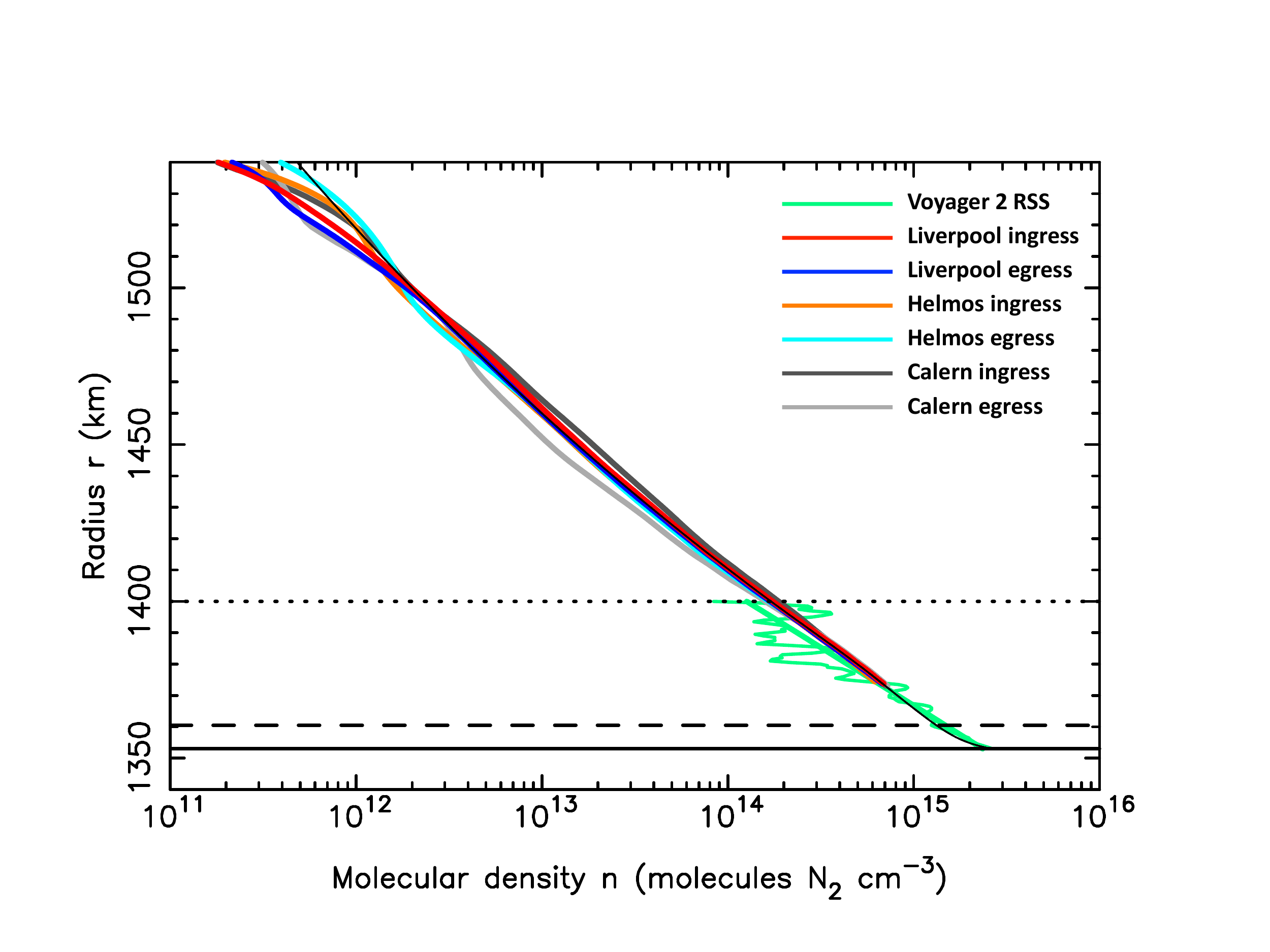}
}
\centerline{
\includegraphics[totalheight=7cm]{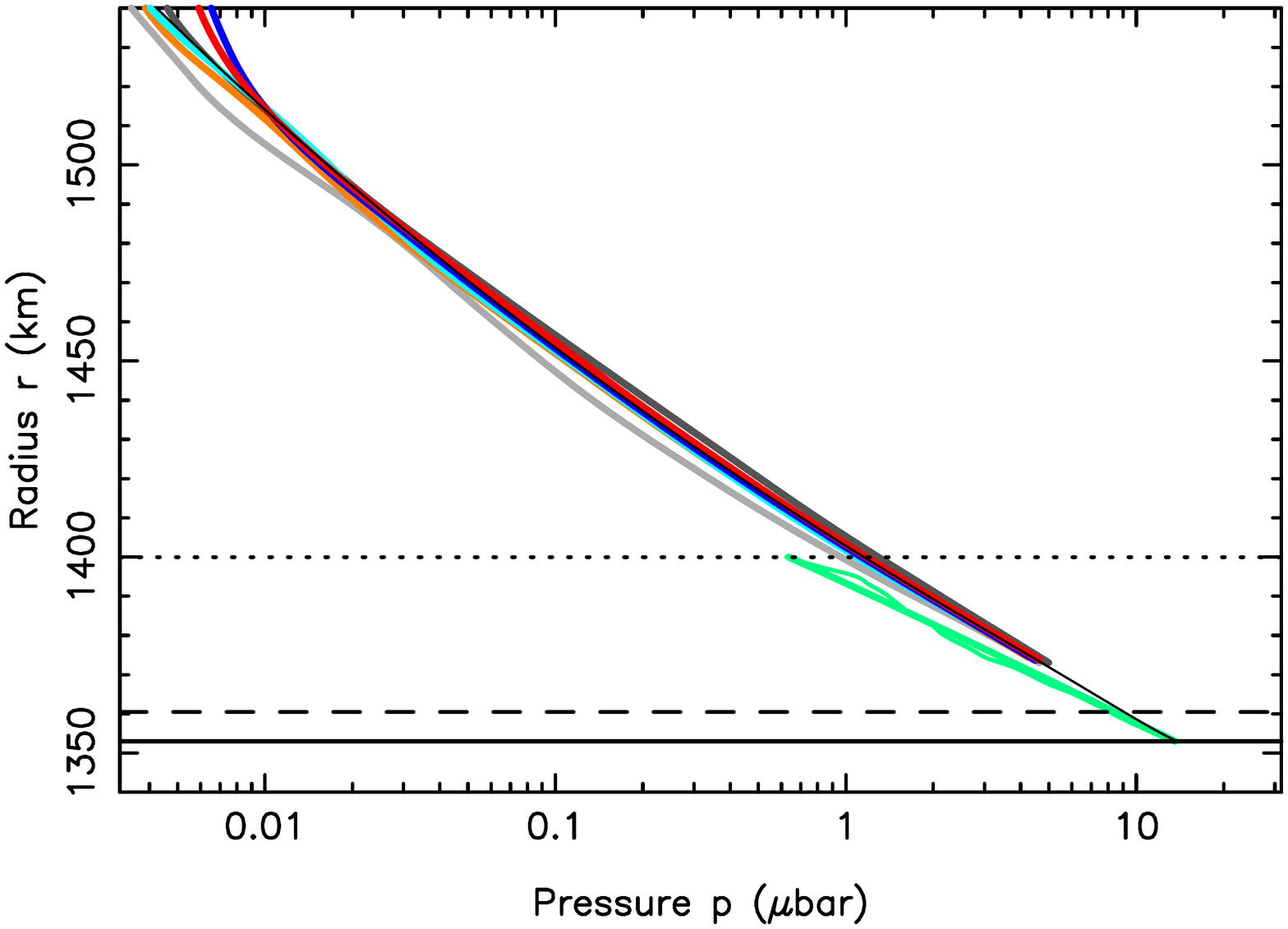}
}
\caption{
Density and pressure profiles of the atmosphere of Triton as a function of radius, $r$ (the distance to Triton's centre).
\textit{Upper panel:} 
Density profiles of Triton's atmosphere as a function of radius, 
retrieved by inverting three light curves obtained during the 5 October 2017 occultation and
from the V2 radio phase delay at 3.6 cm.
The colour codes are indicated in the upper right part of the plot.
The same codes are used in 
Figs.~\ref{fig_T_r_logp}-\ref{fig_dTdr_r} and 
Figs.~\ref{fig_n_p_r_bottom}-\ref{fig_T_logp_bottom}.
The thin black curve is a smooth synthetic density profile that fits the inverted profiles 
and is extrapolated down to the surface. 
It is derived from the smooth temperature profiles shown in Fig.~\ref{fig_T_r_logp}.
The solid horizontal line marks Triton's surface (at radius $R_T = 1353$~km), 
the dashed line indicates the central flash layer (near 1360~km), 
and the dotted horizontal line marks the reference radius, $r_{\rm ref}= 1400$~km.
\textit{Lower panel:}
Corresponding pressure profiles.
%
%
}
\label{fig_n_p_r}
\end{figure}

\begin{figure}[!t]
\centerline{
\includegraphics[totalheight=7cm, trim=0 0 0 0, angle=0]{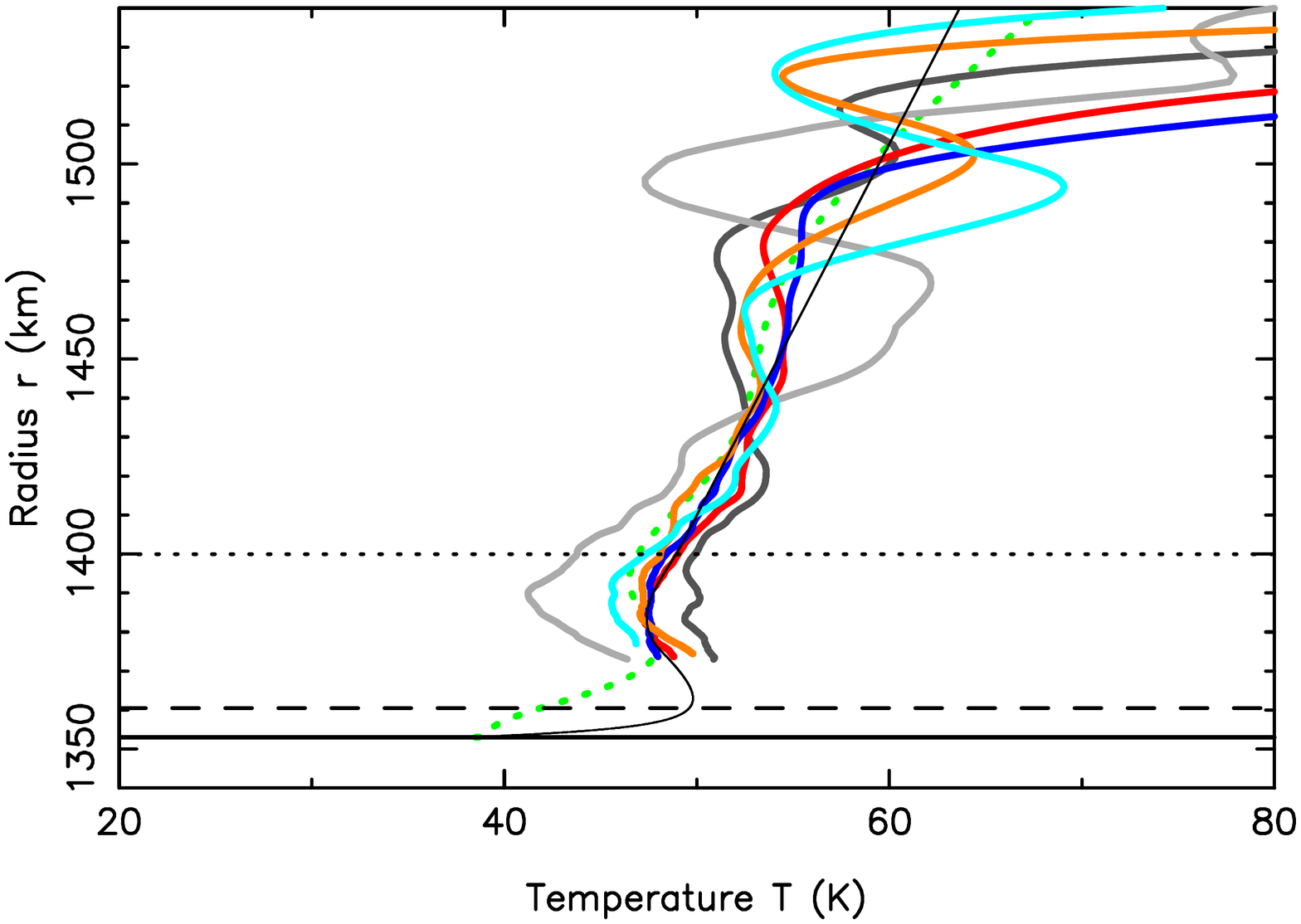}
}
\centerline{
\includegraphics[totalheight=7cm, trim=0 0 0 0, angle=0]{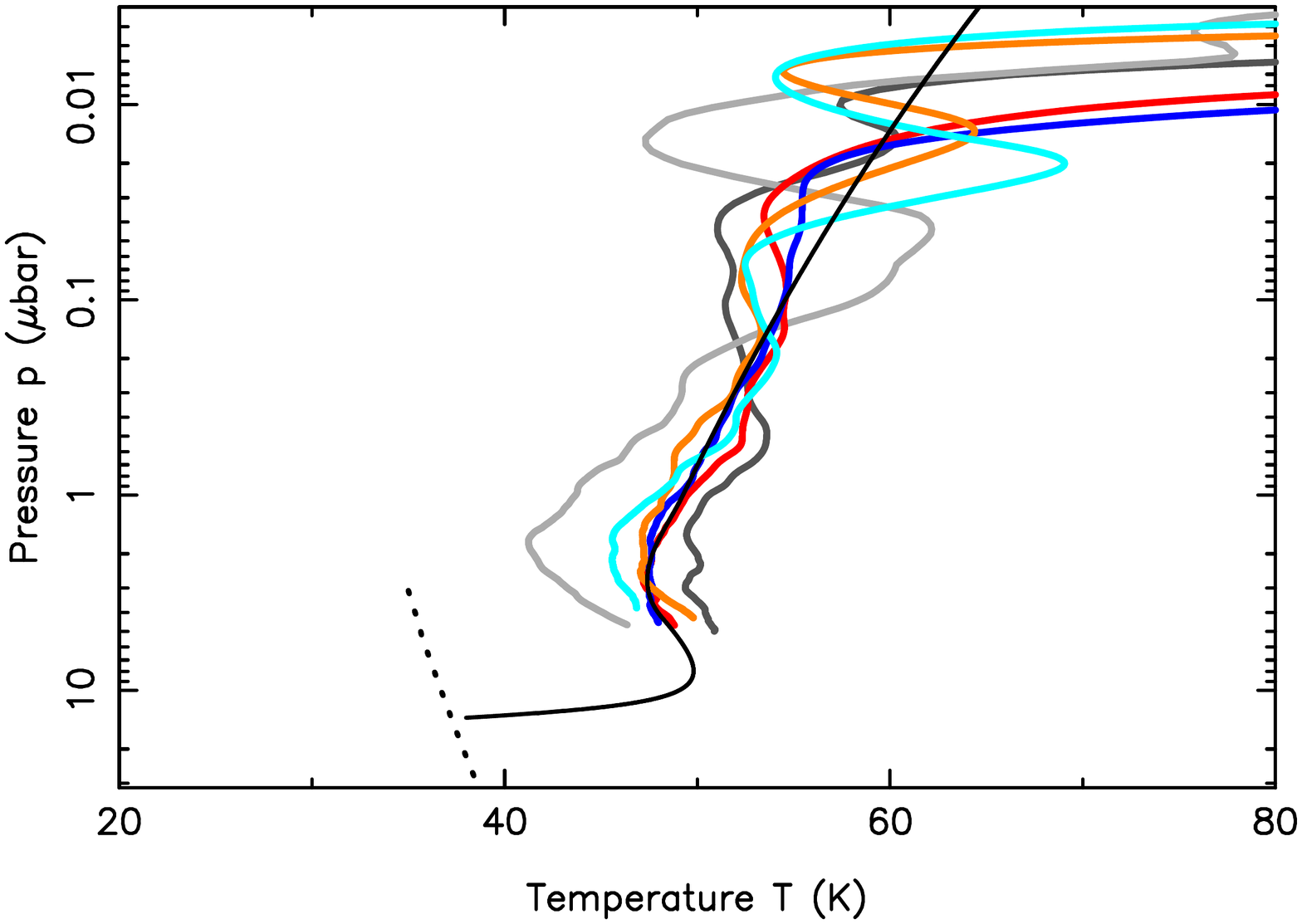}
}
\caption{
Temperature profiles as a function of radius (upper panel) and pressure (lower panel). 
The oblique dotted line in the lower panel is the wet adiabat, 
i.e. the vapour pressure equilibrium line for N$_2$, taken from \cite{fra09}.
}
\label{fig_T_r_logp}
\end{figure}

\begin{figure}[!t]
\centerline{
\includegraphics[totalheight=7cm, trim=0 0 0 0, angle=0]{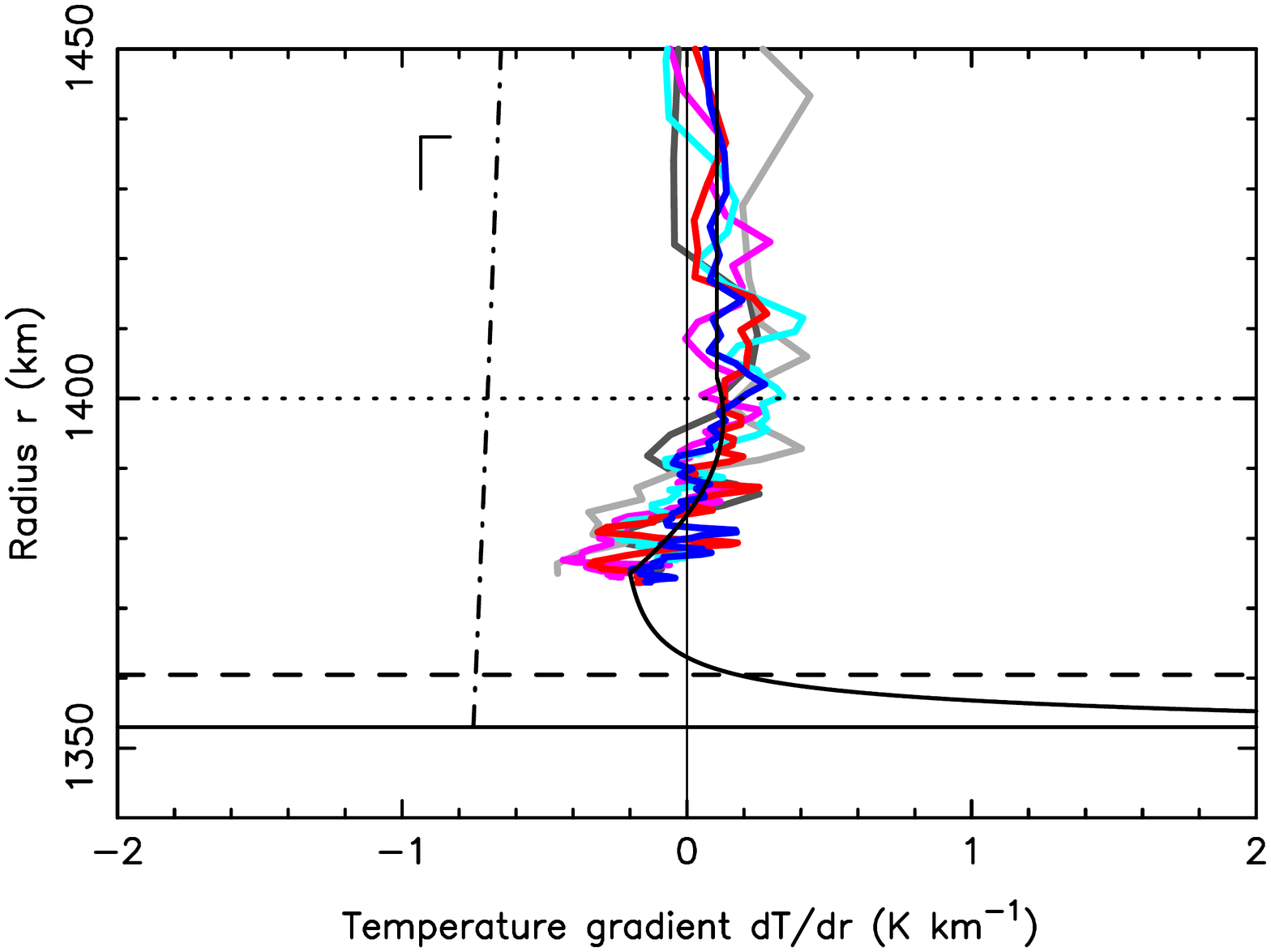}
}
\caption{
Temperature gradient corresponding to the upper panel of Fig.~\ref{fig_T_r_logp}. 
The dot-dashed line is the dry adiabatic temperature gradient $\Gamma= -g/c_p$, i.e. 
the limit of convective instability, where $c_p$ is the specific heat at constant 
pressure for N$_2$ and $g= GM_T/r^2$ is the acceleration due to gravity.
}
\label{fig_dTdr_r}
\end{figure}

The results of the inversions are displayed in Figs.~\ref{fig_n_p_r}-\ref{fig_dTdr_r}.
We also plot in these figures the profiles retrieved from our analysis of 
the RSS occultation (see Sect.~\ref{subsec_v2} for more details, as well as a discussion on the connection of these profiles with our results).
There are five noteworthy features.
First, all six $n(r)$ and $p(r)$ inverted profiles are very similar, showing that the
stations at La Palma, Helmos and Calern probed essentially identical atmospheric 
layers at their ingress and egress points. No significant variations versus local 
time and latitude are observed. 

Second, in their common range of probed altitudes, our density profiles and the 
RSS profile coincide, to within the noise level of the RSS experiment
(the noise level in our retrieved profiles being much smaller).
Third, the pressure profiles from RSS and from our inversions are also close to each other.
However, contrary to the density profile, some small differences appear. 
This is discussed in Sect.~\ref{sec_previous_occs}.

Fourth, the general positive gradient in the upper parts of our retrieved thermal 
profiles is a mere result of the choice of our initial conditions.
This is intended to match the general temperature profiles obtained independently
by \cite{str17}, which were constrained by the RSS occultation data taken in 1989 (see Appendix~\ref{sec_appen_validity}).

Fifth, all of our six retrieved thermal profiles show, however, a marked turning point 
in their deepest parts, where the temperature gradient becomes negative.
This gradient is always well below (in absolute value) the local dry adiabatic
gradient, so that the atmosphere is convectively stable in those parts
(Fig.~\ref{fig_dTdr_r}).

%
%

\subsection{Ray-tracing approach}
\label{sec_ray_tracing_results}

For all the datasets used here, we employed the same procedure as in DO15, which consists of simultaneously fitting the refractive occultation light curves with 
synthetic profiles generated by the ray-tracing code.
For each station, a least-squares fit is performed to adjust the synthetic light curve 
to the observation. Due to the uncertainties in the determination of $\phi_0$ 
(the fraction of the flux attributed to Triton; see Appendix~\ref{sec_appen_validity})
for some stations, and the lack of calibration for most of them, we considered 
$\phi_0$ as a free parameter when performing these fits. We note that this adds one degree 
of freedom per station to the fit, and thus increases the error bars on the retrieved atmospheric parameters.

Our ray-tracing method is mainly sensitive to the half-light level.
It corresponds to a radius of about 1415~km in Triton's atmosphere (altitude $\sim$60~km)
and a pressure level of $\sim$0.55~$\mu$bar. For a prescribed temperature profile $T(r)$,
this method returns two best fitting parameters.
One parameter is the pressure $p_{\rm ref}$ at a reference radius,
here $r_{\rm ref}=1400$~km. 
This particular choice stems from the fact that this reference radius has been used 
in previous works (e.g. \citealp{olk97,ell00b}), thus allowing consistent comparisons.

To proceed forward, we have defined a template $T(r)$ profile that matches
the inverted profiles obtained at the station with the best S/N (La Palma). 
It has the same functional variation with altitude as in \cite{dia15},
where it was applied to Pluto's atmosphere, 
except for the upper branch, which is not isothermal but rather has a constant
thermal gradient that connects the lower atmosphere to an upper thermosphere.
The adopted parameters for the template profile $T(r)$ are provided in Appendix~\ref{sec_appen_phi0}.

The lower part of the profile has been adjusted so as to fit the central flashes (see Appendix~\ref{sec_appen_phi0} for details).
That adjustment provides constraints on the thermal profile between the lowest inverted point 
of La Palma ($\sim$20~km altitude) down to the central flash level ($\sim$8~km altitude).
Finally, below 8~km, the profile has been connected to the surface's temperature at 38~K. 
This is a `blind part' of the profile, as it does not contribute significantly
to the refracted stellar flux received on Earth.

The resulting synthetic temperature profile is shown as a thin black line in Fig.~\ref{fig_T_r_logp}. 
Starting from the surface, the profile first has a strong positive temperature gradient
of 5~K~km$^{-1}$. This gradient decreases rapidly (Fig.~\ref{fig_dTdr_r}) and
the temperature reaches a maximum value of about 50~K at $r=1363$~km (10~km altitude), 
thus implying an average gradient of 1.2~K~km$^{-1}$ in that lower part.
Our data show a hint of a mesosphere with a negative gradient (also seen in \citealp{ell03a}) that reaches $-0.2$~K~km$^{-1}$ at $r=1375$~km (23~km altitude), 
before connecting with the general positive gradient of the upper branch.

The strong surface temperature gradient at the surface derives from the need to connect our 
inverted profiles to the surface at 38~K. 
Since we do not have information in this lower portion of the atmosphere, 
we employed the simple hyperbolic form of DO15 to connect our profile to the surface,
so that our surface gradient does not necessarily reflect the real value at that level.
 
This said, the general positive gradient can be achieved by considering the heating by CH$_4$
stemming from near IR absorbing bands.
For instance, we estimate that a CH$_4$ mixing ratio of 0.0004 yields $T = 52$ K at 1363 km,
and thus could explain our result.
\cite{str17} ran their model for discrete values of the CH$_4$ mixing ratios not included in their paper and found that a CH$_4$ surface mixing ratio $\sim$ 0.00015 would suffice to support a temperature rise $\sim$ 9 K, and a CH$_4$ surface mixing ratio $\sim$ 0.0004 a temperature rise $\sim$ 12 K, in the first 10 km. 
Because CH$_4$ is photochemically destroyed in the lower atmosphere, its scale height is roughly half the N$_2$ scale height and in terms of CH$_4$ column density one needs a higher surface CH$_4$ mixing ratio to compensate for its smaller scale height. 
For remote sensing observations it is the column density that is important and not just the surface mixing ratio that is relevant.
We note that the 0.0004 value is smaller than, but roughly consistent with the range found by 
\cite{lel10} for the CH$_4$ mixing ratio, 0.0005-0.0010. 
Moreover, some complications may arise, like the existence of a troposphere.

The troposphere on Triton has been shown to be controlled by turbulent mixing above the surface, and to be sensitive to surface thermal contrasts between N$_2$ ice and the volatile free bedrock (due to different surface albedo or thermal inertia, \citealp{van13}). 
On Pluto, climate models showed that the sublimation of cold N$_2$ ice and subsequent transport of the cold N$_2$ air in the impact basin Sputnik Planitia yield a km-thick cold troposphere as observed by New Horizons \cite{for17,hin17}.

The negative gradient in the mesosphere, reminiscent of the more extended mesosphere on Pluto \citep{lel17,you18}, calls for the existence of a coolant. It must cool the atmosphere above its peak temperature of $\sim$ 50 K, as well as radiate away the downward thermal heat flux from the upper atmosphere where $T \sim 100$ K. There are a few candidates for this coolant: haze particles and/or influx of dust particles that may either be pure H$_2$O ice or with silicate cores and coated with H$_2$O ice (see \citealp{ohno20} for more details).

The pressure at any level can then be deduced by using the temperature template described above.
In particular, the surface pressure $p_{\rm surf}$ can be obtained by
the relation $p_{\rm surf}= 12.0 \times p_{1400}$.
This ratio will be used to extrapolate $p_{1400}$ from $p_{\rm surf}$ or vice versa.
The other fitted parameter is Triton's DE435/NEP081 ephemeris offset perpendicular 
to its apparent motion projected in the sky, $\Delta \rho$. 
We note that the ephemeris offset along Triton's motion is decoupled from that fit (see DO15 for details).

Error bars are obtained from the classical function 
$\chi^2~=~\sum_1^N~[(\phi_{\rm i,obs}-\phi_{\rm i,syn})/\sigma_i]^2$,
which reflects the noise level $\sigma_i$ of each of the $N$ data points, 
where $\phi_{\rm i,obs}$ and $\phi_{\rm i,syn}$ are the observed and 
synthetic fluxes at the $i^{\rm th}$ data point, respectively.

We simultaneously fitted a selected pool of 52 light curves
obtained during the 5 October 2017 occultation.
Other light curves were not considered at this stage because they are
affected by higher or non-normal noise that would degrade the global fit.
In a first step, we excluded from the fit the parts of the light curves
where a strong central flash is present. 
This is to avoid giving too much weight to those parts, 
while they reflect only the properties of the deepest atmospheric layers.
So, the goal of this fit is to get the global properties of the atmosphere,
and in particular, to constrain $p_{1400}$
and the location of the shadow centre with respect to the occultation chords.
In a second step, we included the central flashes in the fit to assess 
the shape of Triton's atmosphere and to check if the central flash location 
coincides with the centre found by the global approach.

\begin{figure}[!t]
\centerline{
\includegraphics[width=10cm]{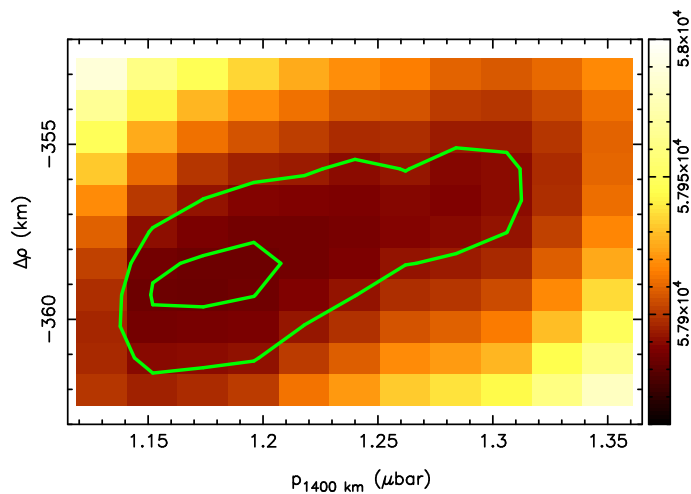}
}
\caption{
 $\chi^2$ map for the simultaneous fit of 52 light curves 
obtained during the occultation of 5 October 2017, using the $\phi_0$ corresponding to the temperature profile in Fig. \ref{fig_T_r_logp}.
The inner green line contour is the 1$\sigma$ limit of the fit,
while the outer green line is the 3$\sigma$ limit.
}
\label{fig_chi2_map_05oct17}
\end{figure}
%
After exploring a grid of values for $\Delta \rho$ and $p_{1400}$,
we obtained the $\chi^{2}(\Delta \rho,p_{1400})$ map displayed in Fig.~\ref{fig_chi2_map_05oct17}.
A satisfactory fit should provide a minimum value $\chi^2_{\rm min}$ 
close to $N-M$, where $M$ is the number of fitted parameters. 
Besides $p_{1400}$ and $\Delta \rho$, we considered Triton's contribution
to the light curve ($\phi_0$) as a free parameter for all light curves. 
This is because no satisfactory values of $\phi_0$ have been obtained for any
of the light curves (see Appendix~\ref{sec_appen_phi0}).
Thus, the fitted parameters are the values of $\Delta \rho$ and $p_{1400}$,
plus the 52~values of the $\phi_0$ (i.e. a total of $M=54$ fitted parameters).
On the other hand, we used $N = 68446$~data points.
We then obtain a global value of $\chi^{2}$ per degree of freedom,
$\chi^{2}_{\rm dof}= \chi^{2}_{\rm min}/(N-M)= 0.85$, indicating a satisfactory 
global fit to the data.
An examination of values of $\chi^{2}_{\rm dof}$ for individual light curves 
also show values near unity for all of them. 
Thus, none of our light curves show significant discrepancies when compared 
to the synthetic light curves derived from the synthetic density model shown in 
Fig.~\ref{fig_n_p_r}.
This confirms the spherical symmetry of Triton's atmosphere on a global scale.

Without considering $\Delta \rho$, the marginal 
distribution\footnote{The marginal distribution is used when we wish to find the probability 
of specific variables of a subset without consideration of other variables.}
for 1$\sigma$ and 3$\sigma$ error contours on $p_{1400}$ are estimated 
by tracing the iso-levels $\chi^{2}_{\rm min}+1$ and $\chi^{2}_{\rm min}+9$, respectively,
as shown in Fig.~\ref{fig_chi2_map_05oct17}. 
The best-fitting value of $p_{1400}$, its 1$\sigma$ error bar and the quality of the fit, 
$\chi^{2}_{\rm dof}$, are listed in Table~\ref{tab_pressure_time}.
The best-fitting value of $\Delta \rho$ (-359.3$\pm$1~km, Fig.~\ref{fig_chi2_map_05oct17})
is used to retrieve the closest geocentric approach distance between Triton and the star
(projected in the sky plane) and its corresponding time (see Table~\ref{tab:OccultationCircumstances}).

Finally, the best simultaneous fit corresponding to the minimum of $\chi^{2}$
is displayed in Figs.~\ref{fig_fit_data_1}-\ref{fig_fit_data_5}.
For the sake of completeness, Figs.~\ref{fig_no_fit_data_1}-\ref{fig_no_fit_data_4} show the
synthetic light curves superimposed on the light curves that have not been included in the fit. Although they have poorer S/N, 
they all confirm that our global model satisfactorily fits these data.

\section{Reanalysis of results from previous events}
\label{sec_previous_occs}

\subsection{The Voyager 2 radio occultation}
\label{subsec_v2}

During its Triton flyby on the 25 August 1989, the V2 spacecraft sent its radio signal (RSS experiment) back to Earth as it passed behind the satellite. Details on the gathering of the V2 RSS data are given in G95.

The main product of this observation was the temperature and pressure at Triton's surface. However, although it becomes quite noisy above the 20 km altitude level, the RSS phase delay still provides useful constraints on Triton's lower atmosphere, with some science left to explore.
Here, we give a summary of Gurrola's work, and describe how we use the V2 phase delay to retrieve Triton's atmospheric structure in the 10-20~km above the surface.

The V2 high gain 3.7-m antenna transmitted to Earth
two radio signals at 3.6 and 13~cm (X band and S band, respectively).
The phases in these bands are related to one another by
\begin{equation}
\Delta \phi = \frac{121}{112} (\phi_{x} - \frac{3}{11} \phi_{s}),
\label{eq_phi_corr}
\end{equation}
where $\phi_{x}$ is the phase in the X band, $\phi_{s}$ 
is the phase in the S band, and $\Delta \phi$ is the corrected radio 
phase corresponding to the neutral atmosphere at the X-band
wavelength. This calculation is done to remove
plasma effects on the phase.
Due to problems in fitting the ingress data, as there seemed to be sudden changes in slope, G95 used only the egress
data for his analysis.

\begin{figure*}[t!]
\centering
\centerline{
\includegraphics[width=70mm]{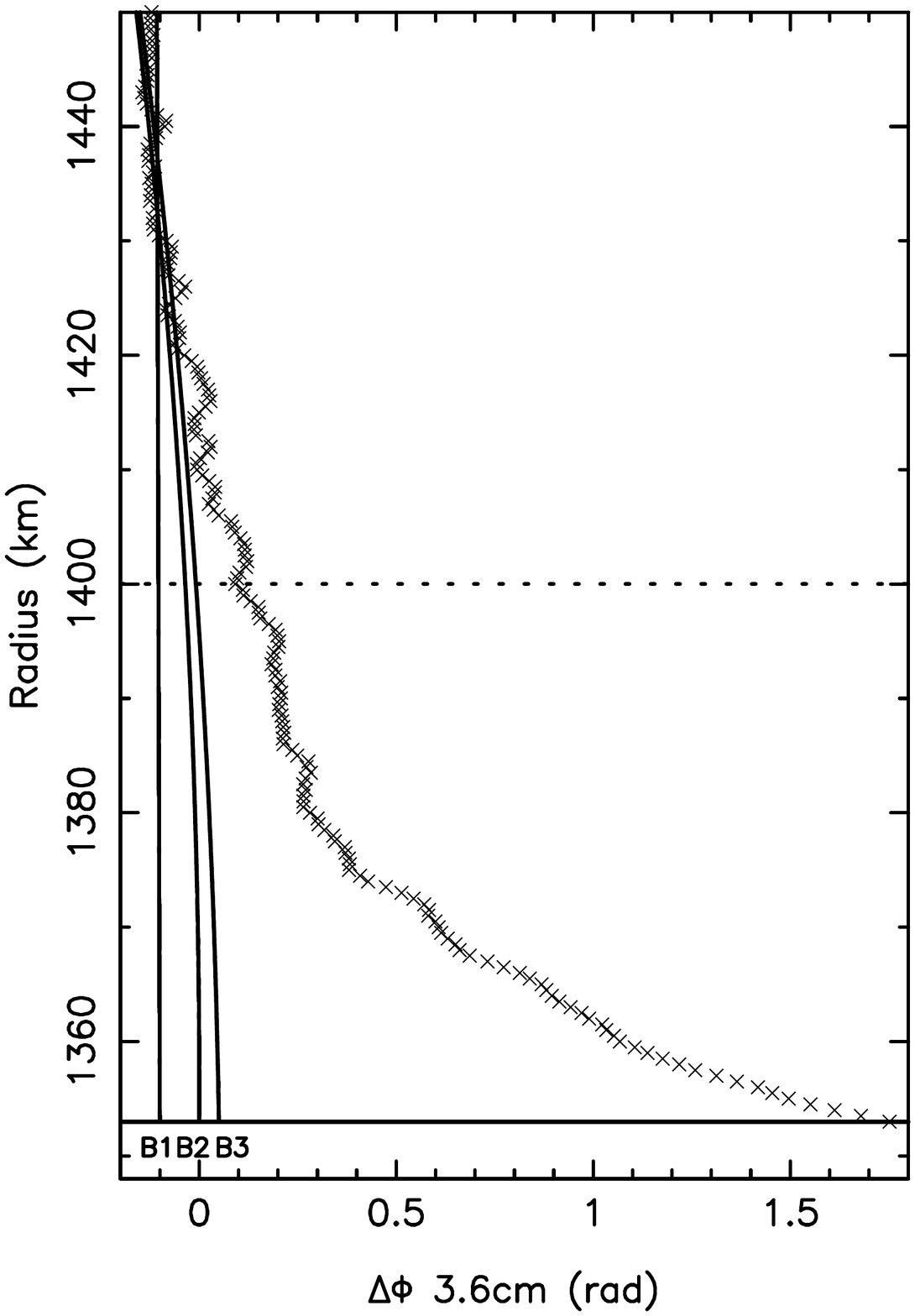}
\includegraphics[width=70mm]{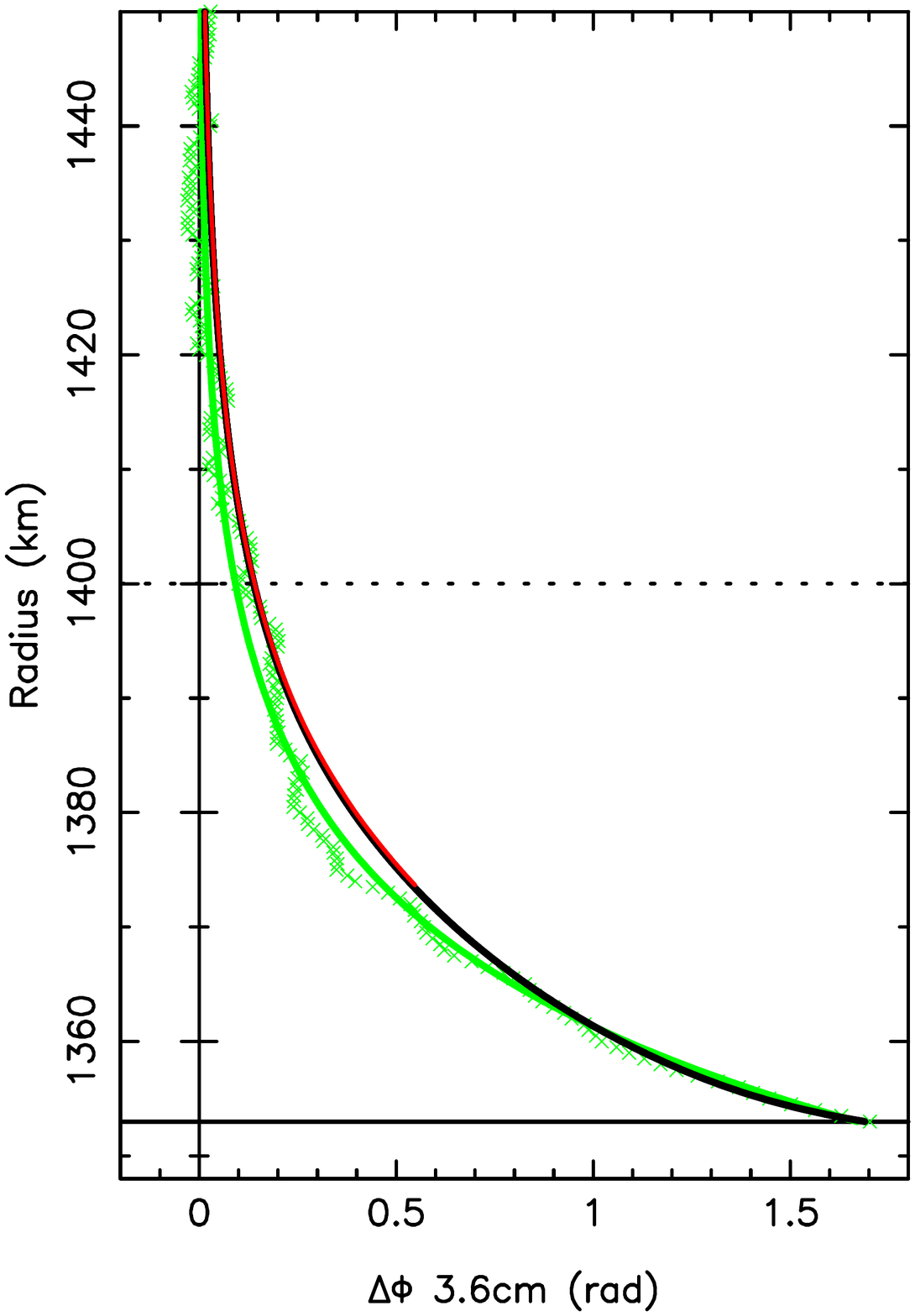}
}
\caption{Radio phase delay during the egress of the V2 RSS occultation on 25 August 1989 in the 3.6 cm X band.
\textit{Left panel:}
Radio phase delay observed (adapted from G95).
The crosses are the data, and the three solid lines (labelled B$_1$, B$_2$, and B$_3$) 
are three polynomial modellings of the phase delay baseline.
As discussed by G95, the preferred baseline solution is B$_2$. 
\textit{Right panel:}
Radio phase after subtraction of the B$_2$ baseline polynomial 
shown in the left panel (green crosses), thus representing the effect of the atmosphere only.
The green line marks
the smooth version of that radio phase delay, as constructed by G95.
The red profile is the phase delay that would be observed at 3.6 cm
from the retrieved density profile of La Palma at immersion (see Fig.~\ref{fig_n_p_r}).
Other phase delay profiles obtained from La Palma (emersion) and Helmos would be
indistinguishable from the red profile and are not plotted here for sake of clarity.
The black profile is the phase delay obtained from our best model of Triton's atmosphere (see text for details).
}
\label{fig_phase_delay_V2}
\end{figure*}

Gurrola provided us with the corrected phase delay 
$\Delta \phi (r)$ versus altitude above Triton's surface, 
as well as the results from his models to obtain only the 
`pure atmosphere phase delay'.
This corresponds to the phase delay once a general
polynomial trend has been subtracted from $\Delta \phi$ 
to account for thermal noise and instabilities in the frequency reference on board V2. 
These polynomials, referred to as baselines in G95, were designated as B$_1$, B$_2$, and B$_3$. 
B$_1$ is the linear baseline used by \cite{tyl89}, determined using 120~km of the data obtained. G95 considered this insufficient to reliably estimate the drift of the instrument over the atmosphere, as it did not extrapolate from high enough altitudes (so that the atmosphere is too thin to affect the signal phase) downwards towards Triton's surface.
On the other hand, baselines B$_2$ and B$_3$ used about 700~km of the data, and are, respectively, the second and third-order polynomials of G95's best fit at egress. The preferred solution of this author is B$_2$.
The resulting $\Delta \phi (r)$ is displayed in green 
in Fig.~\ref{fig_phase_delay_V2}. 

Using the B$_2$ solution, we derive the profiles displayed as green curves in Fig.~\ref{fig_n_p_r}.
In order to compare this result to ours, we generated for
comparison the phase delay at 3.6 cm that would be observed with
our best profiles $n(r)$ (the black line in Fig.~\ref{fig_n_p_r})
as if it were obtained by V2:
\begin{equation}
\Delta \phi(r) = \frac{2\pi}{\lambda} K \sigma_{\rm N_{2}}(r),
\label{eq_phase_shift}
\end{equation}
where 
$\lambda$ is the wavelength (3.6 cm), 
$K$ is the corresponding molecular refractivity of N$_2$ 
(see Table~\ref{tab_param}), and
$\sigma_{\rm N_{2}}(r)$ is now the column density stemming 
from our best model.
The resulting $\Delta \phi(r)$ profile is shown in black 
in Fig.~\ref{fig_phase_delay_V2}, together with the phase
delays deduced from the inversions of La Palma and
Helmos' light curves (in colours).

Conversely, we used the V2 corrected X-band radio phase to
retrieve the refractivity profile from the Abel inversion
\begin{equation}
\nu(r) = 
-\frac{\lambda}{2\pi^2}
\int_{r}^{+\infty} \frac{d (\Delta \phi)}{dR}
\frac{dR}{\sqrt{R^2-r^2}}=
-\frac{\lambda}{2\pi^2}
\int_{0}^{+\infty} \frac{1}{R} \frac{d \Delta \phi}{dR} dl,
\label{eq_Abel_inv}
\end{equation}
using the auxiliary variable $l=\sqrt{R^2-r^2}$ to calculate the integral.
Finally, the density profile $n(r)= \nu(r)/K$ can be deduced.


With this, we can directly compare our results to those of V2. We note that the RSS profiles probe altitude interval levels that overlap our ground-based occultation levels.
This overlapping region extends from the lowest inverted points of the 
La Palma station, $r = 1373$~km (20~km altitude), 
up to roughly the reference level, $r = 1400$~km (47~km altitude), 
at which point the RSS profiles become too noisy to be reliable, reaching a factor of about 2 at that level.

\begin{figure}
\centering
\includegraphics[width=90mm]{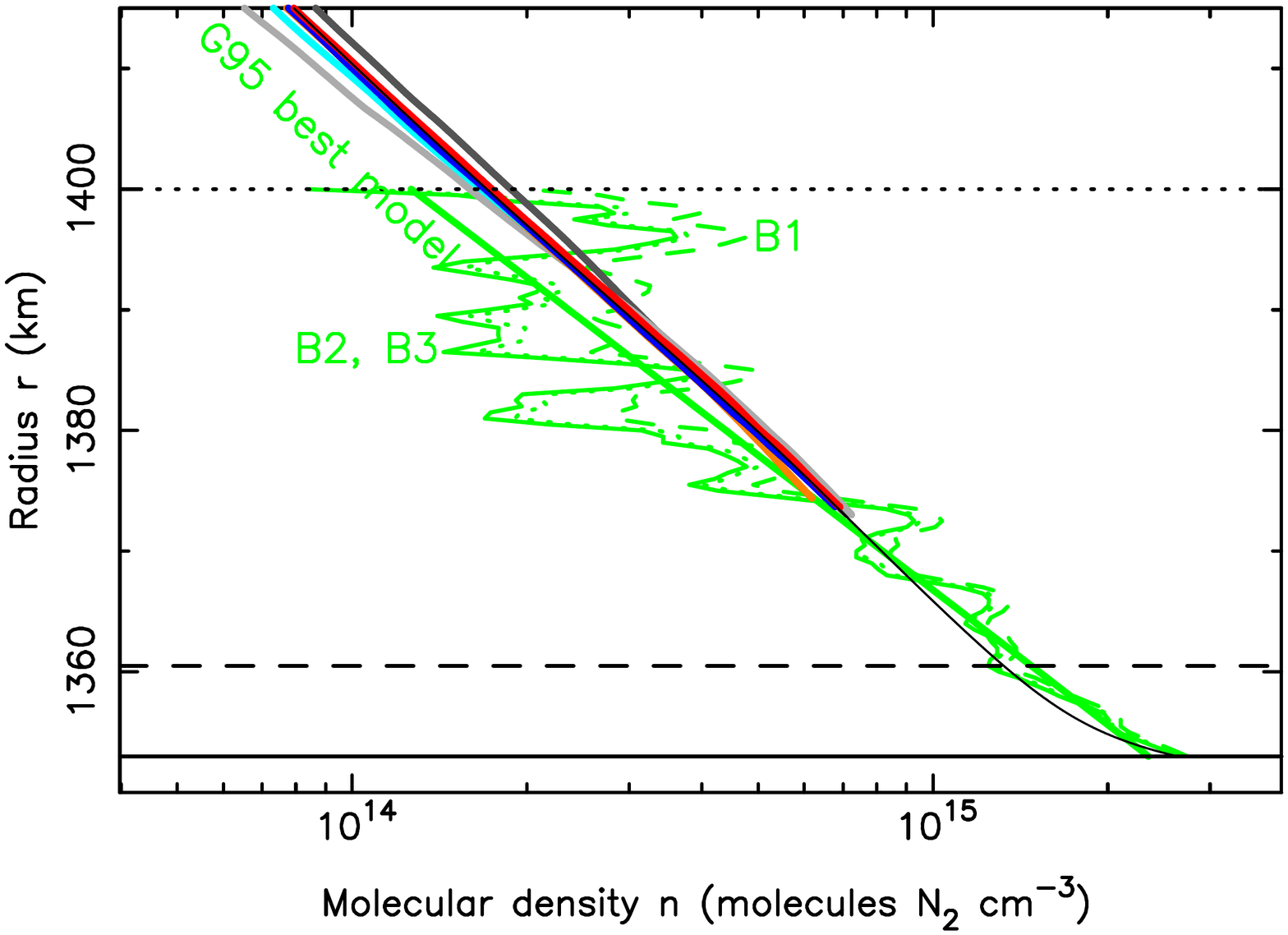}
\includegraphics[width=90mm]{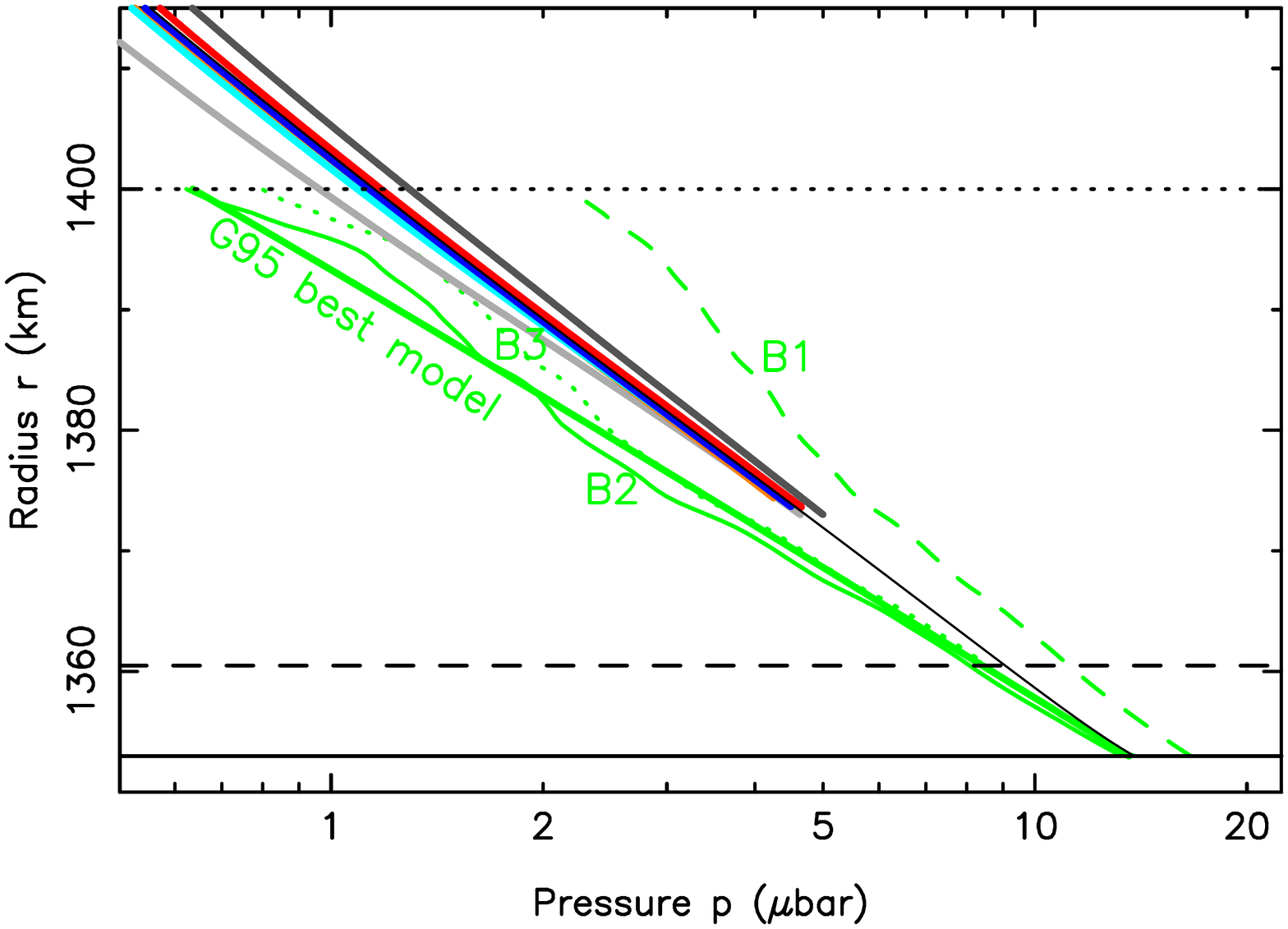}
\caption{Close-up view of Fig.~\ref{fig_n_p_r}.
\textit{Upper panel:} 
Close-up view of the density profiles.
Various profiles derived from the RSS occultation are shown in green.
The dashed line is
the profile retrieved by \cite{tyl89} using a polynomial extrapolation B$_1$ 
to correct for the RSS phase instability.
The thin solid line is
the profile retrieved by G95 using the polynomial extrapolation B$_2$.
The dotted line is 
the same by G95, but using the polynomial extrapolation B$_3$.
The thick solid line is the best model of G95, based on the B$_2$ profile.
\textit{Lower panel:} 
Same for the pressure profiles.
}
\label{fig_n_p_r_bottom}
\end{figure}

The examination of the upper panel of Fig.~\ref{fig_n_p_r_bottom} shows that 
no significant difference in density is detected between the 1989 and 2017 profiles,
especially at the `junction level' at $r = 1373$~km.
We note that the RSS density profile is rather insensitive to the particular solution 
B$_1$, B$_2$, or B$_3$ chosen to retrieve $n(r)$.

Integrating the weight of the atmospheric column provides 
the RSS pressure profile (lower panel of Fig.~\ref{fig_n_p_r}).
However, this profile includes a contribution of the weight of the layers 
above $r = 1400$~km, where the RSS phase delay is very noisy.
We note that the B$_1$ pressure profile is quite offset in slope with respect to 
the B$_2$ and B$_3$ solutions (lower panel of Fig.~\ref{fig_n_p_r_bottom}).
%
%
So, and contrarily to the density profiles, the general slope of
$\log_{10}(p)$ versus radius $r$ is quite sensitive to the particular choice of the 
polynomial baseline.
In this context, it is difficult to conclude if the break in slope between
the RSS profile and our ground-based results of 2017 is real or not.
%

Using the preferred B$_2$ model, and using the Abel inversion (Eq.~\ref{eq_Abel_inv}), 
we obtain $p_{\rm surf,RSS}= 13.6$~$\mu$bar and $p_{\rm 1373,RSS}= 3.77$~$\mu$bar 
as of 1989.
The main result of the inversion is the density profile; to translate this into
pressure, we need an estimate of the surface temperature $T_{\rm surf}$. 
The error bar on $p_{\rm surf,RSS}$ caused by the uncertainties on $T_{\rm surf}$
will be discussed next.

%
In any instance, our results are consistent with the analysis of G95, 
$p_{\rm surf,RSS}= 14 \pm 2$~$\mu$bar.
This is also fully consistent with our estimation of the surface pressure as of 2017, 
$14.1 \pm 0.3$~$\mu$bar (Table~\ref{tab_pressure_time}). 
Thus, no significant variations of surface pressure is found
when comparing the RSS results of 1989 and the results derived from 
the ground-based occultation of 2017.

The value $p_{\rm 1373,RSS}= 3.77$~$\mu$bar that we find is 21\% smaller
than the value we obtained in 2017 at that level, $4.58$~$\mu$bar.
Propagating this 21\% difference to the 1400~km radius then yields 
$p_{\rm 1400,RSS}= 0.97$~$\mu$bar.
Estimating the error bar on that value is difficult, 
because the RSS pressure depends on the (noisy) pressure values obtained above,
as mentioned earlier. 
If we adopt the error bar $p_{\rm surf,RSS}= 14 \pm 2$~$\mu$bar of G95
and propagate it upwards, this yields $p_{\rm 1400,RSS}= 0.97 \pm 0.14$~$\mu$bar.

Another, more robust way to estimate $p_{\rm surf,RSS}$ is to use 
the RSS density profile alone.
The counterpart is that we need an independent measurement of the surface temperature 
$T_{\rm surf}$ in order to derive the pressure from the ideal gas equation 
$p_{\rm surf,RSS}~=~n_{\rm surf,RSS} k_B T_{\rm surf}$.
These temperature measurements (given below) are more accurate than G95's estimation 
($T_{\rm surf}~=~42~\pm~8$~K) and thus reduce the $\pm 2$~$\mu$bar 
uncertainty of G95's value of $p_{\rm surf,RSS}$.
However, this approach is valid only if these temperature measurements
apply to the N$_2$ ice surface that the RSS experiment probed, and if
the vapour pressure equilibrium between the N$_2$ ice and the gas is achieved.

\begin{figure}
\centering
\includegraphics[width=90mm]{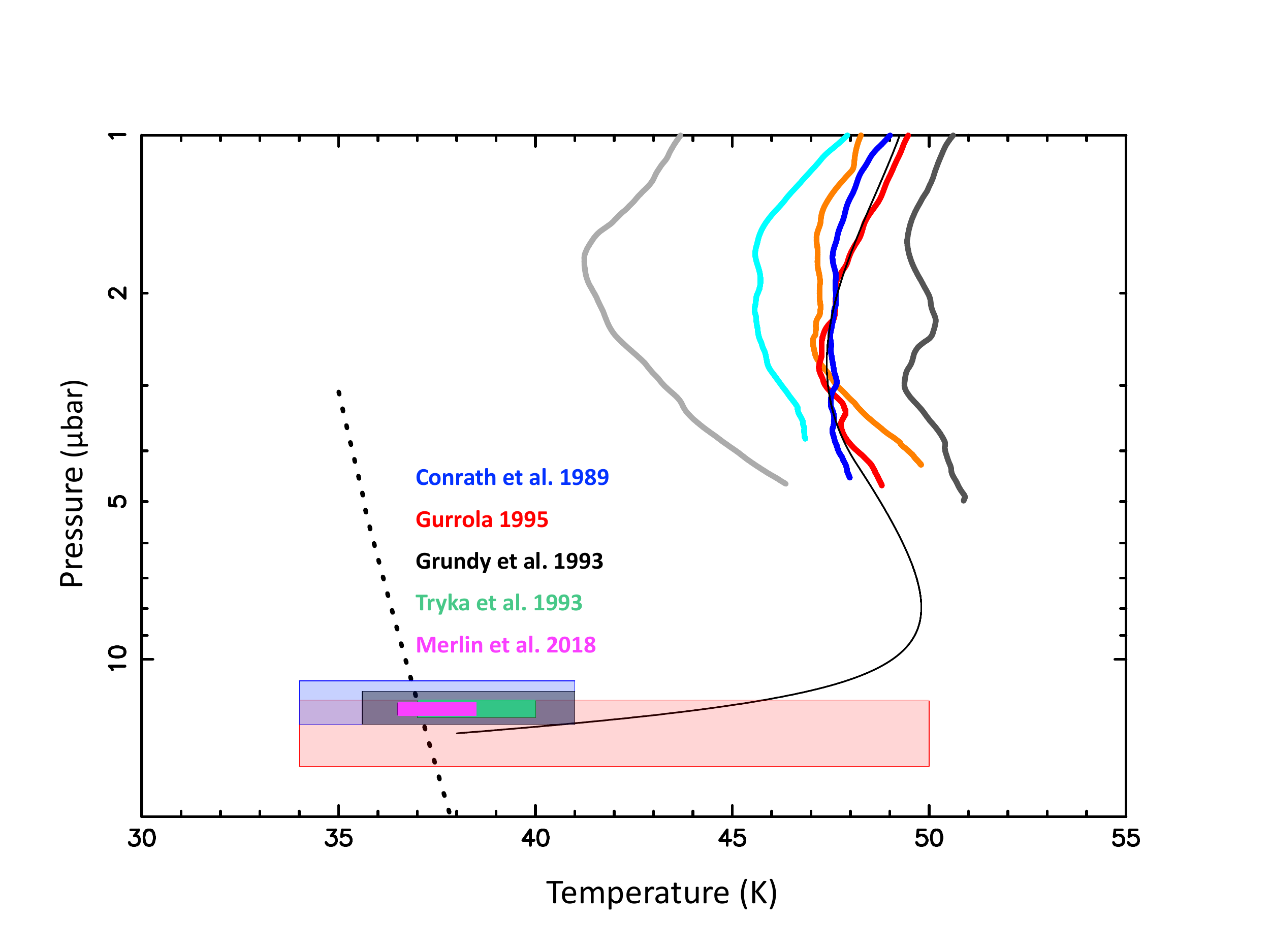}
\caption{
Close-up view of the lower panel of Fig.~\ref{fig_T_r_logp}.
The width of each coloured box is Triton's surface temperature ($T_{\rm surf}$),
as estimated by the various authors mentioned just above the boxes (see text for details).
The heights of the boxes are the range of surface pressures, $p_{\rm surf}$, 
using the ideal gas law $p_{\rm surf}= n_{\rm surf, RSS} k_B T_{\rm surf}$,
where $n_{\rm surf, RSS}= 2.4 \times 10^{15}$~cm$^{-3}$ is the surface molecular
nitrogen density derived from our inversion of the RSS data (see text).
We note that all the boxes intersect the vapour pressure equilibrium, which is plotted as a dotted line.
This shows that the RSS surface density and the estimated surface temperatures
are mutually consistent with a pressure being controlled by the N$_2$ ice sublimation.
}
\label{fig_T_logp_bottom}
\end{figure}

Estimations of $T_{\rm surf}$ are given by various authors
(see also Fig.~\ref{fig_T_logp_bottom}): 
$38^{+3}_{-4}$~K (\citealp{con89}),
$38^{+2}_{-1}$~K (\citealp{try93}),
the range 36.5-41~K (\citealp{gru93}) and
$37.5 \pm 1$~K (\citealp{mer18}).
%
Adopting a value of $n_{\rm surf, RSS}= 2.4 \times 10^{15}$~cm$^{-3}$ derived from our 
RSS phase delay inversion (Fig.~\ref{fig_n_p_r_bottom}), we find surface pressures of
$12.3^{+1.0}_{-1.3}$~$\mu$bar, 
$12.3^{+0.6}_{-0.3}$~$\mu$bar,
a range $11.5~-~13.3$~$\mu$bar and 
$12.4 \pm 0.3$~$\mu$bar,
respectively, for the four choices of surface temperatures.
We note that all these values are consistent with the surface being in vapour pressure 
equilibrium with the atmosphere, as shown in Fig.~\ref{fig_T_logp_bottom}.
This supports the hypothesis that the reported temperatures are indeed representative
of the N$_2$ ice surface.
 
In summary, we estimate from the surface temperatures given above,
a safe surface pressure range of $12.5~\pm~0.5$~$\mu$bar can be derived at the V2 epoch.
In this case, the error bar essentially stems from the uncertainties on the temperatures.
Comparing this value with our estimation $p_{\rm surf}=~14.1~\pm 0.4~\mu$bar in 2017 (Table~\ref{tab_pressure_time}),
and assuming a constant factor (12.5/14.1) throughout the profile, 
we formally obtain $p_{1400}=~1.05~\pm~0.04$~$\mu$bar in 1989. 
This is consistent with our estimate made above, $p_{1400,RSS}=~0.97~\pm~0.14$~$\mu$bar.
We thus estimate a conservative range of 
$p_{1400,RSS}~=~1.0~\pm~0.2$~$\mu$bar for the pressure at 1400~km in 1989.

\subsection{The 18 July 1997 stellar occultation}

\begin{figure}[!t]
\centerline{%
\includegraphics[width=8cm,trim=0 0 0 0,angle=0]{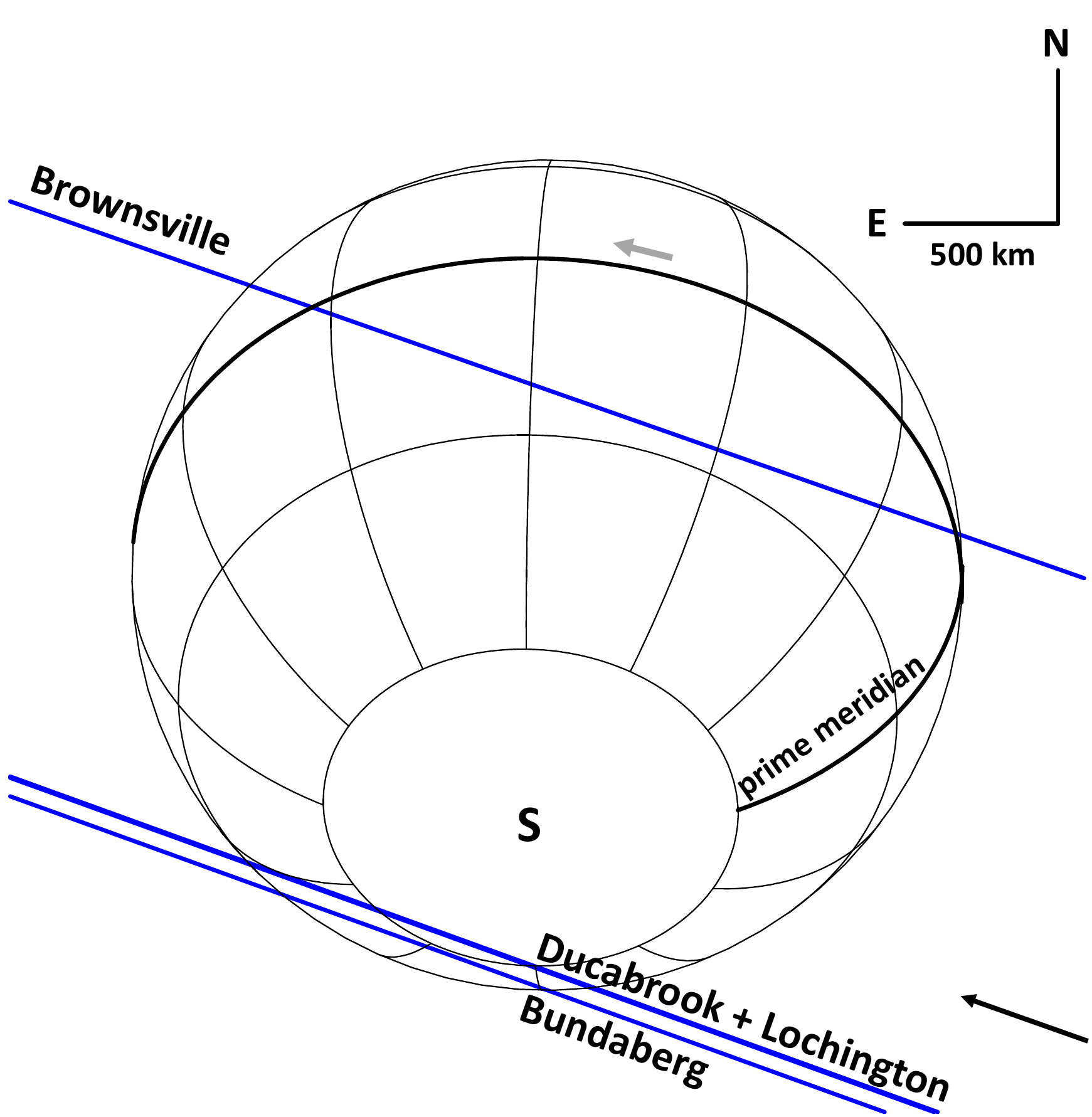}
}%
\caption{
Geometry of the 18 July 1997 occultation, 
with the same conventions as in Fig.~\ref{fig_chords}.
%
}
\label{fig_18jul97_geometry}
\end{figure}

This campaign involved one station in the USA and three stations in Australia. It was a joint effort between two groups, and, therefore, both have access to the data.
The circumstances of observations are listed in Table~\ref{tab_sites} and 
the geometry of the event is displayed in Fig.~\ref{fig_18jul97_geometry}.
More details on these observations and their analysis are given in \cite{ell00}.

\begin{figure}[!t]
\centerline{%
\includegraphics[width=10cm]{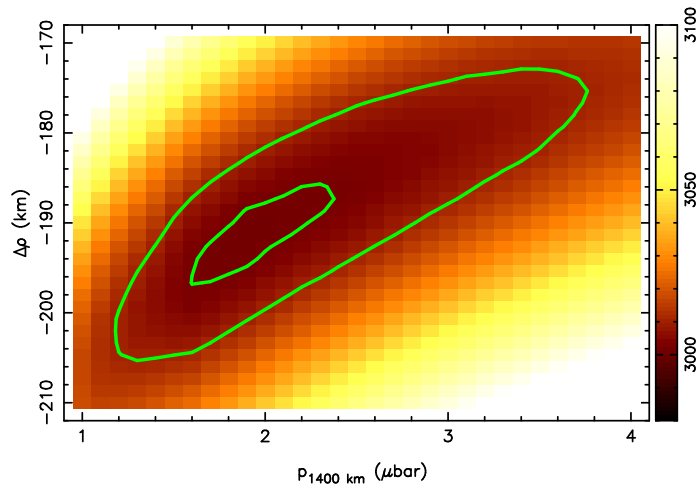}
}%
\caption{
Same as Fig.~\ref{fig_chi2_map_05oct17}, but for the 18 July 1997 occultation.
}
\label{fig_chi2_map_18jul97}
\end{figure}

\begin{figure}[!t]
\centerline{%
\includegraphics[width=7cm,trim=0 0 0 0,angle=0]{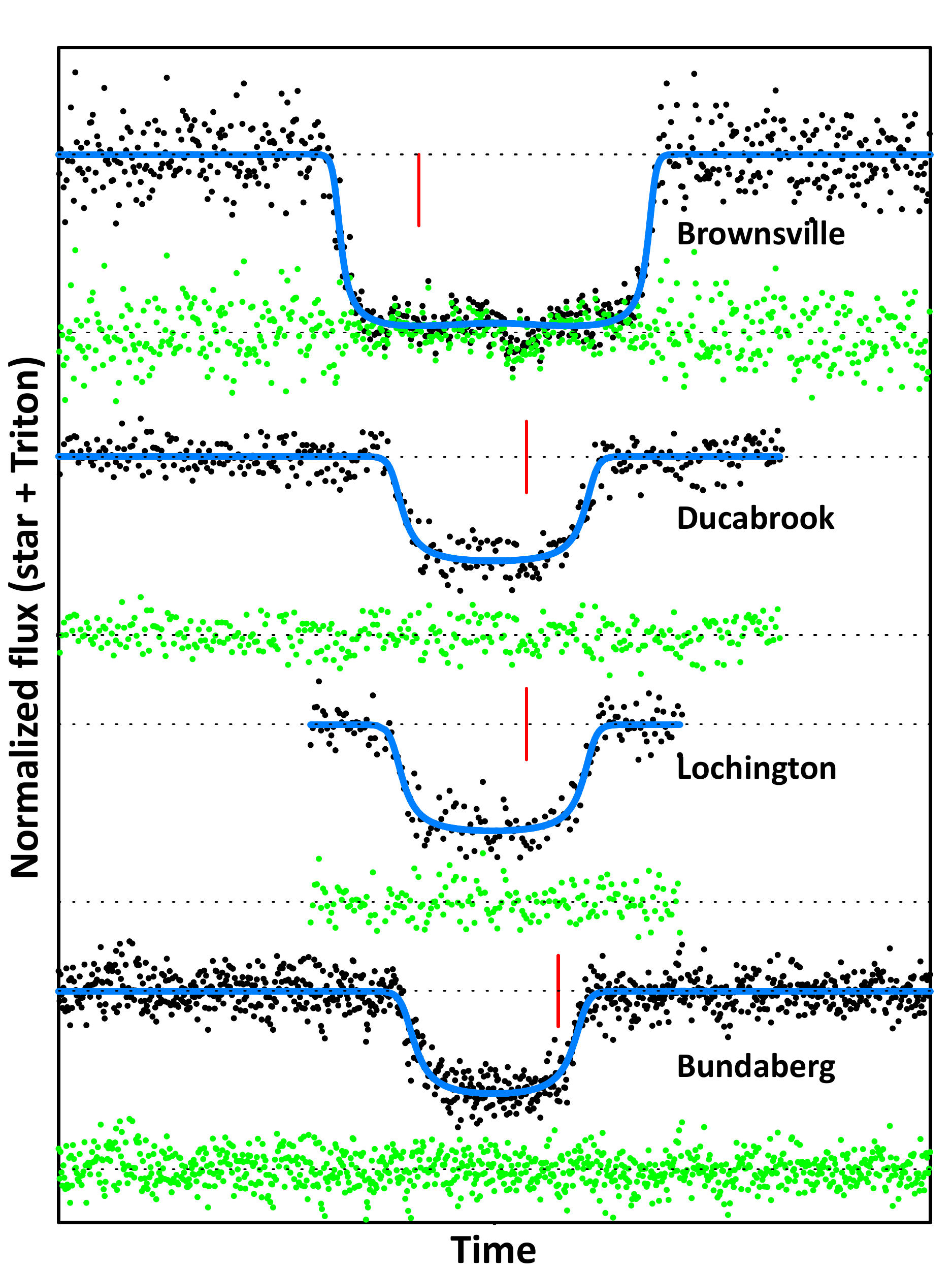}
}%
\caption{
Simultaneous fits to the 18 July 1997 light curves.
The panel covers 300 seconds in time.
All the light curves have been shifted so that 
the mid-occultation times are aligned.
The red vertical tick marks indicate 10:10~UTC at the Brownsville
station (USA) and 10:18~UTC for the three Australian stations.
The blue lines are simultaneous fits to the data (black dots),
using the best value found in Fig.~\ref{fig_chi2_map_18jul97},
$p_{1400}~=~1.9$~$\mu$bar, and the temperature profile shown
in Fig.~\ref{fig_T_r_logp}.
The green dots are the fit residuals.
For each light curve, the upper dotted line is the normalised
value of the star plus Triton flux, and the lower dotted line is
the background flux.
We note that the data from the Brownsville station were normalised during the event, as shown in \citealp{ell00}.
}
\label{fig_fit_18jul97}
\end{figure}

Here we provide the results of our own approach to constrain $p_{1400}$.
In particular, we adopt the same temperature profile $T(r)$ as for 2017 
(see Fig.~\ref{fig_T_r_logp}), but varying $p_{1400}$ to fit the synthetic
light curves to the data.
The $(\chi^2, \Delta \rho)$ map is displayed in Fig.~\ref{fig_chi2_map_18jul97} and the
best fit is shown in Fig.~\ref{fig_fit_18jul97}.
This yields $p_{1400}= 1.90^{+0.45}_{-0.30}$~$\mu$bar
and $\chi^2_{\rm dof}= 0.95$, indicating a satisfactory fit.

\subsection{The 21 May 2008 stellar occultation}

\begin{figure}[!t]
\centerline{%
\includegraphics[width=8cm,trim=0 0 0 0,angle=0]{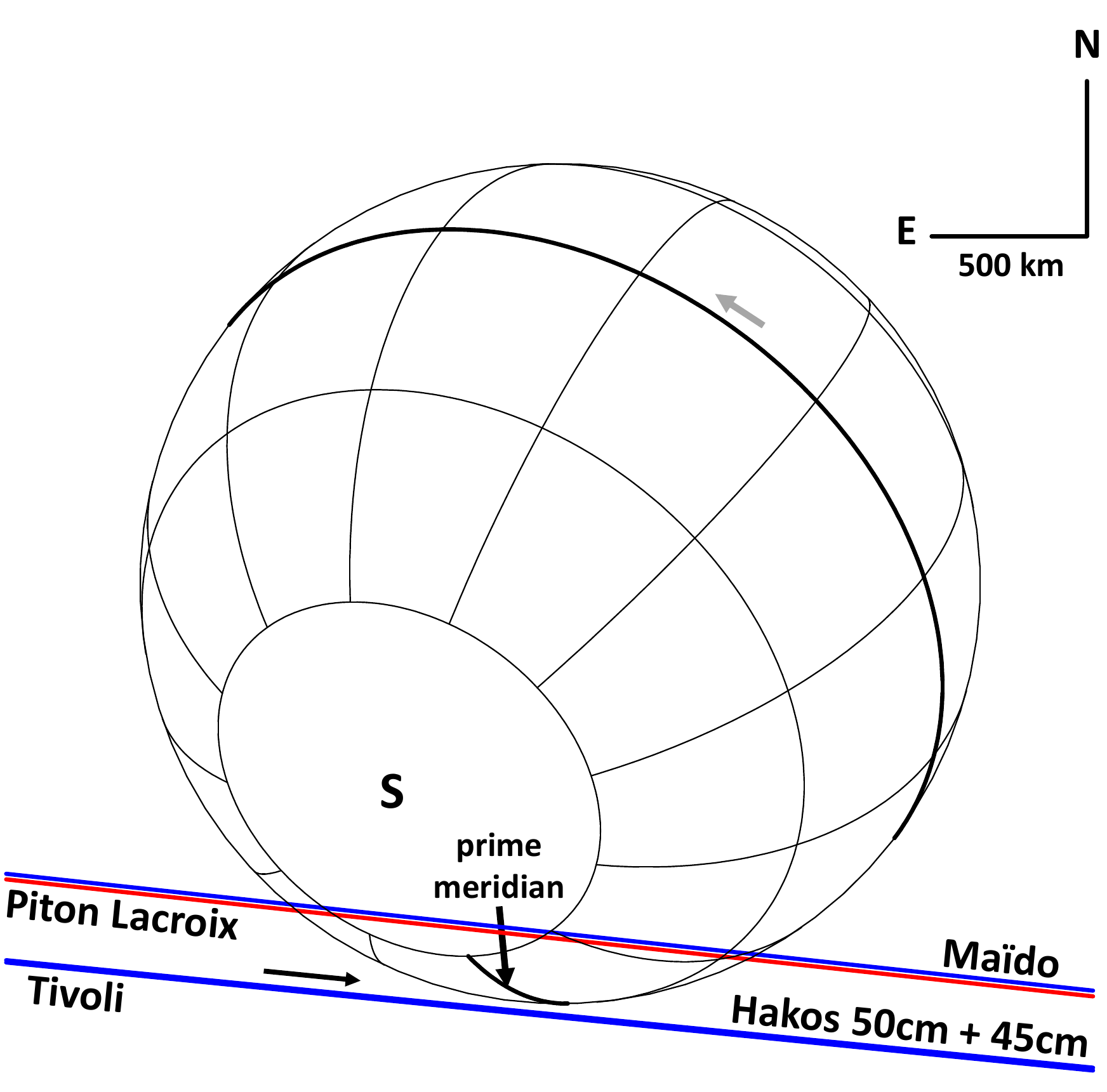}
}%
\caption{
Geometry of the 21 May 2008 occultation,
with the same conventions as in Fig.~\ref{fig_chords}.
}
\label{fig_21may08_geometry}
\end{figure}

\begin{figure}[!t]
\centerline{%
\includegraphics[width=10cm,trim=0 0 0 0,angle=0]{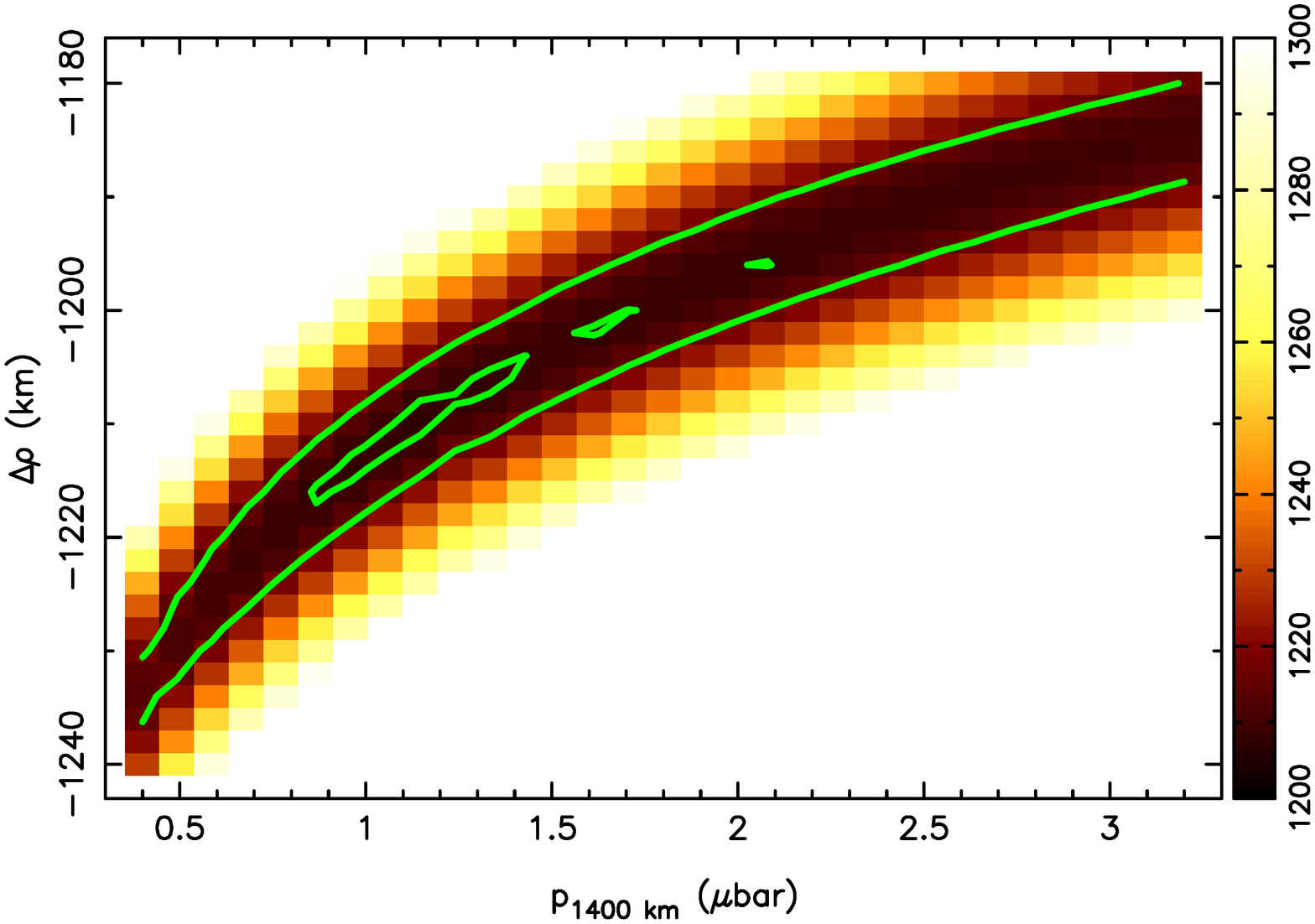}
}%
\caption{
Same as Fig.~\ref{fig_chi2_map_05oct17} but for the 21 May 2008 occultation.
}
\label{fig_chi2_map_21may08}
\end{figure}

\begin{figure}[!t]
\centerline{%
\includegraphics[width=8cm,trim=0 0 0 0,angle=0]{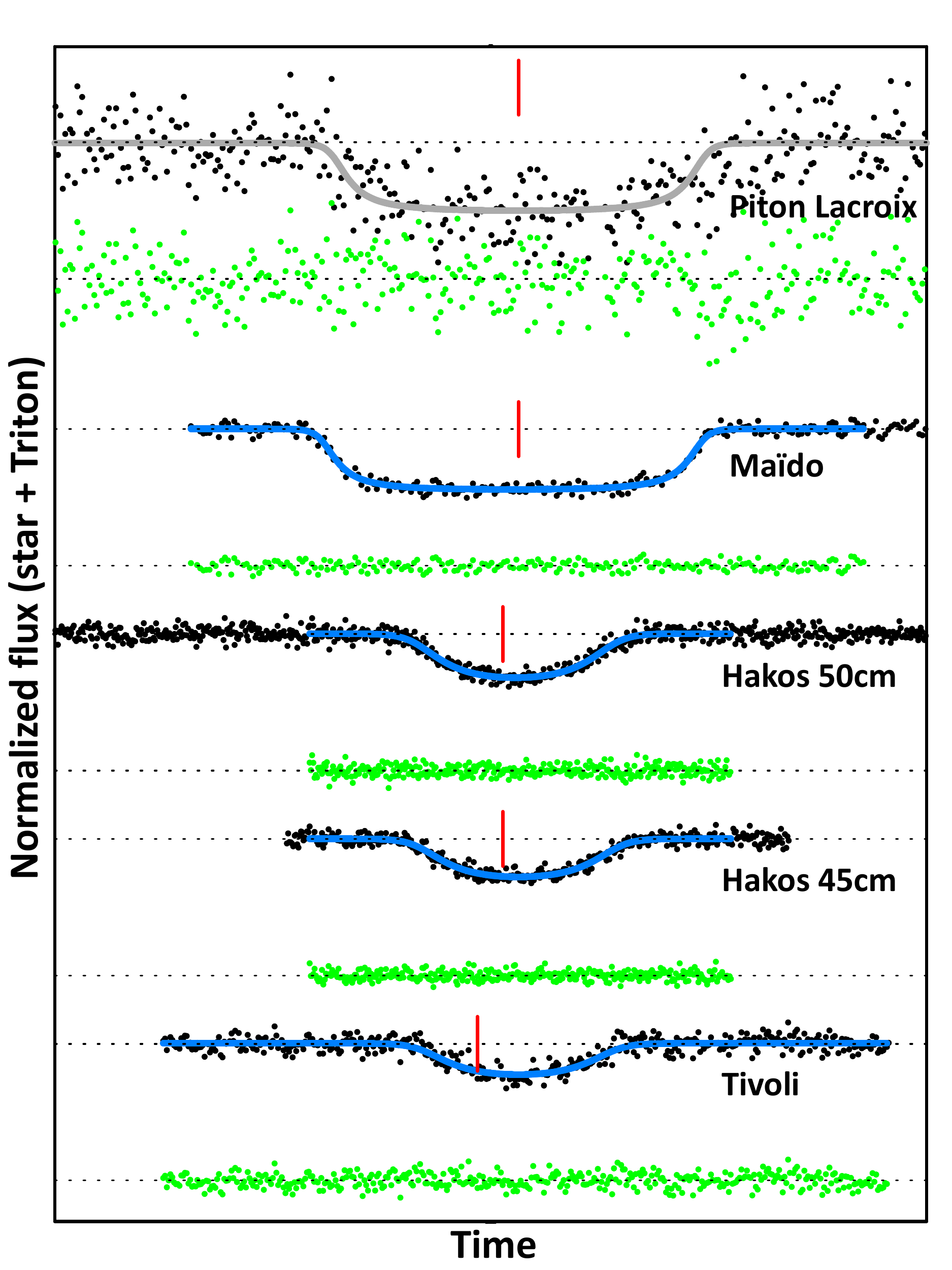}
}%
\caption{
Fit to the 21 May 2008 light curves.
The same conventions as for Fig.~\ref{fig_fit_18jul97} are used,
except that the panel now covers 720 seconds in time.
The synthetic light curve for Piton Lacroix is plotted in grey because it 
is not used in the fit, due to the high noise level.
The red vertical tick marks indicate 01:51 UTC for Piton Lacroix and Ma\"{\i}do,
and 01:41 UTC for Hakos.
}
\label{fig_fit_21may08}
\end{figure}

This event was observed from Namibia (two stations) and from La Réunion island (two stations;  see Table~\ref{tab_sites}). 
Given that each pair of stations are close together, only two effective chords 
have been obtained (see Fig.~\ref{fig_21may08_geometry}).
Moreover, these chords being grazing, there is a strong correlation 
between the closest approach distances of the chords to Triton's shadow centre
and the retrieved reference pressure $p_{1400}$ (see Fig.~\ref{fig_chi2_map_21may08}).
The best fit is also shown in Fig.~\ref{fig_fit_21may08}.
As a consequence, the value of $p_{1400}$ is poorly constrained at the 1$\sigma$
level, $p_{1400}= 1.15^{+1.03}_{-0.37}$~$\mu$bar.
At the 3$\sigma$ level, the value is so unconstrained that 
it does not bring any information on the temporal seasonal variations of the pressure
(Fig.~\ref{fig_t_p_1989_2017}).

\section{Atmospheric seasonal variations}
\label{sec_pres_evolution}

\subsection{Occultation results}

\begin{figure}[!t]
\centerline{
\includegraphics[totalheight=8cm,trim=0 0 0 0, angle=0]{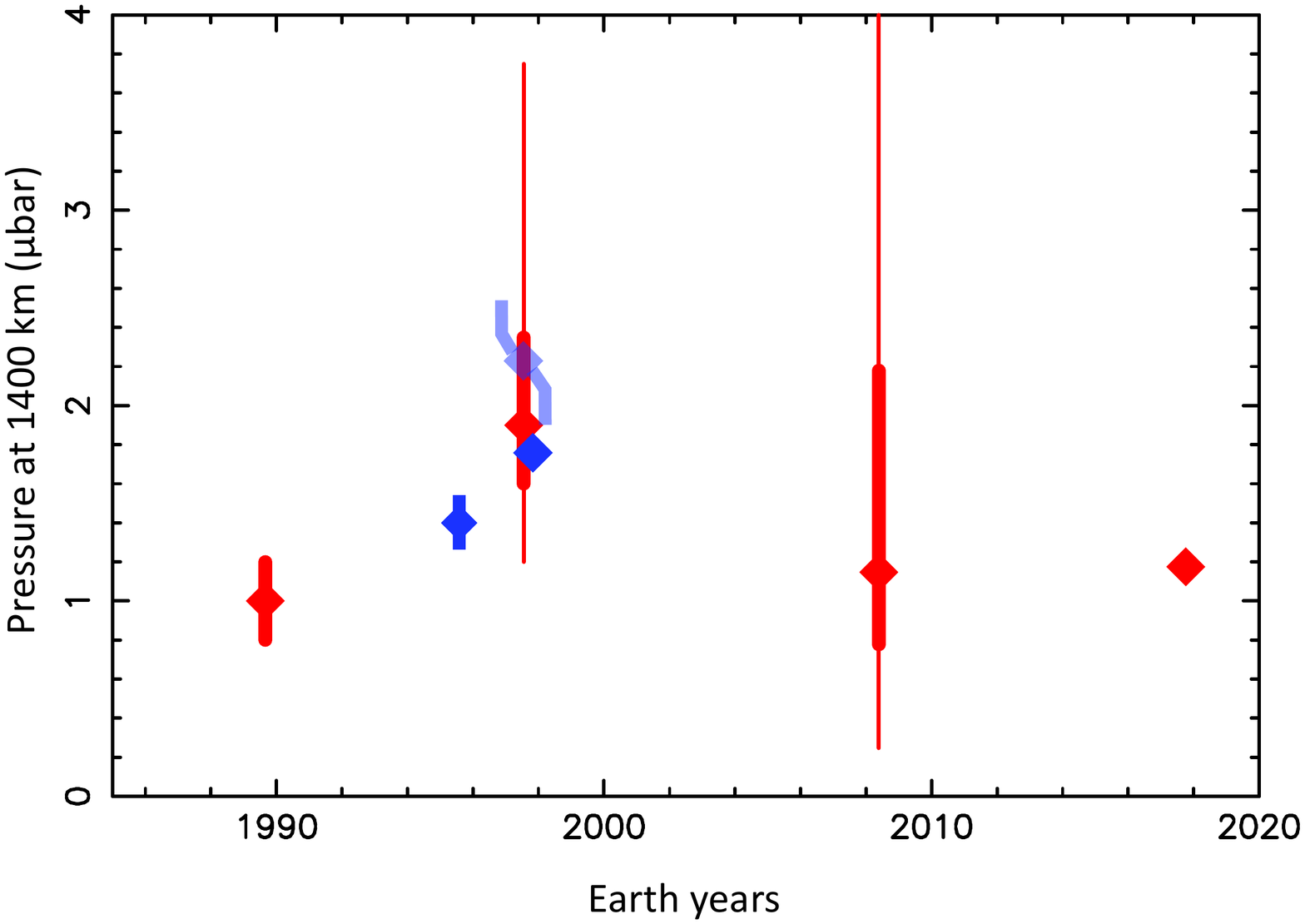}
}
\caption{%
Triton's atmospheric pressure seasonal variations with time, using the values from Table~\ref{tab_pressure_time}.
Our results are in red, and values taken from other works are plotted in blue.
For better viewing, the value derived by \cite{ell00} from the 18 July 1997 occultation is plotted 
in a semi-transparent blue colour, using the same dataset that we are using here for that date,
with our red diamond-shaped point just below it.
For all points, the thick error bars correspond to 1$\sigma$ confidence levels.
We note that for the 4 November 1997 and 5 October 2017 values, 
the 1$\sigma$ error bar is smaller than the diamond-shaped symbol and therefore is not visible. 
For the 18 July 1997 and 21 May 2008 events, 
we have also plotted for information our 3$\sigma$ error bars as thinner lines (see text for discussion).
}
\label{fig_t_p_1989_2017}
\end{figure}

Table~\ref{tab_pressure_time} lists our values of $p_{1400}$ at various epochs, 
as well as values taken from other works.
Extrapolations to the surface have also been included, 
assuming a constant ratio $p_{\rm surf}/p_{1400}= 12.0$.  
The corresponding seasonal variations of $p_{1400}$ with time is displayed in Fig.~\ref{fig_t_p_1989_2017}.

\begin{table}[!b]
\caption{Atmospheric pressure on Triton.}
\label{tab_pressure_time}
\centering
\renewcommand{\arraystretch}{1.2}   
\begin{tabular}{llll}
\hline \hline
        & Pressure              & Pressure       & Fit \\
Date    & at 1400 km            & at the surface & quality\tablefootmark{2} \\
        & $p_{1400}$ ($\mu$bar) & $p_{\rm surf}$ ($\mu$bar)\tablefootmark{1} & $\chi^2_{\rm dof}$\\
\hline
\multicolumn{4}{c}{This work} \\
\hline
25 August 1989\tablefootmark{3}   & $1.0 \pm 0.2$             &  $12.5 \pm 0.5$         & N/A \\
18 July 1997                    & $1.90_{-0.30}^{+0.45}$    & ($22.8_{-3.6}^{+5.4}$)  & 0.95 \\
21 May  2008                    & $1.15_{-0.37}^{+1.03}$    & ($13.8_{-4.4}^{+12.4}$) & 0.93 \\
 5 October 2017                    & $1.18 \pm 0.03$           & ($14.1 \pm 0.4$)        & 0.85 \\
\hline
\multicolumn{4}{c}{Other works} \\
\hline
25 August 1989\tablefootmark{4}   & N/A                & $14 \pm 2$        & N/A \\
14 August 1995\tablefootmark{5}   & $1.4 \pm 0.1$     & ($17 \pm 1$)      & N/A \\
18 July 1997\tablefootmark{6}   & $2.23 \pm 0.28$   & ($26.8 \pm 3.4$)  & N/A \\
 4 November 1997\tablefootmark{7}   & $1.76 \pm 0.02$   & ($21.1 \pm 0.2$)  & N/A \\
\hline
\end{tabular}
\tablefoot{
\tablefoottext{1}{
The values in parentheses assume a constant ratio of 12.0 between 
the pressures at the surface and at 1400~km, 
as derived from our best model (Fig.~\ref{fig_T_r_logp}).
}
\tablefoottext{2}{See discussion in Sect.~\ref{sec_ray_tracing_results}.}
\tablefoottext{3}{Using our own inversion of the V2 RSS phase delay profile (see text).}
\tablefoottext{4}{G95.}
\tablefoottext{5}{\cite{olk97}.}
\tablefoottext{6}{\cite{ell00,ell03a}}
\tablefoottext{7}{This value is the average over ingress and egress obtained by \cite{ell03a}.}
}
\end{table}

The value of \cite{olk97} indicates a 40\% increase in pressure between 1989 and 1995,
but at a low significance level of 1.8$\sigma$.
From the 18 July 1997 event, \cite{ell00} obtained $p_{1400}= 2.23 \pm 0.28$~$\mu$bar,
whereas with the same dataset, we obtain $p_{1400}= 1.90_{-0.30}^{+0.45}$~$\mu$bar.
The difference between the two results amounts to a factor of 0.85 
and stems from the use of a different template model $T(r)$.
This said, this difference remains at the 0.6$\sigma$ level and is statistically insignificant.
Using our value of $p_{1400}$ for 1997 indicates a pressure increase by a factor of 1.9 
between 1989 and 1997, but at a marginally significant 2.5$\sigma$ level only. 

The 4 November 1997 value obtained by \cite{ell00,ell03a}, 
$p_{1400}= 1.76 \pm 0.02$~$\mu$bar, has a much lower error bar due to the high S/N
of the light curve, obtained with the \textit{Hubble Space Telescope}.
Taken at face value, this implies an increase in pressure by a factor of 1.76 between 1989 and 1997, at a 3.8$\sigma$ level. 
However, this observation was a single-chord event, and a model was used to retrieve the astrometry of this event. Consequently, there is an uncertainty that was not accounted for.
Since we do not have access to these data, it is impossible for us to verify their result using our own methods, and therefore, confirm this increase.

Finally, the 21 May 2008 event provided only two grazing chords, 
bringing no new information.\ Thus, no firm conclusion can be drawn 
on any change of pressure between 1989 and 2008.

In summary, we estimate that the surge of pressure reported in the 1990s 
(compared to the V2 epoch) seems to be confirmed by our own analysis, but it remains debatable considering the paucity of data points available,
and the lack of a fully consistent analysis of all the observed events.
We note that the 2017 data rules out the concept of a monotonic increase in Triton’s pressure over time, but does not rule out the observed increase in 1995-1997.
Regardless, the much more accurate value of $p_{1400}$ that we obtain in 2017
is fully compatible with that derived from the V2 RSS experiment.
If we consider the 3$\sigma$ level, Fig.~\ref{fig_t_p_1989_2017} shows that no increase can be claimed between the two measurements.
So, either no surge occurred between 1989 and 2017 
or, if it did, the pressure was back to its V2 value in 2017.

From high-resolution spectroscopy in July 2009, \cite{lel10} obtained the first detection of methane gas in Triton's atmosphere since V2, and the first CO gas detection. Their analysis yielded a CH$_4$ gas number density at the surface $4.0_{-2.5}^{+5.0}$ larger than inferred from V2 \citep{her91,str95}.
Assuming that the N$_2$ pressure would qualitatively follow a similar seasonal variation, they estimated a 40~$\mu$bar pressure in 2009. This value (which did not represent a direct measurement of the N$_2$ pressure) is clearly at odds with the picture shown in Fig.~\ref{fig_t_p_1989_2017}, in particular with the 21 May 2008 point.
%
%

\subsection{Climatic context from numerical volatile transport modelling}
\label{subsec_vtm}

The climatic context of Triton is described and analysed in detail in a recent paper by \cite{ber22}. 
\cite{ber22} employed the VTM of Triton, developed at the Laboratoire de Météorologie Dynamique (LMD), to investigate the long-term and seasonal volatile cycles of N$_2$ and CH$_4$ on Triton. 
Their simulations are constrained by the surface pressure derived from the stellar occultations presented in this paper. 
In this section, we summarise the main results of this paper that are relevant for the interpretation of our observations.

\begin{figure}[!t]
\centering
\includegraphics[totalheight=48mm,angle=0, trim=0 0 0 0]{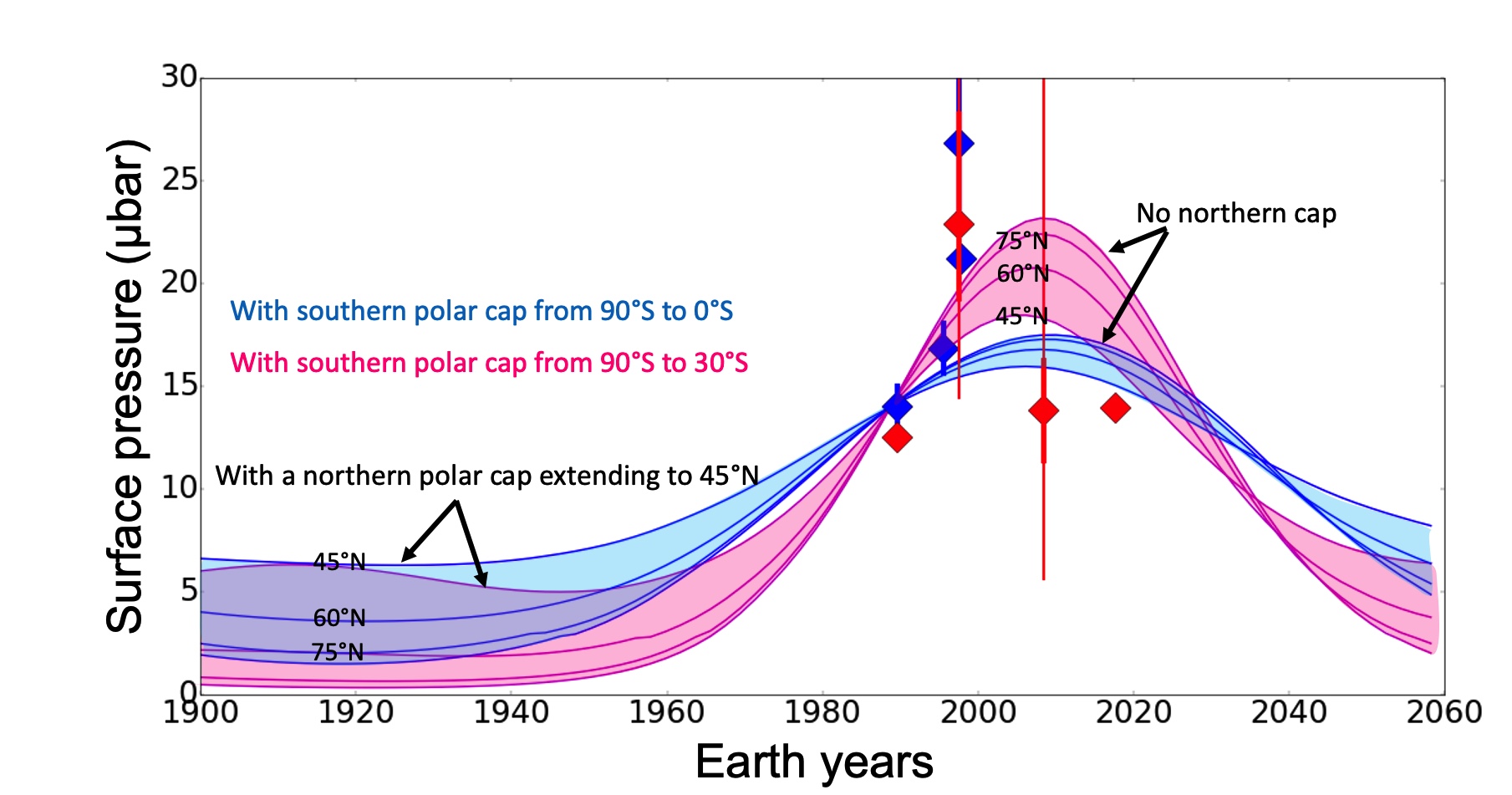}
\caption{%
Surface pressure cycle on Triton as simulated with the VTM assuming a different 
fixed N$_2$ ice distribution in both hemispheres. 
A thermal inertia of 1000~J~s$^{-1/2}$~m$^{-2}$~K$^{-1}$ (SI) was assumed.
\cite{ber22} include more simulations in their paper, and their Fig. 9 shows cases with different thermal inertias. It is of note that their simulations show that a lower thermal inertia would delay the peak of the surface pressure, in opposition with the occultation data points.
The blue lines refer to a southern cap extending to the equator, while the pink lines are for a southern cap extending to 30$^\circ$ S. 
Each line, marked with its corresponding value, refers to a different extension of the northern cap: 45$^\circ$ N, 60$^\circ$ N, 75$^\circ$ N, and no cap.
}%
\label{fig_tanguy}
\end{figure}

In VTM simulations, the surface pressure peak occurs slightly after the southern summer solstice (2000) between years 2000-2010 (see Fig.~\ref{fig_tanguy}). 
The surface pressure seasonal variations obtained by \cite{ber22} is similar to that obtained by \cite{spmo92}, when they artificially maintained a permanent large southern cap of bright N$_2$ (see their Fig. 7).
The larger the northern cap, the more it can serve as a condensation area and buffer N$_2$ sublimation in the southern hemisphere, which results in a lower and earlier surface pressure peak. 
The amplitude of the surface pressure peak is strongly attenuated if N$_2$ ice remains between 30$^\circ$~S - 0$^\circ$, because condensation will dominate over sublimation between the years 1980-2020.

According to the model, Triton’s atmospheric surface pressure will remain at 5~$\mu$bar during the next solstice season if the north polar cap extends to 60$^\circ$~N and the south polar cap extends to 0$^\circ$. 
The amplitude of the pressure peak is attenuated if N$_2$ ice deposits remain between 30$^\circ$~S - 0$^\circ$ because these latitudes are dominated by condensation rather than sublimation after the year 2000.

These results suggest that a northern cap extending down to at least 45$^\circ$~N – 60$^\circ$~N is needed in 2017 to restore the surface pressure back to the V2 measured value $\sim14~\mu$bar.
Otherwise, the surface pressure will remain higher than $16~\mu$bar in 2017 with no northern cap.
A strong increase in surface pressure cannot occur before 2000 if N$_2$ ice remains between 30$^\circ$~S - 0$^\circ$. 
To ensure that the surface pressure remains greater than $5~\mu$bar during the opposite season (southern winter) a permanent northern cap extending down to 45$^\circ$~N is required.

In their simulations, \cite{ber22} also investigated the CH$_4$ cycle by taking into account a small amount of pure CH$_4$ ice at the surface in addition to the N$_2$-rich mixture \citep{mer18}. 
In the case where this pure CH$_4$ ice is placed at the south pole of Triton, covering 2\% of the surface of the visible projected disk, they obtain a large increase in the CH$_4$ gas abundance from 1990 to 2005, without any significant change in N$_2$ surface pressure.
Since CH$_4$ is not completely mixed with N$_2$ ice, it implies that the large increase in CH$_4$ (with relation to V2) reported by \citealp{lel10} could be decoupled from the N$_2$ seasonal variations and, therefore, does not necessarily represent a measurement of the global pressure of the atmosphere.

For more details on the N$_2$ and CH$_4$ seasonal variations as simulated by the VTM, the reader is referred to \cite{ber22}.

\section{Triton's lower atmosphere: Central flash}
\label{sec_lower_atmo}

The detection of a central flash during the 5 October 2017 occultation offered 
a unique opportunity to study Triton's lower atmosphere.
Our ray-tracing code shows that the flash is caused by 
a layer having a typical thickness 2~km, lying at about 8~km above Triton's surface 
(radius of 1361~km).
In that altitude range, the Abel inversion method is no longer valid, 
due to the co-existence of two stellar images along Triton's limb (see Fig.~\ref{fig_effect_secondary_image}).
This problem arises at altitude levels of about 20~km,
corresponding to the deepest layers probed by the light curve obtained at La Palma.
Consequently, the central flash allows us to gain about 12~km downwards 
(about 0.6~scale height) compared to the inversion method. 
The explanation for this is shown in Fig. \ref{fig_T_r_various_phi0} and further detailed in Appendix \ref{sec_appen_phi0}.

\subsection{The central flash: Observations}

The central flash swept Europe along the lines shown in grey in Fig.~\ref{fig:prediction}.
Among the 90 light curves shown in Figs.~\ref{fig_fit_data_1}-\ref{fig_fit_data_5}
and Figs.~\ref{fig_no_fit_data_1}-\ref{fig_no_fit_data_2},
42 show evidence of a stellar flux increase near mid-occultation, 
and 23 of them have enough S/N to be used in the central flash modelling.

\begin{figure}[!h]
\centerline{\includegraphics[totalheight=76mm]{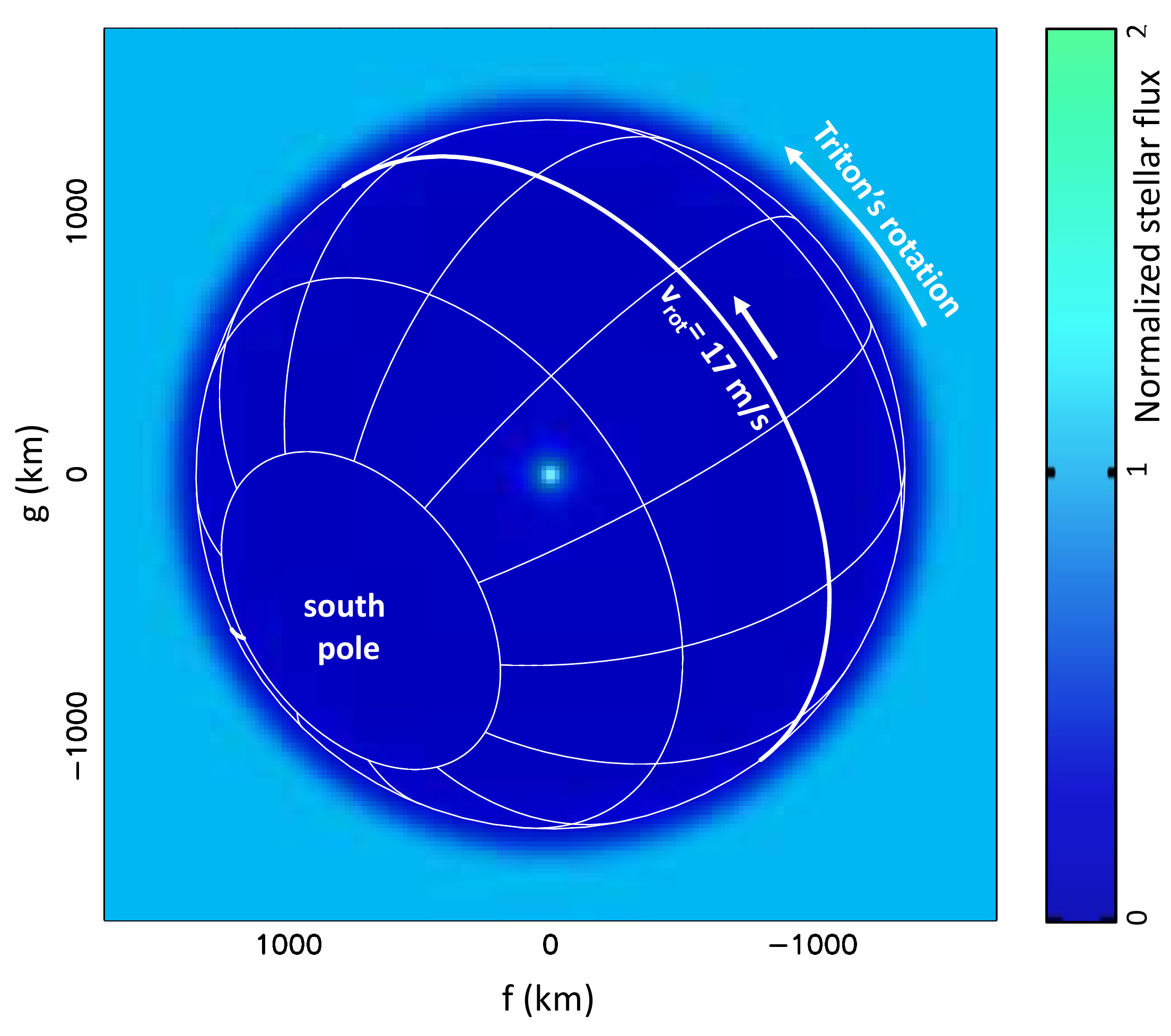}}
\caption{%
Stellar flux in Triton's shadow for the 5 October 2017 event, normalised to unity outside the body (light blue region).
The flux reaches a minimum of about 7\% of the unocculted flux inside the shadow and
then rises sharply at the shadow centre. 
The direction of Triton's rotation is indicated, 
as well as the equatorial rotation velocity, $v_{\rm rot}= 17$~m~s$^{-1}$ (in an inertial frame),
using the parameters of Table~\ref{tab_param}.
}%
\label{fig_map_2D_flash_gen}
\end{figure}

Figure~\ref{fig_map_2D_flash_gen} displays the reconstructed intensity map of Triton's shadow,
with in particular the presence of a bright dot (central flash) near the shadow centre.

\begin{figure*}
\centerline{\includegraphics[width=135mm]{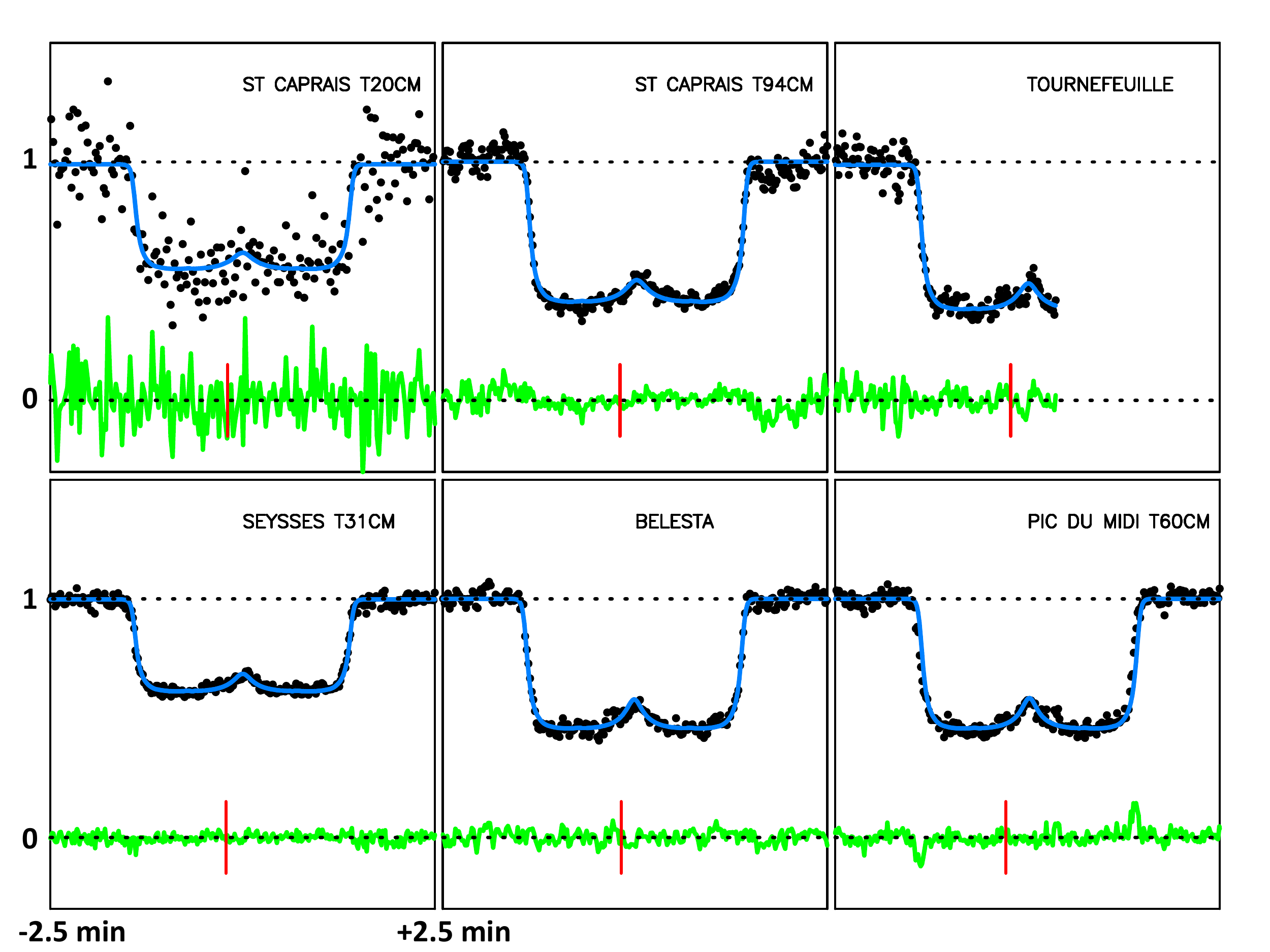}}
\centerline{\includegraphics[width=135mm]{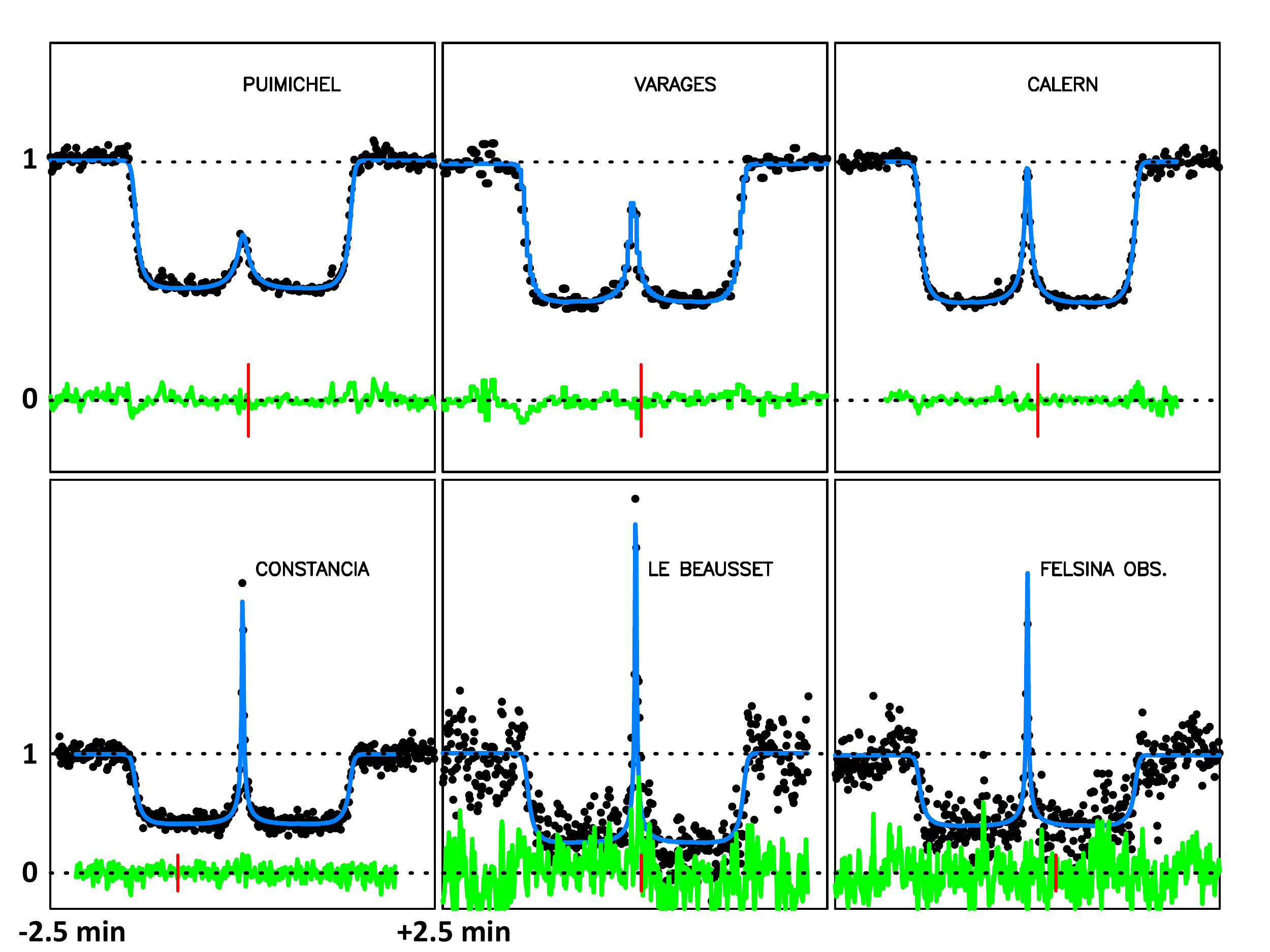}}
\caption{%
Simultaneous fits of the data (black dots) by synthetic light curves 
(blue lines), based on the temperature profile displayed in Fig.~\ref{fig_T_r_logp} 
(black line) and the pressure boundary condition $p_{1400}= 1.18$~$\mu$bar
(Table~\ref{tab_pressure_time}).
The green curves are the residuals of the fits.
The lower and upper horizontal dotted lines mark the zero-flux level and the total 
star plus Triton unocculted flux, respectively.
We also note that the three bottom light curves are plotted at a different vertical scale from the others
to accommodate the presence of a strong central flash.
The stations are sorted from left to right and top to bottom from the northernmost track
(St Caprais) to the southernmost track (Calar Alto; see next figure).
Each panel has a duration of five minutes and is centred around the time of closest approach 
(or mid-occultation time) of the station to Triton's shadow centre, 
as indicated under the lower left panel in each block of six light curves.
For reference, the vertical red line marks the time 23:48 UTC.
The stations with exposure times of less than 1~s have been smoothed to have a sampling 
time as close as possible to 1~s, for easier S/N comparison of the various datasets.
We note that in this approach, the sampling of the Const\^ancia, Le Beausset, and Felsina Observatory
light curves (0.64~s) is kept at its original value so that full resolution versions of the 
corresponding strong flashes at those stations are displayed here.
The same kinds of plots showing all the stations, but with the flashes excluded from the fits,
are displayed in Figs.~\ref{fig_fit_data_1}-\ref{fig_fit_data_5}.
}%
\label{fig_fit_flash_1}
\end{figure*}

\begin{figure*}
\centerline{\includegraphics[width=135mm]{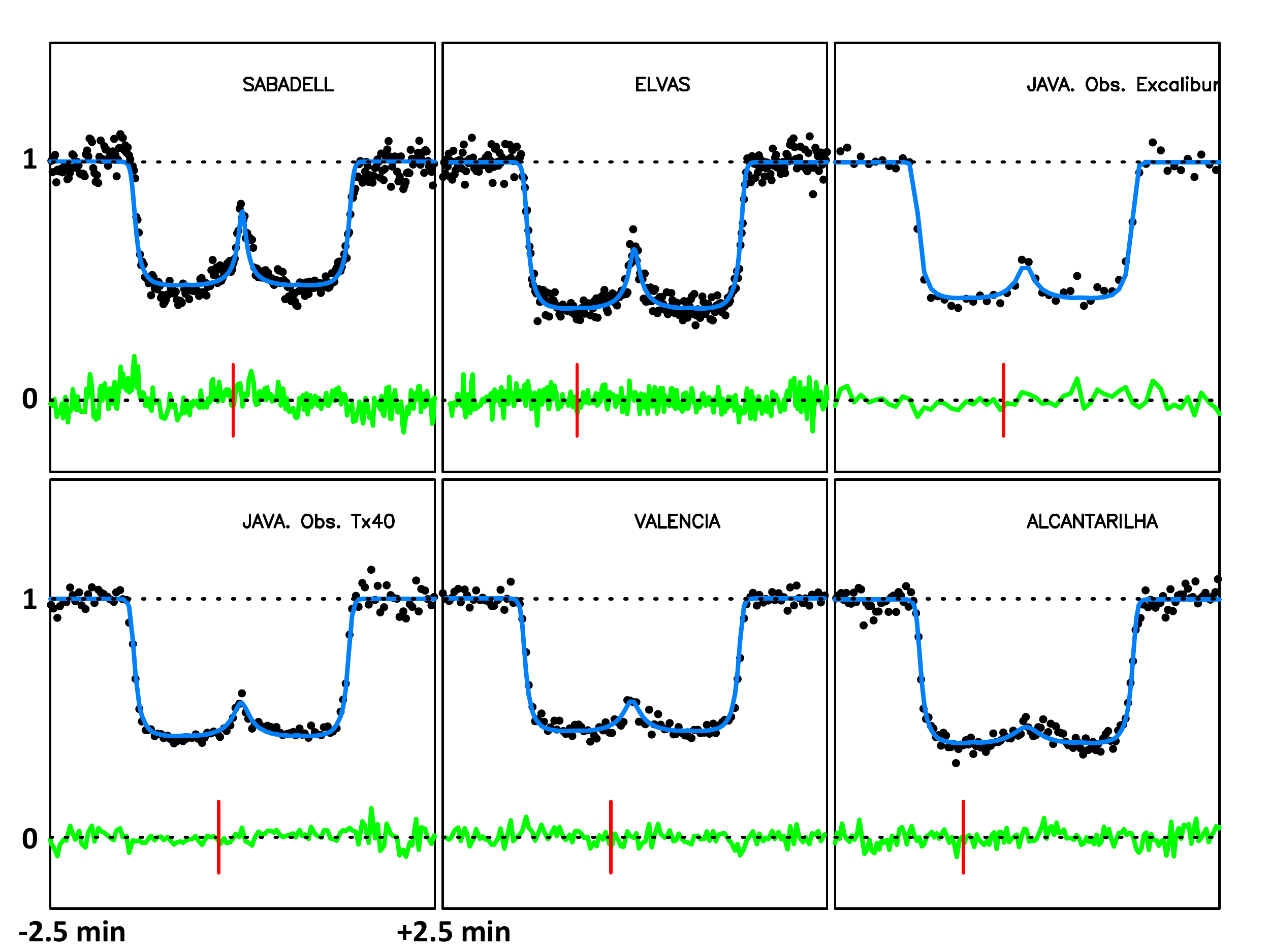}}
\centerline{\includegraphics[width=135mm]{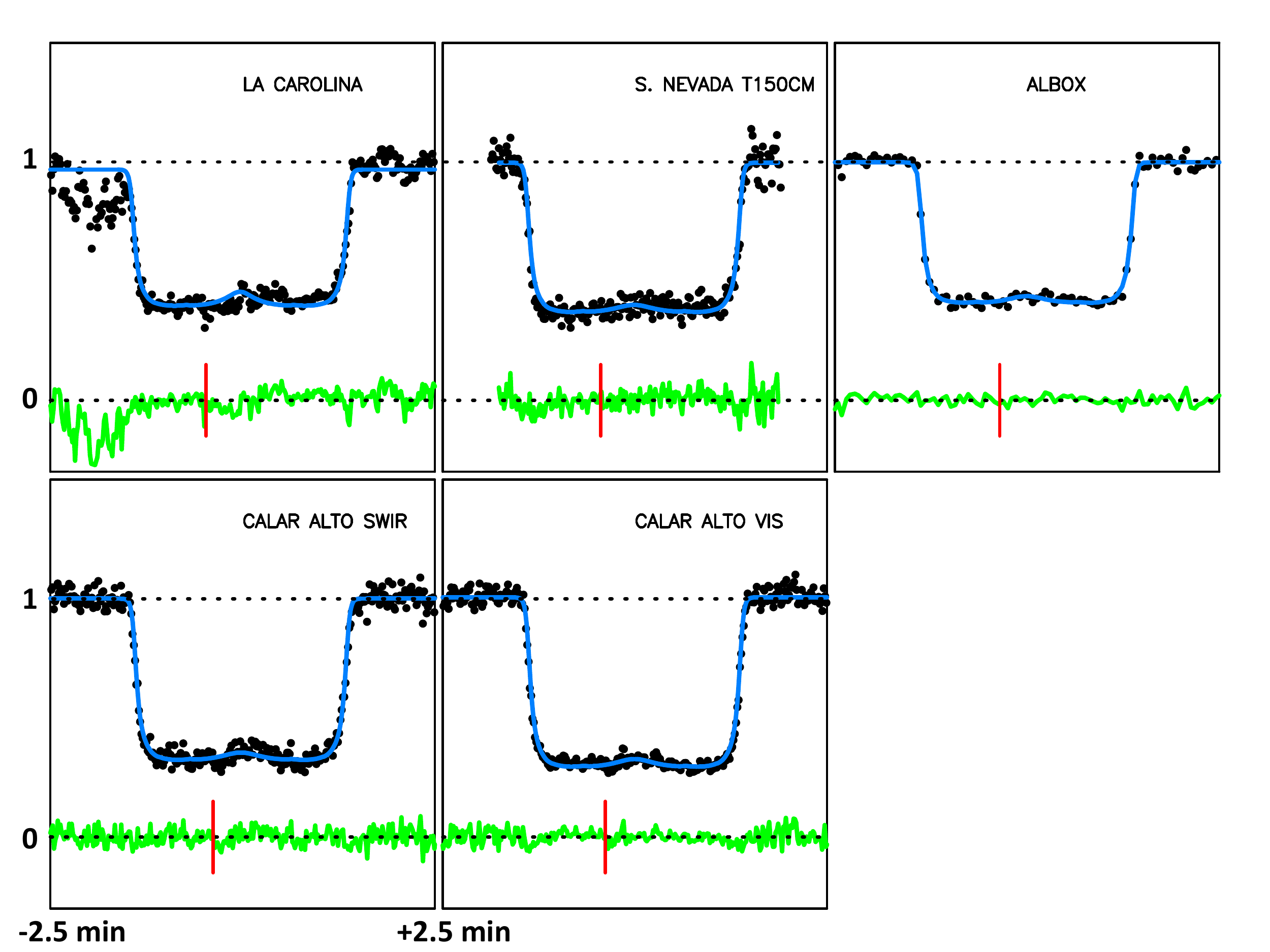}}
\caption{%
Continuation of Fig.~\ref{fig_fit_flash_1}.
NB. `JAVA.' is the abbreviation of Javalambre, 
used so that the name of the station fits into the plot.
}%
\label{fig_fit_flash_2}
\end{figure*}

\begin{figure}
\centerline{\includegraphics[width=80mm]{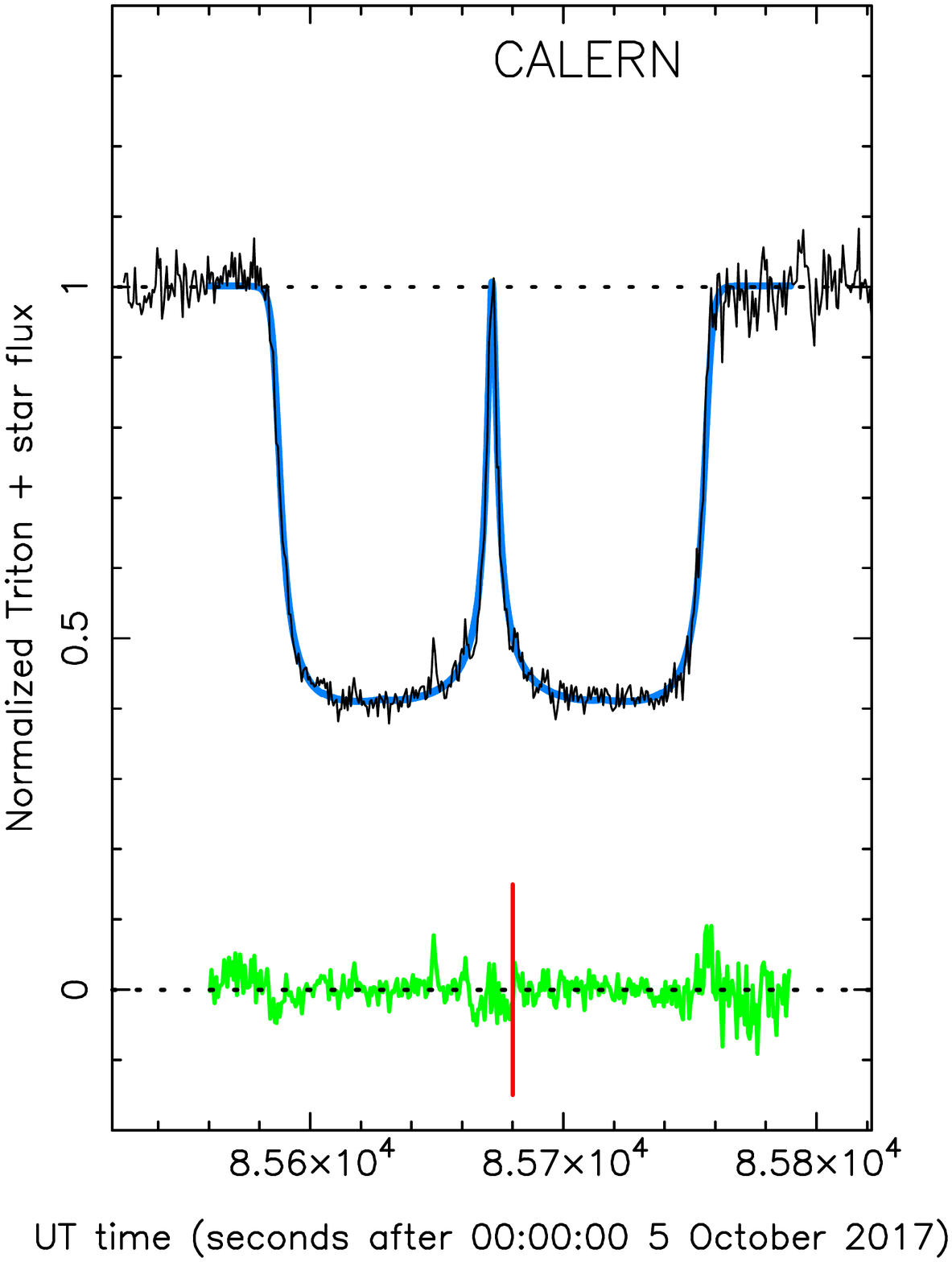}}
\centerline{\includegraphics[width=80mm]{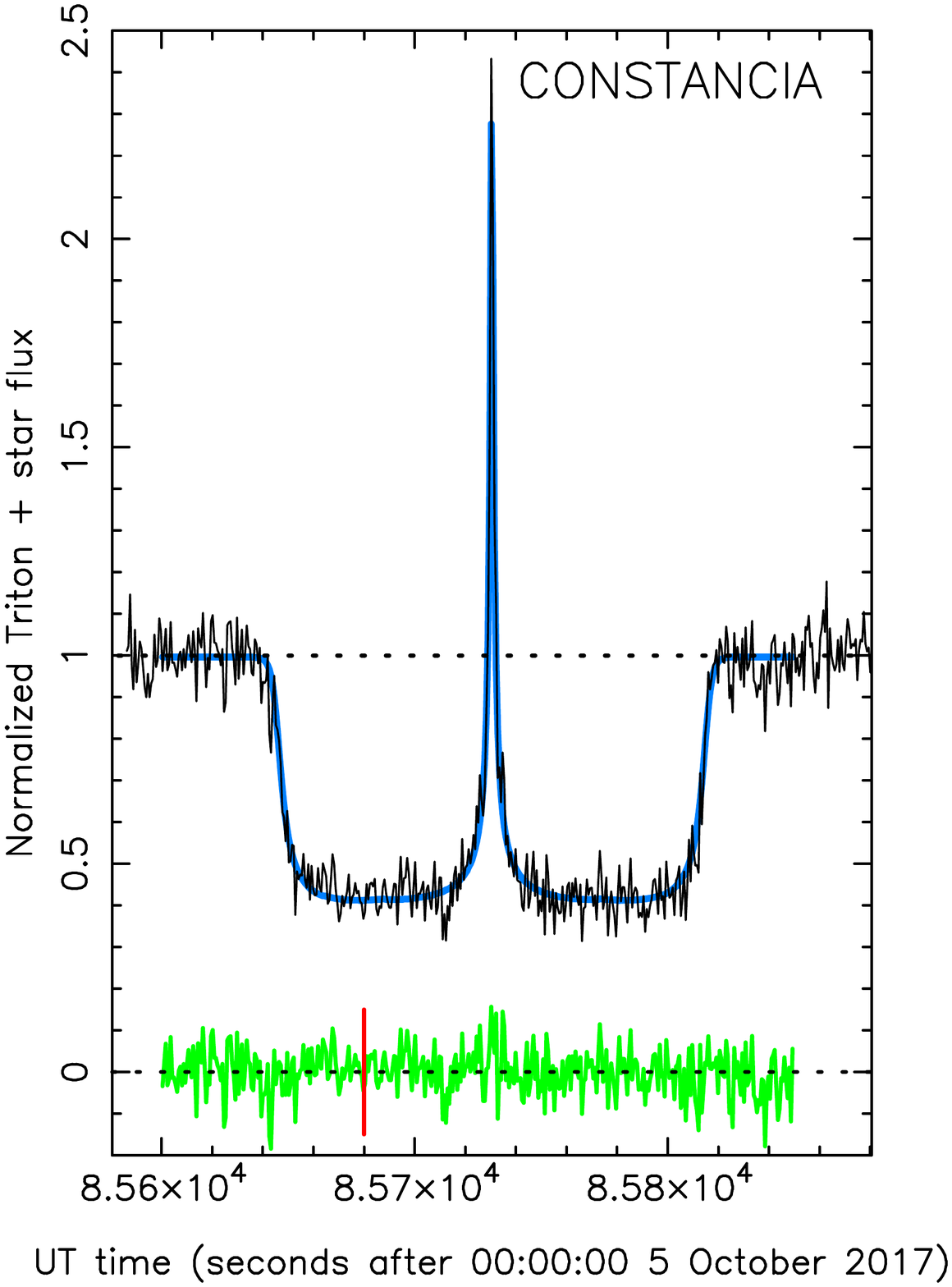}}
\caption{%
More detailed views taken of the flashes observed at Calern (top panel) and Const\^ancia 
(bottom panel), with the same setup as in the previous figure.
}%
\label{fig_fit_flash_cal_con}
\end{figure}

At Calar Alto, which passed at about 300~km from the shadow centre at closest approach (C/A),
the increase in stellar flux is barely noticeable (Fig.~\ref{fig_fit_flash_2}),
while it reaches the full unocculted stellar flux at Calern, 
which passed at 29~km from the centrality. 
At Const\^ancia (C/A 8.4~km), 
the maximum of the flash peaks at three times the unocculted stellar flux, and 
about 3.4 times the unocculted flux at Le Beausset (C/A 6.7~km, the closest of all stations; see Figs.~\ref{fig_fit_flash_1} and \ref{fig_fit_flash_cal_con}).

The fit of the central flash is described in Sect.~\ref{sec_ray_tracing_results},
except that we now allow a departure from sphericity of the layer responsible for the flash
(see \citealp{sic06} for details).
We note that the ray-tracing code accounts for both the primary and secondary stellar images.
Thus, we are not restricted in using this code as would be the case for the Abel inversion scheme (Sect.~\ref{sec_appen_secondary_image}).

%

\subsection{The central flash: Spherical fit}
\label{sec_spherical_fit}
 
Assuming a spherical flash layer, we obtain the best simultaneous fits 
(now including the central flashes) displayed in Figs.~\ref{fig_fit_flash_1}-\ref{fig_fit_flash_2}.
The quality of the fit ($\chi^2_{\rm dof}= 0.80$) is comparable to the
quality obtained without the flashes ($\chi^2_{\rm dof}= 0.85$),
showing that no departure from sphericity is detected.
A more quantitative assessment for the upper limit for such a departure (some 1.5~km along the limb,
as projected in the sky plane) is provided in the next subsection.
A closer visual examination of the residuals for the strongest flashes with best S/N
reveal, however, some minor and localised features, possibly due to atmospheric waves  
(Fig.~\ref{fig_fit_flash_cal_con}), but no global departure from the spherical model.

There is another argument supporting the spherical nature of Triton's atmosphere.
The centre of Triton's shadow, as determined by the simultaneous fit to all the flashes,
while excluding the ingress and egress parts of the light curves, 
coincides to within 0.1~km with the shadow centre determined by a global fit to all 52 light curves, 
by excluding the central flashes but including the ingress and egress parts.
This 0.1 km offset is not significant, considering that the global fit centre
has a typical 1$\sigma$ error of 1~km cross-track (Fig.~\ref{fig_chi2_map_05oct17}).
In other words, the centre of the central flash layer, which is sensitive to the 8~km altitude level, 
coincides with the global shadow centre, which is sensitive to the 60 km altitude level.
It could be that both atmospheric levels are close to spherical, but displaced in the same
way with respect to Triton's centre, but this configuration seems unlikely.

\subsection{The central flash: Limit on atmospheric distortions and winds}

\begin{figure*}
\centerline{
\includegraphics[totalheight=80mm,trim=0 0 70 0]{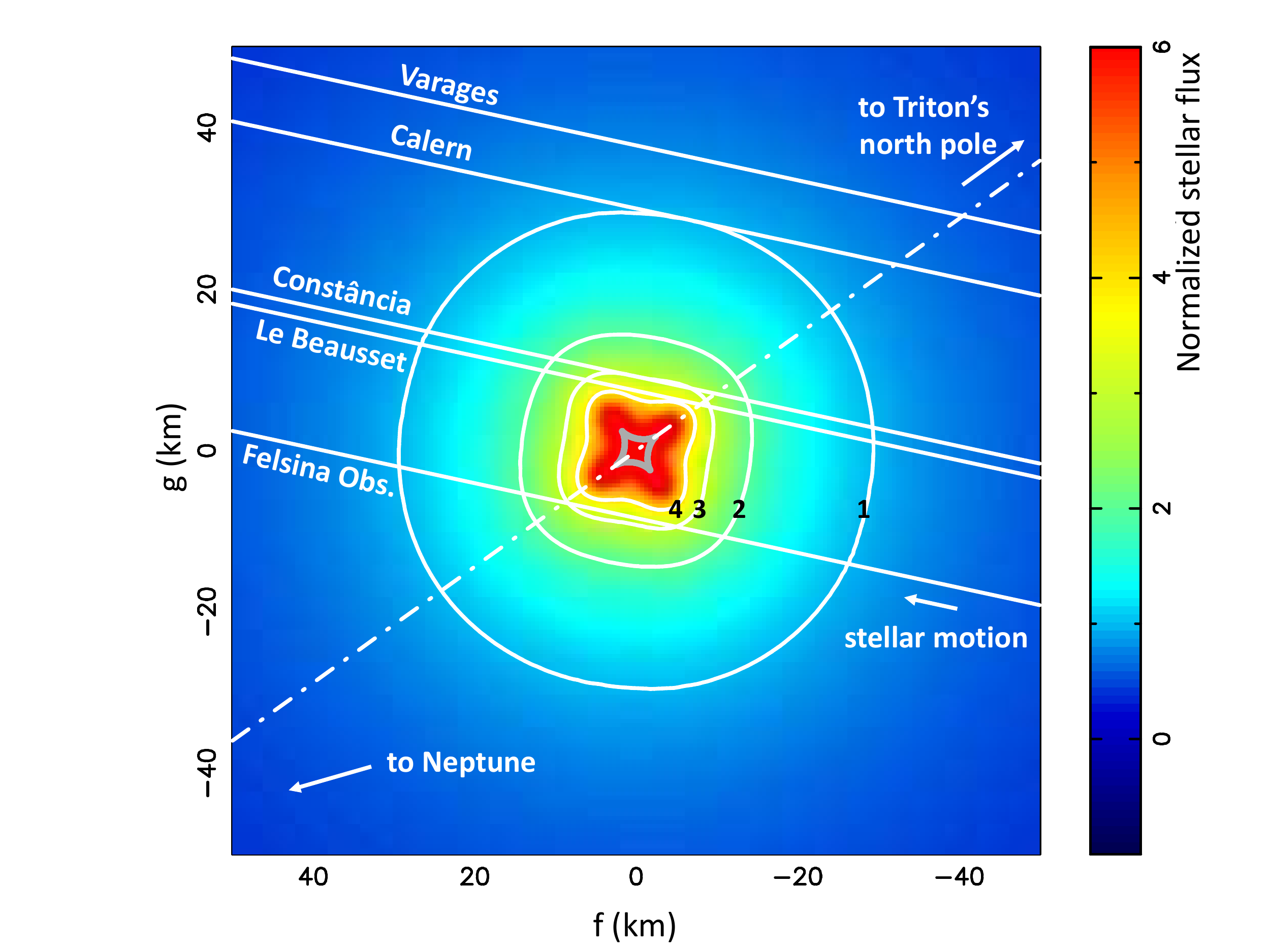}
\includegraphics[totalheight=80mm,trim=70 0 0 0]{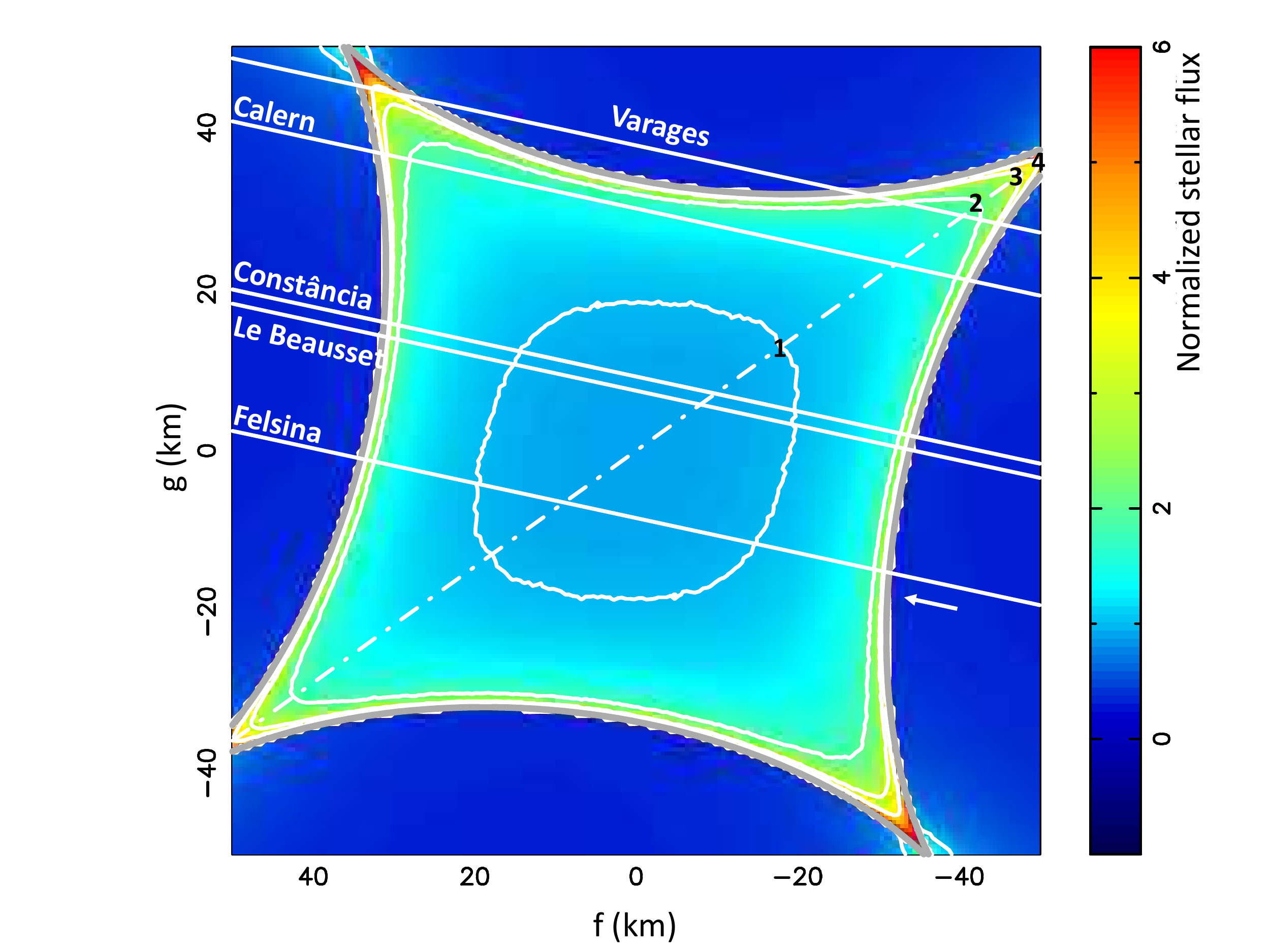}
}
\caption{Maps of the central flash intensity.
\textit{Left panel:}
Map of the central flash intensity, adopting the 1$\sigma$ upper limit 
$\epsilon' = 0.0011$ for the apparent oblateness of the flash layer 
(corresponding to a deprojected oblateness $\epsilon = 0.0019$) (see text for details).
The black labels along the iso-intensity contours (in curves) 
indicate the received stellar flux in units of its unocculted value.
The grey diamond-shaped feature near the centre is the caustic curve described 
by Eq.~\ref{eq_caustic} and corresponding to $\epsilon' = 0.0011$.
In the case considered here, the flash layer is assumed to be aligned with the 
apparent direction of Triton's poles, indicated by the dash-dotted line.
Neptune's direction is determined from the position angle 286$^\circ$ of
Triton with respect to the planet at the moment of the occultation.
\textit{Right panel:}
Same, but adopting the oblate solution $\epsilon = 0.042$ found by \cite{ell97}
from the shape of a central flash observed during the 14 August 1995 occultation.
}%
\label{fig_map_2D_flash_upper_limit_0.0014}
\end{figure*}

We now assess a possible departure of Triton's lower atmosphere from sphericity,
restricting ourselves to the simple model of a globally oblate flash layer.
Testing more complex shapes will be performed once Triton's 3D GCMs
are available, something that is beyond the scope of this paper.
Once projected in the sky plane, an oblate layer appears as an ellipse with 
apparent semi-major and semi-minor axes $a'$ and $b'$, respectively. 
The centres of curvature of that ellipse form a diamond-shaped caustic curve 
where abrupt flux variations are observed (see examples in Fig.~\ref{fig_map_2D_flash_upper_limit_0.0014}). 
The equation of the caustic is \citep{ell77}:
\begin{equation}
(a' x)^{2/3} + (b' y)^{2/3} = (a'^2 -  b'^2)^{2/3},
\label{eq_caustic}
\end{equation}
where $Oxy$ is a Cartesian reference system 
whose origin $O$ is fixed at the ellipse centre, and
where $Ox$ (resp. $Oy$) is aligned with $a'$ (resp. $b'$).
Since the flash layer lies at $\sim 8$~km altitude, we have $a'\sim 1360$~km.
We note that the orientations of the $a'$ and $b'$ axes are still to be specified.

We define the apparent oblateness of the flash layer as $\epsilon'~=~(a'-b')/a'$.
We have explored values of $\epsilon'$ from zero (i.e. a spherical flash layer)
to some maximum value, and tracked the corresponding variations of $\chi^2$
stemming from a simultaneous fit to central flashes.
In this study, only the central flashes of the
Varages, Calern, Const\^ancia, Le Beausset, and Felsina Observatory stations have been considered.
The other stations are too far away from centrality and/or with lower quality 
to usefully constrain $\epsilon'$.
This is because the four cusps of the diamond-shaped caustic curve
extend up to $\sim 2 \epsilon' a$ from the shadow centre according to Equation~\ref{eq_caustic}.
As we obtain upper limits of $\sim$0.002 for $\epsilon'$ (see below), 
we have $2 \epsilon' a < \sim 5$~km for $a \sim 1360$~km.
Thus, only the immediate vicinity of the shadow centre (typically less than 20~km)
is sensitive to departures from sphericity. More distant stations essentially probe 
flashes that are indistinguishable from a spherical solution.

We first assume that the semi-minor axis $b'$ is aligned with Triton's pole.
This corresponds to an oblate flash layer maintained by an axisymmetric zonal wind regime
that has a constant angular velocity around that axis.
By using the $\chi^2 < \chi^2_{\rm min} + 1$ (resp. $\chi^2 < \chi^2_{\rm min} + 9$) 
criterion, we find 1$\sigma$-level (resp. 3$\sigma$-level) upper limits of
$$
\epsilon' < 0.0011 {\rm~~(resp.~~}\epsilon' < 0.0014)
$$
for the apparent oblateness of the flash layer.
The flash intensity map corresponding to this limit is displayed in 
Fig.~\ref{fig_map_2D_flash_upper_limit_0.0014}.
%
%
This apparent oblateness must be `deprojected' to obtain the actual oblateness,
$\epsilon$, through the relation
$$
\epsilon= 1 - \frac{\sqrt{(1-\epsilon')^2 - \sin^2 B}}{\cos B}
\sim \frac{\epsilon'}{\cos^2(B)},
$$
where $B= 40.5^\circ$ S is the sub-observer latitude (Table~\ref{tab_param}), and 
where the approximation holds for $\epsilon' \ll 1$.
Using $\epsilon' < 0.0011$, 
this yields a 1$\sigma$-level upper limit $\epsilon < 0.0019$ for the deprojected oblateness.
This corresponds to a difference between the equatorial and polar radii 
$r_e$ and $r_p$ of the layer, respectively, of $r_{\rm e}- r_{\rm p} \sim 3$~km,
using $r_e= a'= 1360$~km.

We assume that the flash layer shape is entirely supported by zonal winds.
In particular, we assume the absence of a horizontal temperature gradient,
so that the isobar level also corresponds to the isopycnic (constant density) layer.
The radius $r$ of the flash layer is given as a function of the latitude $\varphi$
by the equation \citep{hub93,sic06}
$$
\frac{1}{r} \frac{dr}{d\varphi} = - \frac{f\cos(\varphi) \sin(\varphi)}{1-f\cos^2(\varphi)},
$$
where $f= rv^2(\varphi)/GM\cos^2(\varphi)$. 
This equation states that the isobar is locally perpendicular to the effective gravity field, 
where both the gravity field of the (spherical) body and centrifugal forces are accounted for. 
Introducing in that expression the polar equation of an oblate flash layer,
$$
r= \frac{r_{\rm e}r_{\rm p}}{[r_{\rm e}^2 \sin^2(\varphi) + r_{\rm p}^2 \cos^2(\varphi)]^{1/2}},
$$
we obtain to lowest order in $\epsilon= (r_{\rm e}- r_{\rm p})/r_e$ the velocity
\begin{equation}
v = \sqrt{\epsilon} \sqrt{\frac{2GM_T}{r_{\rm e}}} \cos{\varphi} \sim 
1450 \sqrt{\epsilon} \cos{\varphi}~~{\rm m~s}^{-1},
\label{eq_wind_vel}
\end{equation}
the value of $GM_T$ is listed in Table~\ref{tab_param}.
Using $\epsilon < 0.0019$, this provides a 1$\sigma$-level upper limit for the zonal wind at the equator of $|v_{\rm e}| < 63$~m~s$^{-1}$.

We note that this motion can be prograde (positive sign) or retrograde (negative sign), 
and that it is measured in an inertial frame. 
Thus, noting $v'_{\rm e}$ the zonal wind in a frame rotating with Triton
(which thus measures the atmospheric circulation at that level), 
we have $v_{\rm e}= v'_{\rm e} + v_{\rm rot}$, 
where $v_{\rm rot}= 17$~m~s$^{-1}$ is the equatorial velocity stemming from Triton's rotation (see Fig.~\ref{fig_map_2D_flash_gen}).
Consequently, a retrograde zonal wind regime ($v'_{\rm e} < 0$) implies a 1$\sigma$ limit 
$|v'_{\rm e}| < 63+17 = 80$~m~s$^{-1}$,
while a prograde regime ($v'_{\rm e} > 0$) implies 
$v'_{\rm e} < 63-17 = 46$~m~s$^{-1}$.
Those values are respectively 87~m~s$^{-1}$ and 53~m~s$^{-1}$ if the 
3$\sigma$ upper limit is considered.

Information on the atmospheric circulation of Triton was obtained in 1989 by V2: while surface wind streaks suggested eastward surface winds between latitudes of 15$^\circ$ S and 45$^\circ$ S \citep{han90}, the deflection of plumes showed that in the atmosphere above, at 8 km near 49$^\circ$ S and 57$^\circ$ S, the wind was westward and prograde \citep{han90,yel95}. 
On the basis of theoretical consideration, \cite{ing90} proposed that this could result from a temperature contrast between the cold frost-covered pole and the warm un-frosted equator. 
More realistic GCMs simulations, including the N$_2$ condensation-sublimation cycle, have been reported in \cite{van13}, and additional relevant simulations have been performed to explore the circulation on Pluto, which is similar to Triton in terms of rotation rate and atmospheric composition (see \citealp{for21} and reference therein). 
These models show that if N$_2$ significantly sublimes in the southern hemisphere and condenses in the northern hemisphere, the circulation should be dominated by a retrograde circulation resulting from the conservation of angular momentum of the flow, with velocities that cannot be higher than the rotation of the planet (17~m~s$^{-1}$). 
This is significantly less than the upper limits that we derive from our observations. 
In any case, as mentioned above, global retrograde winds were not observed in 1989: to get prograde rotation in the mid southern latitude as suggested by the V2 plume observations, the inter-hemispheric condensation must be weak. 
In that case a thermal gradient could create a weak prograde wind as suggested by \cite{ing90}, reaching a few metres per second in GCM simulations. 
However, modelling Pluto suggests that in some conditions a regime of super-rotation (like on Venus or Titan) could occur \citep{for17}. 
This could explain the plume direction on Triton. 
Such a super-rotation is thought to initially result from the formation of a high-mid-latitude jet (due to thermal balance between a warm equator and a colder pole, or condensation flow from low latitudes to the pole). 
Then barotropic waves can transport angular momentum to and from the equator and seriously accelerate the entire atmosphere. 
In their Pluto GCM, \cite{for17} found mean equatorial zonal wind up to 15~m~s$^{-1}$. However, this could be model-dependant. 
It is not easy to set a theoretical limit to such a super-rotation. 
Our upper limit on prograde wind near 50~m~s$^{-1}$ provides a constraint for such a hypothetical super-rotation in 2017.

We have considered other orientations for the central flash layer, as projected in the 
sky plane, by relaxing the condition that $b'$ should be aligned with Triton's pole.
This might be the case if other causes of distortion than zonal winds are at work,
for example tidal forces from Neptune or Triton's potential anomalies.
Those orientations provide more stringent upper limits of the apparent
oblateness $\epsilon'$ because the cusps of the caustic can then get closer
to the paths of the central-most stations (Fig.~\ref{fig_map_2D_flash_upper_limit_0.0014}).
For instance, rotating the caustic by 45~degrees imposes the more stringent 
1$\sigma$ upper limit, $\epsilon < 0.00074$ (instead of 0.0019).
%
%
This requires an equatorial wind of $\sim 40$~m~s$^{-1}$, which is still quite a bit larger 
than the values $\sim 10$~m~s$^{-1}$ expected from GCMs, and thus
not a constraining limit as far as GCMs are concerned.

We now compare our upper limit for the deprojected oblateness of the central flash layer
($\epsilon < 0.0024$) with the value obtained by \cite{ell97} from a single cut inside
the central flash region during the 14 August 1995 occultation. 
Two solutions are considered by those authors, 
an oblate flash layer with $\epsilon= 0.042$ and a prolate one with $\epsilon= -0.032$.
Adopting the $\epsilon= 0.042$ value, we obtain the central map displayed in 
the right panel of Fig.~\ref{fig_map_2D_flash_upper_limit_0.0014}.
This would imply the crossing of the caustic by the five stations shown in the map,
and result in strong flux variations at those crossings that are not observed in the data (see Fig.~\ref{fig_fit_flash_cal_con}).
A similar conclusion would be drawn by adopting the prolate value $\epsilon= -0.032$ 
of \cite{ell97}. 
We note that the closest approach distance to the shadow centre during 
the central observation flash of 1995 was about 100~km, which is well outside 
the diamond-shaped caustic displayed in the right panel of
Fig.~\ref{fig_map_2D_flash_upper_limit_0.0014}.
Thus, caustic crossings could not be tested at that epoch.

The main problem of the large oblateness values obtained by \cite{ell97} is that
they imply unrealistically large wind velocities to maintain such distortions.
For instance, taken at face value, $\epsilon= 0.042$ results in an equatorial
wind velocity of $v_{\rm e} \sim 300$~m~s$^{-1}$, 
more than twice the sonic velocity near Triton's surface 
($\sim 130$~m~s$^{-1}$ at $\sim 40$~K) and 
much larger than predicted by GCMs (see above).
A possibility considered by \cite{ell97} was that Triton's atmospheric distortion was
restricted to mid-latitude regions (i.e. it was local rather than global). 
This permits lower values of wind speeds (110-170~m~s$^{-1}$) depending on whether 
prolate or oblate solutions are considered. 
This is still essentially supersonic and not expected from circulation models.
Moreover, this distortion would also be detected in our dataset, which densely sampled 
the central flash region.

An alternative explanation proposed by \cite{ell97} is the presence of hazes that 
absorbed part of the stellar flux at some specific locations along Triton's limb, 
thus altering the central flash shape. However, according to those authors, 
neither the optical depth obtained for those hazes at the corresponding altitude
levels, nor its dependence with wavelength, were consistent with the V2 results.
The haze problem is discussed in more detail in Sect.~\ref{sec_discuss}.

Finally, the shape of Triton's solid body as observed in V2 images indicates 
an oblateness smaller than 0.0014 \citep{tho00}, which is too small to explain the
claimed distortions.
Moreover, non-radial components of Triton's inner gravitational field might
also cause atmospheric distortions, but they would then be permanent and thus, 
should be observed also in 2017. 
From all this, we conclude that the large oblatenesses reported by \cite{ell97} are both theoretically unexpected and inconsistent with our observations.

At this point, we think that the mismatches between the flash models and its observation in 1995 
(as well as small departures from our spherical model in Fig.~\ref{fig_fit_flash_cal_con})
could be caused by small local corrugations of the flash layer induced by gravity waves.
As the caustic is the locus of the limb centre of curvatures, 
its shape is very sensitive to local (but small) corrugations of that layer.
Examples of such effects have been investigated for explaining flash shapes in the cases
of stellar occultations by Neptune \citep{hub88} and Titan \citep{sic06}.
As mentioned earlier, this approach remains beyond the scope of this paper, as long as 
zonal wind and gravity wave regimes are not available for Triton's lower atmosphere.


\section{Pending issues}
\label{sec_discuss}

\subsection{Hazes}

In the present work, we assumed that Triton's atmosphere is clear (i.e. free of absorbing material). 
However, V2 observations in 1989 revealed two kind of absorbing features
in the lower atmosphere: hazes and clouds.
These features were detected in the visible through the imaging system \citep{rag92} 
and during the UV occultation experiment \citep{kra92,kra93,krcr95}.

The general picture that emerges from the V2 observations is reviewed in \cite{yel95}
and is described as follows:
hazes are detected up to an altitude of about 30~km, 
they were observed around the entire Triton's limb, except for a small clear
region near east longitude 280$^\circ$ and between latitudes 4 and 18$^\circ$~S.
Thus, unless drastic changes in haze formation occurred, they should also be 
present during our observation of 5 October 2017.
Clouds are observed closer to the surface compared to hazes (i.e. below an altitude 
level of about 8~km). Contrary to hazes, they exhibit a patchier distribution
along the limb.

Due to their low altitudes, absorbing materials should be best detected in the 
central flash structures, caused by a layer at about 8~km altitude.
Estimation of the integrated (down to the surface) vertical optical depth 
of the hazes at 0.47~$\mu$m is $\tau_{\rm vis}= 0.005 \pm 0.001$, 
with a typical scale height of $H_{\rm h} \sim 12$~km \citep{kra93,rag92}.
Thus, the integrated vertical optical depth down to the 8~km altitude level
should be reduced by a factor of $\exp(-8/H_{\rm h})~\sim~0.5$.
Moreover, the slant optical depth  (along the line of sight) is amplified by
a factor of $\sqrt{2\pi R_T/H_{\rm h}} \sim 27$,
yielding a slant optical depth of hazes at 8~km of the order of
$0.005~\sqrt{2\pi R_T/H_{\rm h}}~\exp(-8/H_{\rm h})~\sim~0.07$, and a reduction in the flash amplitude by 5-10\% if no changes have occurred since 1989.

Clouds are much denser than hazes, with vertical optical depth to the surface 
of 0.1 or higher. Cloud particles have a vertical distribution with a 
scale height $H_{\rm c}$ comparable to or larger than the atmospheric 
scale height (about 20~km).
Thus, the slant optical depth of clouds at 8~km should be in the order of
$0.1~\sqrt{2\pi R_T/H_{\rm c}}~\exp(-8/H_{\rm c})~\sim~1.4$ 
or larger, corresponding to a decrease in the flash amplitude by a factor of 4 or more.

To summarise, hazes are expected to have a mild effect on the central flash heights,
with an expected reduction of only 5-10\%.
We do not see any departure from the model at that level on the best flash profiles.
Moreover, we do not detect trends on the observed flash amplitudes versus wavelength. 
For instance, the strong flash at Const\^ancia (Fig.~\ref{fig_fit_flash_cal_con})
observed at an effective wavelength of $\sim$0.6~$\mu$m agrees with the model
at the same satisfaction level as the Varages flash,
observed at $\sim$1.3~$\mu$m (Fig.~\ref{fig_fit_flash_1}). 
The same is true with the Calar Alto observation, which was made simultaneously
in the visible (0.4-1.0~$\mu$m) and the near IR (1.0-1.7~$\mu$m) (see Fig.~\ref{fig_fit_flash_2} and Table~\ref{tab_sites}).
A flash is observed at that station, but it is too faint to reveal a difference
between the two channels.
Finally, the dual observation made at Kryoneri (R and I bands) was too far away
from the centre line to show a central flash, but does not show significant
differences anyway in the fits by the synthetic light curve (Fig.~\ref{fig_fit_data_5}).

Taken at face value, these results indicate that hazes have no detectable effect
on the flash shapes.
This result is true \textit{a fortiori} for the clouds, which would reduce the flash amplitudes 
by a factor of 4 or greater; this is not observed in the data.
In other words, our model consistently explains all the flashes using 
a clear atmosphere.

However, a difficulty arises at this point since the height of the flash actually 
depends on the assumed template temperature profile. 
In order to disentangle the haze versus temperature effects, independent information on the thermal profile of the lower atmosphere is required.
This requirement could be met with the ALMA results reported by \cite{gur19}.
This issue remains, for the moment, beyond the scope of the present paper.

\subsection{Troposphere}

\cite{yelle91} inferred a troposphere from the V2 observations of geysers and clouds.
We do not observe this, as our deepest layer probed, at the central flash level, which coincides with the expected altitude of the tropopause. Therefore, our model is consistent with our central flashes, as discussed in Sect. \ref{sec_lower_atmo}, and there is no need to include a troposphere to accommodate for the data. However, this does not mean that we can exclude a troposphere as we do not have information down to this part of the lower atmosphere.

The absorption, caused by hazes and clouds, impacts our data and the model. For a troposphere to be included, we need independent measurements of the lower atmosphere.

\subsection{Gravity waves}

In Sect. \ref{sec_lower_atmo} and Fig. \ref{fig_fit_flash_cal_con} we mention that the residuals in these light curves show minor features that are probably attributed to atmospheric waves.
They present themselves as small fluctuations, with little effect on the overall light curve shape, and therefore we did not study them in detail.
This topic needs further analysis that is beyond the scope of the present paper.

\subsection{$\phi_0$ of inverted profiles}

The determinations of the baseline levels $\phi_0$ used for inverting the profiles 
of La Palma and Helmos remain an open issue, as they provide inconsistent results
in the deepest parts of the $T(r)$ profiles (Appendix \ref{sec_appen_phi0}).
This matter must be analysed further and needs independent results, in particular from ALMA, 
to confirm which $\phi_0$ should be used.


\section{Concluding remarks}
\label{sec_conclusion}

In this paper we present results obtained from the ground-based stellar occultation 
by Triton observed on 5 October 2017. The main goals were
$(i)$ retrieve the general structure of Triton's atmosphere between altitude levels
of $\sim$8~km ($\sim$9~$\mu$bar) and $\sim$190~km (few-nanobar level) and
$(ii)$ compare these results with other ground-based occultations 
and the V2 radio occultation to assess the pressure seasonal variation over the
last three decades.

The 2017 event yielded 90 positive observations, 42 of which showed a central flash.
We used Abel inversions to retrieve density, pressure, and temperature profiles from 
our three best S/N light curves.
We find a hint of a mild negative temperature gradient (reaching -0.2~K~km$^{-1}$) 
at the deepest part of our profiles (i.e. below the altitude of $\sim$30~km).
This constitutes a mesosphere just above an expected stratosphere with a positive temperature gradient 
that connects the atmosphere to the cold surface.

A ray-tracing approach was used for a global fit to the best 52 light curves, providing
a pressure $p_{1400}= 1.18 \pm 0.03$~$\mu$bar at radius 1400~km.
It also provides a synthetic and smoothed model to extrapolate the density, pressure, and
temperature down to the surface.

A new analysis of the V2 radio experiment, with useful information extracted from the surface 
up to around 1400~km, shows that the pressure retrieved in the 2017 event is consistent with the pressure
obtained in 1989.
A survey of pressure values obtained between 1989 and 2017 was conducted.
The two past occultations (in 1997 and 2008), reanalysed using our methods,
indicate that the surface pressure reported in the 1990s is real, 
but this remains debatable due to the scarcity of high S/N light curves and 
the lack of a fully consistent analysis of the best datasets used by other teams.

The pressure that we obtain from the 2017 occultation is consistent with that derived from the Voyager radio experiment, meaning that the pressure is back to its 1989 level. 
Results from a VTM of Triton, described in detail in \cite{ber22}, do not support a strong increase in surface pressure in the last few decades but instead a modest increase, with a surface pressure reaching up to $20~\mu$bar in the intervening 28 years. 
The VTM simulations also suggest that
(1)~a strong increase in surface pressure before 2000 cannot be obtained if N2 is present between latitude 30$^\circ$~S - 0$^\circ$, and 
(2)~a northern polar cap should have extended down to at least 45$^\circ$~N - 60$^\circ$~N in 2017 to have the surface pressure back at the V2 level from 1989.

Finally, the central flash analysis does not reveal any evidence of an atmospheric distortion. 
The atmosphere appears as globally spherical, 
with a 1$\sigma$ upper limit of 0.0011 for its apparent oblateness near the 8~km altitude.
This corresponds to a global difference of less than 1.5~km 
between the largest and smallest atmospheric radii at that altitude.
This is much smaller than values reported in the literature. In particular, this does not
support the existence of supersonic winds previously claimed by \cite{ell97}.

Open issues, requiring a more specific analysis of our dataset, will be addressed elsewhere.
These include the possible presence (or absence) of hazes and of a troposphere just above Triton's surface, 
as well as the possible detection of gravity waves.

\begin{acknowledgements}
J.M.O. acknowledges financial support from the Portuguese Foundation for Science and Technology (FCT) and the European Social Fund (ESF) through the PhD grant SFRH/BD/131700/2017.

The work leading to these results has received funding from the 
European Research Council under the European Community's H2020 2014-2021 ERC grant Agreement n$^\circ$ 669416 ``Lucky Star''.

We thank S. Para who supported some travels to observe the 5 October 2017 occultation.

T.B. was supported for this research by an appointment to the National Aeronautics and Space Administration (NASA) Post-Doctoral Program at the Ames Research Center administered by Universities Space Research Association (USRA) through a contract with NASA.

We acknowledge useful exchanges with Mark Gurwell on the ALMA CO observations.

This work has made use of data from the European Space Agency (ESA) mission {\it Gaia} (\url{https://www.cosmos.esa.int/gaia}), processed by the {\it Gaia} Data Processing and Analysis Consortium (DPAC, \url{https://www.cosmos.esa.int/web/gaia/dpac/consortium}).
Funding for the DPAC has been provided by national institutions, in particular the institutions participating in the {\it Gaia} Multilateral Agreement. 

J.L.O., P.S.-S., N.M. and R.D. acknowledges financial support from the State Agency for Research of the Spanish MCIU through the ``Center of Excellence Severo Ochoa'' award to the Instituto de Astrofísica de Andalucía (SEV-2017-0709), they also acknowledge the financial support by the Spanish grant AYA-2017-84637-R and the Proyecto de Excelencia de la Junta de Andalucía J.A. 2012-FQM1776. The research leading to these results has received funding from the European Union's  Horizon  2020  Research  and  Innovation  Programme,  under Grant Agreement no. 687378, as part of the project ``Small Bodies Near and Far'' (SBNAF).
P.S.-S. acknowledges financial support by the Spanish grant AYA-RTI2018-098657-J-I00 ``LEO-SBNAF''. 

The work was partially based on observations made at the Laboratório Nacional de Astrofísica (LNA), Itajubá-MG, Brazil. 
The following authors acknowledge the respective CNPq grants: 
F.B.-R. 309578/2017-5; 
R.V.-M. 304544/2017-5, 401903/2016-8; 
J.I.B.C. 308150/2016-3 and 305917/2019-6; 
M.A. 427700/2018-3, 310683/2017-3, 473002/2013-2. 
This study was financed in part by the Coordenação de Aperfeiçoamento de Pessoal de Nível Superior - Brasil (CAPES) - Finance Code 001 and the National Institute of Science and Technology of the e-Universe project (INCT do e-Universo, CNPq grant 465376/2014-2). 
G.B.R. acknowledges CAPES-FAPERJ/PAPDRJ grant E26/203.173/2016 and CAPES-PRINT/UNESP grant 88887.571156/2020-00, M.A. FAPERJ grant E-26/111.488/2013 and A.R.G.Jr. FAPESP grant 2018/11239-8. 
B.E.M. thanks CNPq 150612/2020-6 and CAPES/Cofecub-394/2016-05 grants.

Part of the photometric data used in this study were collected in the frame of the photometric observations with the robotic and remotely controlled telescope at the University of Athens Observatory (UOAO; \citealp{gaz16}).

The 2.3 m Aristarchos telescope is operated on Helmos Observatory by the Institute for Astronomy, Astrophysics, Space Applications and Remote Sensing of the National Observatory of Athens. Observations with the 2.3 m Aristarchos telescope were carried out under OPTICON programme. This project has received funding from the European Union's Horizon 2020 research and innovation programme under grant agreement No 730890. This material reflects only the authors views and the Commission is not liable for any use that may be made of the information contained therein.

The 1.2 m Kryoneri telescope is operated by the Institute for Astronomy, Astrophysics, Space Applications and Remote Sensing of the National Observatory of Athens.

The Astronomical Observatory of the Autonomous Region of the Aosta Valley (OAVdA) is managed by the Fondazione Clément Fillietroz-ONLUS, which is supported by the Regional Government of the Aosta Valley, the Town Municipality of Nus and the ``Unité des Communes valdôtaines Mont-Émilius''. The 0.81 m Main Telescope at the OAVdA was upgraded thanks to a Shoemaker NEO Grant 2013 from The Planetary Society. D.C. and J.M.C. acknowledge funds from a 2017 'Research and Education' grant from Fondazione CRT-Cassa di Risparmio di Torino.

P.M. acknowledges support from the Portuguese Fundação para a Ciência e a Tecnologia ref. PTDC/FISAST/29942/2017 through national funds and by FEDER through COMPETE 2020 (ref. POCI010145 FEDER007672).

F.J. acknowledges Jean Luc Plouvier for his help.

S.J.F. and C.A. would like to thank the UCL student support observers: Helen Dai, Elise Darragh-Ford, Ross Dobson, Max Hipperson, Edward Kerr-Dineen, Isaac Langley, Emese Meder, Roman Gerasimov, Javier Sanjuan, and Manasvee Saraf. 

We are grateful to the CAHA, OSN and La Hita Observatory staffs. This research is partially based on observations collected at Centro Astronómico Hispano-Alemán (CAHA) at Calar Alto, operated jointly by Junta de Andalucía and Consejo Superior de Investigaciones Científicas (IAA-CSIC). This research was also partially based on observation carried out at the Observatorio de Sierra Nevada (OSN) operated by Instituto de Astrofísica de Andalucía (CSIC). 

This article is also based on observations made with the Liverpool Telescope operated on the island of La Palma by Liverpool John Moores University in the Spanish Observatorio del Roque de los Muchachos of the Instituto de Astrofisica de Canarias with financial support from the UK Science and Technology Facilities Council.

Partially based on observations made with the Tx40 and Excalibur telescopes at the Observatorio Astrofísico de Javalambre in Teruel, a Spanish Infraestructura Cientifico-Técnica Singular (ICTS) owned, managed and operated by the Centro de Estudios de Física del Cosmos de Aragón (CEFCA). Tx40 and Excalibur are funded with the Fondos de Inversiones de Teruel (FITE).

A.R.R. would like to thank Gustavo Román for the mechanical adaptation of the camera to the telescope to allow for the observation to be recorded.

R.H., J.F.R., S.P.H. and A.S.L. have been supported by the Spanish projects AYA2015-65041-P and PID2019-109467GB-100 (MINECO/FEDER, UE) and Grupos Gobierno Vasco IT1366-19.

Our great thanks to Omar Hila and their collaborators in Atlas Golf Marrakech Observatory for providing access to the T60cm telescope.

TRAPPIST is a project funded by the Belgian Fonds (National) de la Recherche Scientifique (F.R.S.-FNRS) under grant PDR T.0120.21. TRAPPIST-North is a project funded by the University of Liège, and performed in collaboration with Cadi Ayyad University of Marrakesh. E.J. is a FNRS Senior Research Associate.
\end{acknowledgements}

\bibliographystyle{aa}
\bibliography{references}




\longtab{
\begin{longtable}{lllll}
\caption{
\label{tab_sites}
Circumstances of observations.
} \\
\hline \hline
\textbf{Site} & \textbf{Coordinates} & \textbf{Telescope aperture (m)} & \textbf{Exp. Time/Cycle (s)} & \textbf{Observers} \\
              & \textbf{Altitude (m)} & \textbf{Instrument/filter} &        &            \\
\hline
\endfirsthead
\multicolumn{5}{c}
{\tablename\ \thetable\ -- \textit{Continued from previous page}} \\
\textbf{Site} & \textbf{Coordinates} & \textbf{Telescope aperture (m)} & \textbf{Exp. Time/Cycle (s)} & \textbf{Observers} \\
              & \textbf{Altitude (m)} & \textbf{Instrument/filter} &        &            \\
\hline
\hline
\endhead
\hline \multicolumn{5}{r}{\textit{Continued on next page}} \\
\endfoot
\hline
\endlastfoot

\multicolumn{5}{c}{\textbf{Observations that provided light curves that are used in Triton's atmospheric fit}} \\
\hline
Agerola         &  40 37 26.2 \ N   &  0.28         &  0.86/0.86    &  L. Morrone                 \\
Italy           &  14 33 50.6 \ E   &  CCD/clear    &               &                                    \\
                &  660.0            &               &               &                                            \\

Albox           &  37 24 20.0 \ N   &  0.41         &  2.00/3.34    &  J. L. Maestre Garc\'ia   \\
Spain           &  02 09 07.0 \ W   &  CCD/clear    &               &                           \\
                &  485.0            &               &               &                           \\

Alcantarilha    &  37 07 58.8 \ N   &  0.36         &  0.33/0.64    &  F. Marques Dias          \\
Portugal        &  08 21 54.0 \ W   &  CCD/red      &               &                           \\
                &  65.0             &               &               &                           \\

Algiers         &  36 47 52.1 \ N   &  0.81         &  0.04/0.04    &  D. Baba Aissa,           \\
Algeria         &  03 01 56.2 \ E   &  video/clear  &               &  Z. Grigahcene            \\
                &  331.0            &               &               &                           \\

Ariana          &  36 53 03.0 \ N   &  0.20         &  0.50/0.60    &  S. Kamoun                \\
Tunisia         &  10 11 42.0 \ E   &  CCD/clear    &               &                           \\
                &  5.0              &               &               &                           \\

Athens          &  37 58 06.8 \ N   &  0.40         &  1.00/3.50    &  K. Gazeas,               \\
Greece          &  23 47 00.1 \ E   &  CCD/clear    &               &  E. Karampotsiou,         \\
                &  250.0            &               &               &  L. Tzouganatos           \\

Belesta         &  43 26 39.4 \ N   &  0.30         &  0.64/0.64    &  P. André,                \\
France          &  01 48 58.6 \ E   &  video/clear  &               &  M. Llibre,               \\
                &  235.0            &               &               &  F. Pailler               \\

Boissettes      &  48 31 41.0 \ N   &  0.30         &  0.50/0.50    &  M. Irzyk                                   \\
France          &  02 37 28.0 \ E   &  CCD/clear    &                   &                                                 \\
                &  75.0             &               &               &                                            \\

Calar Alto      &  37 13 24.7 \ N   &  1.23         &  VIS: 0.063/0.063 &  R. Hueso,            \\
(2 channels) & 02 32 44.9 \ W & SCMOS/VIS: (0.4-1.0 $\mu$m) & SWIR: 0.11/0.11 & S. Pérez-Hoyos,  \\
Spain           &  2160.0           &  SWIR: (1.0-1.7 $\mu$m)) &    &  A. Sánchez-Lavega        \\

Calern          &  43 45 13.2 \ N   &  1.04         &  0.10/0.11    &  J. Ferreira, P. Machado, \\
France          &  06 55 22.4 \ E   &  CCD/I'       &               &  P. Tanga, J.-P. Rivet    \\
                &  1268.0           &               &               &                           \\

Catania         &  37 41 35.8 \ N   &  0.80         &  0.80/1.70    &  G. Leto,                 \\
Italy           &  14 58 29.1 \ E   &  CCD/clear    &               &  R. Zanmar Sanchez,       \\
                &  1725.0           &               &               &  P. Bruno, G. Occhipinti  \\

Clanfield       &  50 56 19.2 \ N   &  0.61         &  0.16/0.16    &  D. Briggs,               \\
England         &  01 01 10.6 \ W   &  video/clear  &               &  S. Broadbent             \\
                &  155.0            &               &               &                           \\

Const\^ancia    &  39 29 41.6 \ N   &  0.51         &  0.64/0.64    &  R. Gon\c{c}alves,        \\
Portugal        &  08 19 25.2 \ W   &  video/clear  &               &  M. Ferreira              \\
                &  147.0            &               &               &                           \\

C\^otes de Meuse & 49 00 07.2 \ N   &  0.83         &  0.06/0.06    &  S. Renner,               \\
France          &  05 41 06.6 \ E   &  EMCCD/clear  &               &  M. Kaschinski            \\
                &  284.0            &               &               &                           \\

Cuq les Vielmur &  44 29 05.0 \ N   &  0.20         &  2.00/4.00    &  A. Cailleau,             \\
France          &  01 55 05.0 \ E   &  CCD/clear    &               &  V. Pic,                  \\
                &  300.0            &               &               &  L. Granier               \\

Dark Sky Obs.   &  36 15 09.6 \ N   &  0.36         &  1.50/2.50    &  J. Pollock,                                \\
United States   &  81 25 01.2 \ W   &  CCD/clear    &               &  D. B. Caton                                \\
                &  960.0                    &               &               &  V. Kouprianov                     \\

Elvas           &  38 50 47.6 \ N   &  0.27         &  0.30/0.30    &  W. Beisker               \\
Portugal        &  07 12 27.4 \ W   &  CCD/clear    &               &                           \\
                &  285.0            &               &               &                           \\

Felsina Obs.    &  44 21 22.0 \ N   &  0.40         &  0.64/0.64    &  R. Di Luca,              \\
Italy           &  11 09 09.0 \ E   &  video/clear  &               &  D. Alboresi              \\
                &  652.0            &               &               &                           \\

Forcarei        &  42 36 38.4 \ N   &  0.51         &  0.64/0.64    &  R. Iglesias-Marzoa,      \\
Spain           &  08 22 15.1 \ W   &  video/clear  &               &  H. Gonz\'alez            \\
                &  673.0            &               &               &                           \\

Hartley Wintney &  51 18 23.2 \ N   &  0.36         &  1.00/3.31    &  G. Thurston              \\
England         &  00 54 44.9 \ W   &  CCD/Johnson V &              &                           \\
                &  65.0             &               &               &                           \\

Helmos          &  37 59 08.1 \ N   &  2.28         & 0.60/0.674    &  E. M. Xilouris, I. Alikakos, \\
Greece          &  22 11 54.6 \ E   &  CCD/V+R      &               &  A. Gourzelas,            \\
                &  2323.0           &               &               &  V. Charmandaris          \\

Hornchurch      &  51 33 06.4 \ N   &  0.24         &  0.64/0.64    &  P. Denyer                \\
England         &  00 11 38.8 \ E   &  video/clear  &               &                                               \\
                &  14.0             &               &               &                                               \\
                           
Ithaca          &  42 27 29.4 \ N   &  0.60         &  0.20/0.36    &  J. Lloyd,                          \\
United States   &  76 23 05.5 \ W   &  CCD/clear    &               &  M. El Moutamid,                    \\
                &  530.0            &               &               &  C. Lamarche                                \\

Javalambre Astrophysical & 40 02 30.6 \ N & 0.40 Tx40 CCD/clear & 2.00/2.50 & R. Iglesias-Marzoa,           \\
Observatory (2 telescopes) & 01 00 57.6 \ W & 0.28 Excalibur CCD/clear & 2.00/3.00 & J. Abril Ib\'a\~nez,                \\
Spain           &  1955.0           &               &               &  M. Chioare D\'iaz Mart\'in \\

Kryoneri        &  37 58 19.0 \ N   &  1.20         &  0.19/0.20    &  E. M. Xilouris,                    \\
(2 channels)    &  22 37 07.0 \ E   & sCMOS/R and I &               &  A. Liakos,                         \\
Greece          &  930.0            &               &               &  V. Charmandaris                    \\

La Palma        &  28 45 43.2 \ N   &  2.00         & 0.6/0.635     &  J. Marchant,                       \\
Spain               &  17 52 39.3 \ W   &  Liverpool Tel. &             &  B. Sicardy                             \\
                &  2340.0               &  CCD/I+Z      &               &                                                 \\

La Carolina     &  38 16 27.0 \ N   &  0.64         &  0.15/0.15    &  S. Alonso,                 \\
Spain           &  03 36 55.0 \ W   &  CCD/clear    &               &  A. Rom\'an Reche                   \\
                &  595.0            &               &               &                                            \\

La Roche-Sur-Yon & 46 32 48.4 \ N   &  0.40         &  0.20/0.20    &  J. Desmars,                                \\
France          &  01 19 37.8 \ W   &  CMOS/clear   &               &  R. Tanguy,                         \\
                &  48.0             &               &               &  J. David                                   \\

Le Beausset     &  43 11 38.1 \ N   &  0.25         &  0.64/0.64    &  S. Lisciandra,                     \\
France          &  05 48 05.9 \ E   &  video/clear  &               &  J. F. Coliac                               \\
                &  197.0            &               &                       &                                                \\

Marrakech       &  31 35 16.2 \ N   &  0.60         &  2.5/5.5      &  A. Daassou, K. Barkaoui,   \\
Morocco         &  08 00 46.9 \ W   &  CCD/clear    &               &  Z. Benkhaldoun,            \\
                &  494.0            &               &               &  M. Guennoun, J. Chouqar  \\

Nancy           &  48 39 51.6 \ N   &  0.25         &  1.50/2.10    &  D. Lavandier,                      \\
France          &  06 09 28.7 \ E   &  CCD/luminance &              &  D. Walliang                                \\
                &  284.0            &               &               &                                            \\

Newark          &  43 00 23.7 \ N   &  0.25         &  0.53/0.53    &  B. Timerson                                \\
United States   &  77 07 06.5 \ W   &  video/clear  &               &                                            \\
                &  165.0                    &               &               &                                            \\

Ouka\"imeden    &  31 12 23.0 \ N   &  0.50         &  2.00/6.80    & C. Rinner                              \\
Morocco         &  07 51 59.0 \ W   &  CCD/clear    &               &                                            \\
                &  2727.0           &               &               &                                    \\

Paris           &  48 52 17.0 \ N   &  0.15         &  0.25/0.25    &  R. Chauvet                         \\
France          &  02 23 07.0 \ E   &  CCD/clear    &               &                                            \\
                &  92.0                     &                   &                           &                                            \\

Pic du Midi Obs.&  42 56 12.0 \ N   &  0.60         &  0.05/0.05    &  D. Berard                          \\
France          &  00 08 31.9 \ E   &  EMCCD/visible &              &                                    \\
                &  2862.0                   &               &               &                                            \\

Puimichel       &  43 58 53.1 \ N   &  0.60         &  0.10/0.10    &  S. Moindrot                \\
France          &  06 02 10.0 \ E   &  EMCCD/visible &              &                                    \\
                &  714.0            &               &               &                                            \\

Reading         &  51 30 23.8 \ N   &  0.30         &  0.64/0.64    &  T. V. Haymes                       \\
England         &  00 48 58.1 \ W   &  video/red    &               &                                            \\
                &  80.0                     &               &               &                                            \\

Sabadell        &  41 33 00.2 \ N   &  0.50         &  0.16/0.16    &  A. Selva,                          \\
Spain           &  02 05 24.8 \ E   &  video/clear  &               &  C. Perello,                                \\
                &  224.0                    &               &               &  V. Cabedo                         \\

Saint Caprais   &  43 52 25.9 \ N   & 0.20 CCD/Johnson V & 1.00/1.58 & A. Klotz,                          \\
(2 telescopes)  &  01 43 05.8 \ E   & 0.94 SWIR/clear    & 0.05/0.08 & Y. Rieugnie,                       \\
France          &  185.0                    &                    &           & A. N. Klotz                               \\

Seysses         &  43 30 06.3 \ N   &  0.31         &  0.32/0.32    &  M. Boutet                          \\
France          &  01 17 20.3 \ E   &  video/UV-IR block &          &                                            \\
                &  183.0                    &               &               &                                            \\
        
Sierra Nevada   &  37 03 51.0 \ N   &  1.50         &  0.04/0.04    &  J. L. Ortiz,                               \\
Spain           &  03 23 05.0 \ W   &  video/clear  &               &  P. Santos-Sanz,                    \\
                &  2925.8                   &               &               &  V. Casanova                               \\

Southampton     &  50 55 18.9 \ N   &  0.40         &  0.25/0.25    &  N. J. Haigh                                \\
England         &  01 22 28.1 \ W   &  CMOS/clear   &               &                                            \\
                &  16.0                     &               &               &                                            \\

Sternwarte Stuttgart &  48 46 56.7 \ N   &  0.41         &  0.64/0.64    &  A. Eberle                            \\
Germany         &  09 11 47.4 \ E   &  video/clear  &                      &  K. Rapp                                    \\
                &  346.0                    &               &               &                                            \\

Steyning        &  50 53 16.0 \ N   &  0.28         &  2.00/4.00    &  N. Quinn                           \\
England         &  00 19 57.0 \ W   &  CCD/luminance &              &                                            \\
                &  20.0                     &               &               &                                            \\

Tournefeuille   &  43 34 58.8 \ N   &  0.32         &  0.25/0.25    &  M. Delcroix                                \\
France          &  01 19 35.4 \ E   &  CCD/clear    &               &                                            \\
                &  163.0                    &               &               &                                            \\

UCL Observatory &  51 36 47.6 \ N   &  0.35         &  1.00/2.38    &  S. J. Fossey,                      \\
London          &  00 14 32.3 \ W   &  CCD/luminance &              &  C. Arena                               \\
England         &  82.0             &               &               &                                            \\

Valencia        &  39 56 42.0 \ N   &  0.50         &  1.00/2.50    &  V. Peris,                          \\
Spain           &  01 06 05.4 \ W   &  CCD/clear    &               &  O. Brevia                          \\
                &  1280.0                   &               &               &                                            \\

Varages         &  43 36 44.6 \ N   &  0.50                             &  2.30/2.30    &  F. Jabet       \\
France          &  05 57 49.1 \ E   &  InGaAs/RG1000 &                  &                                                 \\
                &  300.0                    &  (effective $\lambda$=1.3 $\mu$m) &               &                   \\
                   
\hline
\multicolumn{5}{c}{\textbf{Observations that provided light curves that are not used in Triton's atmospheric fit}} \\
\hline
Abingdon        &  51 37 53.1 \ N   &  0.30         &  0.16/0.16    &  J. Talbot                          \\
England         &  01 16 55.2 \ W   &  video/clear  &               &                                            \\
                &  59.0             &               &               &                                            \\

Agerola         &  40 37 26.2 \ N   &  0.50         &  0.50/0.50    &  A. Noschese,                       \\
Italy           &  14 33 50.6 \ E   &  CMOS/clear   &               &  A. Vecchione                       \\
                &  660.0                    &               &               &                                                \\

Calar Alto      &  37 13 24.7 \ N   &  0.36         & 1.00/1.00     &  J. F. Rojas,               \\
Spain           &  02 32 44.9 \ W   &  CCD/clear     &               &  A. S\'anchez-Lavega        \\
                &  2160.0                   &  SWIR          &               &                                       \\

Caserta         &  41 16 11.3 \ N   &  0.24         &  2.00/2.46    &  L. Cupolino                                \\
Italy           &  13 56 28.9 \ E   &  CCD/clear    &               &                                            \\
                &  407.0                    &               &               &                                            \\

Catania         &  37 30 24.0 \ N   &  0.28         &  0.64/0.64    &  C. Scalia,                                 \\
Italy           &  15 05 35.5 \ E   &  video/clear  &               &  R. Lo Savio,                       \\
                &  50.0             &               &                   &  G. Giardina                            \\

Charles Fehrenbach      & 50 05 01.5 \ N &  0.36         &  0.28/0.30    &  P. Morel                     \\
Observatory, La Biette  &  03 46 34.4 \ E   &  CCD/clear &               &                               \\
France                   &  191.0                   &            &               &                               \\

Glyfada-Athens  &  37 52 33.6 \ N   &  0.35         &  1.00/1.15    &  E. I. Kardasis,            \\
Greece          &  23 46 12.0 \ E   &  CCD/clear    &               &  A. Christou                                \\
                &  100.0            &               &               &                                            \\

Guirguillano    &  42 42 42.2 \ N   &  0.31         &  2.56/2.56    &  I. Ord\'o\~nez-Etxeberria, \\
Spain           &  01 51 54.3 \ W   &  video/clear  &               &  P. Martorell,                      \\
                &  595.0                    &               &               &  J. Salamero                               \\

Hamsey Green    &  51 19 09.4 \ N   &  0.28         &  1.00/1.00 &  M. Jennings                         \\
England         &  00 04 01.4 \ W   &  video/clear  &          &                                                     \\
                &  170.0                    &               &               &                                            \\

Houten          &  52 01 59.3 \ N   &  0.36         &  0.41/0.41    &  J. Sussenbach                      \\
Netherlands     &  05 09 44.2 \ E   &  CCD/         &               &                                            \\
                &  2.0              &  685 nm long pass &           &                                            \\

La Hita Observatory & 39 34 06.8 \ N & 0.77 CCD/luminance & 1.00/3.55 & N. Morales, J. L. Ortiz, \\
(2 telescopes)  &  03 11 09.5 \ W   &  0.40 video/clear & 0.30/0.30 & F. Organero, L. Ana,               \\
Spain               &  695.0                &                           &                      & F. Fonseca, P. Santos-Sanz \\

Lamandine       &  44 12 35.0 \ N   &  0.30         &  0.06/0.06    &  M. Miniou                          \\
France          &  01 42 01.7 \ E   &  CCD/clear    &               &                                            \\
                &  297.0                    &               &               &                                            \\

La Sagra Observatory & 37 58 58.1 \ N & 0.36 each   &  1.00/3.79 (tetra1) &  N. Morales,          \\
(4 telescopes) & 02 34 05.1 \ W & tetra1, tetra2: CCD/clear & 1.00/3.56 (tetra2) & J. L. Ortiz, \\
Spain           &  1530.0           & tetra3, tetra4: video/clear & 0.20/0.20 & P. Santos-Sanz        \\

Leeds           &  53 50 15.6 \ N   &  0.20         &  1.28/1.28    &  A. Pratt                                   \\
England         &  01 36 27.8 \ W   &  video/IR-UV block &          &                                                   \\
                &  113.0                    &               &               &                                            \\

Lias            &  43 33 33.3 \ N   &  0.40         &  0.13/0.13    &  F. Metz,                           \\
France          &  01 06 34.4 \ E   &  CCD/clear    &               &  D. Erpelding,                      \\
                &  300.0                    &               &               &  J.-P. Nougayrède              \\

Louargat            &  48 32 16.7 \ N   &  0.30         &  0.20/0.20    &  B. Reginato,                           \\
France          &  03 21 30.5 \ W   &  CMOS/L       &               &  E. Reginato                                \\
                &  196.0                &                       &                           &                           \\

Massa       & 44 01 17.2 \ N & 0.20 & 5.12/5.12 (1\textsuperscript{st} half) & P. Baruffetti    \\
Italy           &  10 07 56.7 \ E & video/clear & 2.56/2.56 (2\textsuperscript{nd} half) &      \\
                &  30.0             &               &               &                                            \\

Montigny le Bretonneux & 48 45 54.0 \ N &  0.28     &  0.50/0.51    &  O. Dechambre                       \\
France          &  02 00 52.0 \ E   &  CCD/clear    &                   &                                                 \\
                &  168.0            &               &               &                                               \\

Mount Agliale Observatory &  43 59 43.1 \ N   &  0.50         &  1.50/3.23    &  F. Ciabattari     \\
Italy           &  10 30 53.8 \ E   &  CCD/clear    &                   &                                                 \\
                &  750.0                    &               &               &                                            \\

Nerpio          &  38 09 56.0 \ N   &  0.32         &  1.00/16.11   &  E. Briggs                          \\
Spain           &  02 19 35.0 \ W   &  CCD/clear    &                   &                                                 \\
                &  1650.0           &               &               &                                            \\

Neutraubling    &  48 59 23.1 \ N   &  0.28         &  0.16/0.16       &  B. Kattentidt                  \\
Germany         &  12 12 57.3 \ E   &  video/clear  &                  &                                                 \\
                &  333.0            &               &                  &                                                 \\

Nice            &  43 43 32.9 \ N   &  0.13         &  2.00/2.00    &  M. Conjat                          \\
France          &  07 17 59.4 \ E   &  video/clear  &               &                                            \\
                &  350.0            &               &                       &                                            \\

Observatoire de Biscarmiau &  43 08 40.4 \ N & 0.31 &  0.64/0.64    &  G. Vaudescal                   \\
France              &  00 03 31.8 \ E   &  video/clear  &               &                                                 \\
                &  488.0                    &                   &                           &                                            \\
                
Observatoire des & 44 24 29.7 \ N   &  0.80         &  2.00/2.00    &  M. Bretton                         \\
Baronnies Proven\c{c}ales & 05 30 54.4 \ E & CCD/clear &            &                                           \\
France          &  820.0            &               &                       &                                            \\

Observatoire Monplaisir &  43 40 58.3 \ N & 0.28    &  1.03/1.49    &  J. Ardissone                               \\
France          &  04 38 32.6 \ E   &  CCD/clear    &               &                                    \\
                &  5.0                      &               &               &                                            \\

Orcemont        &  48 35 28.0 \ N   &  0.10         &  10.00/10.15  &  P. Delay                                   \\
France          &  01 48 45.0 \ E   &  CCD/IR cut   &                   &                                                 \\
                &  165.0                    &                   &                           &                                            \\

Overath         &  50 57 11.4 \ N   &  0.20         &  0.32/0.32    &  B. Klemt                       \\
Germany         &  07 14 53.1 \ E   &  video/clear  &               &                                            \\
                &  200.0            &                   &                           &                                            \\

Pf\"{u}nz               &  48 53 26.0 \ N   &  0.28             &  1.00/1.20    &  B. G\"{a}hrken                    \\
Germany         &  11 16 23.0 \ E   &  CCD/clear        &               &                                                 \\
                &  488.0                    &                   &                           &                                            \\

Pic du Midi Obs.&  42 56 12.0 \ N   &  1.05             &  0.10/0.10    &  F. Colas                               \\
France          &  00 08 31.9 \ E   &  sCMOS/clear  &                       &                                            \\
                &  2862.0                   &                   &                           &                                            \\

Rosenfeld-Brittheim & 48 17 17.8 \ N & 0.80             &  0.40/0.40    &  S. Kowollik,               \\
Germany             &  08 40 38.9 \ E   & CMOS/UV-IR block &            &  R. Bitzer                              \\
                &  700.0                &                       &                           &                                            \\

Saint Ch\'eron  &  48 32 16.6 \ N   &  0.18             &  0.08/0.08        &  J. Berthier                       \\
France          &  02 07 51.7 \ E   &  video/clear      &               &                                         \\
                &  160.0                &                       &                           &                                            \\

Saint-Maurice-Navacelles & 43 50 39.3 \ N & 0.21    &  0.64/0.64    &  J. Lecacheux             \\
France          &  03 33 47.3 \ E   &  CCD/clear    &               &                                            \\
                &  583.0                &                       &                           &                                                \\

Saint Michel    &  43 54 23.3 \ N   &  0.60             &  0.05/0.05    &  E. Meza,                       \\
France          &  05 43 34.7 \ E   &  EMCCD/clear  &                       &  O. Labrevoir              \\
                &  564.0                    &                   &                           &                                                \\

Selztal Obs.    &  49 50 00.0 \ N   & 0.51          & 0.8/0.8       &   G. M. Piehler                  \\
Friesenheim     &  08 15 00.0 \ E   & CMOS/clear    &                       &                                    \\
Germany         &  110.0                &                       &                           &                                            \\

Seysses             &  43 30 06.3 \ N   &  0.23         &  0.32/0.32    &  J. Sanchez                             \\
France              &  01 17 20.3 \ E   &  video/UV-IR block &          &                                                 \\
                &  183.0                    &                   &                           &                                            \\

Sierra Nevada   &  37 03 51.0 \ N   &  0.90             &  1.00/3.33    &  J. L. Ortiz,                       \\
Spain               &  03 23 05.0 \ W   &  CCD/clear    &               &  P. Santos-Sanz,                        \\
                &  2925.8                   &                   &                           &  V. Casanova                               \\

Valle d'Aosta   &  45 47 23.1 \ N   &  0.81         & 0.2/0.2       &  D. Cenadelli,              \\
Italy           &  07 28 42.3 \ E   &  EMCCD/clear  &                       &  J.-M. Christille,         \\
                &  1675.0                   &                   &                           &  B. Sicardy                                \\ 

\hline
\multicolumn{5}{c}{\textbf{Stations with technical or weather problems that provided no light curves}} \\
\hline
Bilbao          &  43 15 43.6 \ N   &  0.5          &               &  S. P\'erez-Hoyos,      \\
Spain           &  02 56 54.6 \ W   &  CCD/clear    &               &  A. S\'anchez-Lavega                \\
                &  47.0             &                   &                           &                                                \\

Buthiers            &  48 17 30.0 \ N   &  0.59         &  0.10/0.10    &  J. L. Dauvergne                    \\
France          &  02 26 18.0 \ E   &  CMOS/L       &               &                                            \\
                &  87.0             &                   &                           &                                                \\
 
Cavarc Observatory &  44 40 35.2 \ N &  0.50        &               &  J. Rudelle,                                \\
France          &  00 37 43.1 \ E   &  CMOS/clear   &               &  B. Tregon                          \\
                &  100.0                &                       &                           &                                            \\

Comthurey           &  53 15 57.8 \ N   &  0.18         &                   &  K. Guhl                                   \\
Germany             &  13 11 24.6 \ E   &                               &               &                                                 \\
                &  74.0                 &                       &                           &                                            \\

Emmendingen         &  48 06 14.2 \ N   &  0.18         &  3.00/3.33    &  K.-L. Bath                             \\
Germany         &  07 51 28.6 \ E   &  CMOS/clear       &               &                                                 \\
                    &  210.0            &                       &                           &                                            \\

Eppstein-Bremthal &  50 08 17.4 \ N &  0.25         &               &  O. Klös                                \\
Germany         &  08 21 50.4 \ E   &  video/clear  &               &                                            \\
                &  256.0                &                       &                           &                                            \\

Gnosca              & 46 13 53.2 \ N    & 0.40          &               & S. Sposetti                             \\
Switzerland     & 09 01 26.5 \ E    &               &               &                                            \\
                & 260.0             &                   &                           &                                            \\
                
Grapfontaine    &  49 48 54.2 \ N   &  0.28             &                   &  F. Van Den Abbeel         \\
Belgium         &  05 23 57.5 \ E   &  video/clear      &               &                                                 \\
                &  445.0            &                   &                           &                           \\

Hamburg-Bergedorf &  53 28 50.0 \ N &  0.60         &               &  V. Perdelwitz              \\
Germany         &  10 14 29.0 \ E   &  CCD/clear    &               &                                            \\
                &  25.0             &                   &                           &                                                \\

Handeloh                &  53 14 06.4 \ N   &  0.35         &               &  K. von Poschinger                 \\
Germany         &  09 49 46.7 \ E   &  CCD/clear    &               &                                            \\
                &  60.0             &                   &                           &                                                \\

Observatory Hoher List &  50 09 42.0 \ N   &  1.00  &            &  M. Miller, G. Herzogenrath, \\
Eifel           &  06 50 55.0 \ E   &  CCD/clear    &           &  D. Frangenberg, L. Brandis,  \\
Germany         &  549.0                &                       &                           &  I. P\"{u}tz                               \\
                          
Labastide-Murat &  44 38 43.2 \ N   &  0.20             &  0.32/0.32    &  E. Frappa                              \\
France          &  01 34 13.9 \ E   &  video/clear      &               &                                                 \\
                &  445.0                &                       &                           &                                            \\

Lauenbr\"{u}ck  &  53 12 26.6 \ N   &  0.50         &               &  M. Kretlow                 \\
Germany         &  09 34 36.1 \ E   &  CCD/clear    &               &                                            \\
                &  31.0             &                   &                           &                                            \\

Malvilliers             &  47 01 58.5 \ N   &  0.30             &                   &  L. Falco                                  \\
Switzerland     &  06 52 03.0 \ E   &  CCD/clear        &               &  R. Leiva                               \\
                &  845.0                &                       &                           &                                            \\

EXPO Observatory &  52 15 45.0 \ N  &  1.12         &  0.04/0.04    &  P. Riepe,                          \\
Melle           &  08 20 12.0 \ E   &  video/clear  &               &  E. Bredner                         \\
Germany         &  185.0                &                       &                           &                                            \\

Meudon              &  48 48 18.3 \ N   &  1.00         &                  &  T. Widemann, W. Thuillot, \\
France          &  02 13 51.6 \ E   &  CMOS/clear   &               &  D. Hestroffer,                     \\
                &  157.6                &                       &                           &  E. Lellouch                               \\

Ond\v{r}ejov    &  49 54 33.8 \ N   &  0.50         &  0.04/0.04    &  M. Jel\'{\i}nek,         \\
Czech Republic  &  14 46 52.9 \ E   & EMCCD/R       &               &  J. \v{S}trobl            \\
                &  530.0                &                       &                           &                                    \\

Saint-Luc               & 46 13 41.8 \ N    & 0.30          &               & E. Bouchet,               \\
Obs. F.-X. Bagnoud & 07 36 45.6 \ E & CCD/clear     &               & M. Cottier                \\
Switzerland         & 2200.0            &                       &                           &                                \\

Saulges             &  47 59 02.0 \ N   &  0.25         &  0.20/0.20    &  T. Midavaine                           \\
France          &  00 24 30.0 \ W   &  video/clear  &               &                                            \\
                &  97.0             &                   &                           &                                            \\

TRAPPIST-North  &  31 12 22.0 \ N   &  0.60             &               &  E. Jehin                               \\
Ouka\"imeden    &  07 51 59.0 \ W   &  CCD/clear    &               &                                    \\
Morocco         &  2751.0           &               &               &                                            \\

Uranoscope      &  48 44 32.0 \ N   &  0.28 video/clear &  0.25/0.25 &  A. Leroy,                      \\
(2 telescopes)  &  02 44 32.0 \ E   &  0.36 CMOS/clear &  0.25/0.25 &  S. Bouley                          \\
France          &  110.0                &                       &                           &                                            \\

Val\'ee de Joux &  46 37 06.0 \ N  & 0.61           &  1.00/1.15        & R. Barbosa,         \\
Switzerland     &  06 13 10.0 \ E  & CMOS/clear     &                   & R. Behrend        \\
                &  1145.0               &                       &                               & M. Spano                   \\

Vierzon             &  47 13 23.7 \ N   &  0.25         &                   &  L. Rousselot                              \\
France          &  02 03 09.8 \ E   &  CMOS/clear   &               &                                            \\
                &  97.0             &                   &                           &                                            \\

Zurich          &  47 24 27.8 \ N   &  0.50             &  2.00/4.00    &  S. Gallego,                            \\
Switzerland     &  08 30 39.5 \ E   &  CCD/Johnson R &              &  L. Tortorelli                      \\
                &  553.0                &                       &                           &                                            \\

\hline \hline
\multicolumn{5}{c}{\textbf{Previous occultations}} \\
\hline \hline
\multicolumn{5}{c}{\textbf{18 July 1997}} \\
\hline

Brownsville     & 25 58 40.9 \ N    & 0.35          & 0.5/0.5       & W. B. Hubbard,        \\
United States   & 97 32 11.3 \ W    & CCD/clear     &               & H. J. Reitsema,       \\
                &  9.0              &               &               & R. Hill               \\

Bundaberg       &  24 56 37.5 \ S   & 0.48          & 0.33/0.33     & E. Hummel, M. Moy,    \\
Australia       & 152 22 35.4 \ E   & CCD/clear     &               & I. Pink, R. Walters   \\
                & 10.0              &               &               &                       \\


Ducabrook       &  23 53 55.0 \ S   & 0.35          & 0.66/0.66     & L. Ball,              \\
Australia       & 147 26 40.0 \ E   & CCD/clear     &               & G. Neilsen            \\
                &  320.0            &               &               &                       \\

Lochington      &  23 56 42.5 \ S   & 0.35          & 0.66/0.66     & W. Beisker,           \\
Australia       & 147 31 24.8 \ E   & CCD/clear     &               & S. Hutcheon           \\
                &  270.0            &               &               &                       \\

\hline
\multicolumn{5}{c}{\textbf{Observations that provided data that are not used in Triton's atmospheric fit}} \\
\hline

Ballandean      & 28  49 05.0 \ S   & 0.48          &               & K. Lay                \\
Australia       & 151 48 36.0 \ E   & Visual        &               &                       \\
                & 710.0             &               &               &                       \\

Ipswich         & $\sim$27  38 \ S  & 0.30          & 10            & B. Downs              \\
Australia       & $\sim$152 45 \ E  & CCD/clear     & grazing       &                       \\
                & $\sim$40          &               &               &                       \\
                
The Gap         & 27  27 42.3 \ S   & 0.40          &               & P. Anderson           \\
Australia       & 152 55 58.0 \ E   & Visual        &               &                       \\
                & 176.0             &               &               &                       \\

\hline \hline
\multicolumn{5}{c}{\textbf{21 May 2008}} \\
\hline
 & & & & \\
\hline

Hakos           & 23 14 11.0 \ S    & 0.5          & 0.67/1.49     & K.-L. Bath     \\
Namibia         & 16 21 41.5 \ E    & CCD/clear    &               &                \\
                & 1825.0            &              &               &                \\

Hakos           & 23 14 11.0 \ S    & 0.45         & 0.84/1.0       & B. Sicardy     \\
Namibia         & 16 21 41.5 \ E    & CCD/clear    &                &                \\
                & 1825.0            &              &                &                \\

Ma\"{\i}do          & 21 04 15.5 \ S    & 0.23      & 2.56/2.56     & J. Lecacheux  \\
La R\'eunion Island & 55 23 14.2 \ E    & CCD/clear &               & T. Payet      \\
France              & 2200.0            &           &               &               \\

Tivoli          & 23 27 40.2 \ S    & 0.35          & 1.48/1.48     & H.-J. Bode    \\
Namibia         & 18 01 01.2 \ E    & CCD/clear     &               &               \\
                & 1344              &               &               &               \\

\hline
\multicolumn{5}{c}{\textbf{Observations that provided light curves that are not used in Triton's atmospheric fit}} \\
\hline

Piton Lacroix       & 21 12 54.4 \ S & 0.28         & 2/2            & E. Frappa     \\
La R\'eunion Island & 55 38 38.5 \ E & CCD/clear    &                &               \\
France              & 2330.0         &              &                &               \\

\hline
\multicolumn{5}{c}{\textbf{Stations with technical or weather problems that provided no light curves}} \\
\hline

Gr\"unau        & $\sim$27 44 \ S       & 0.30      & clouded out       & W. Beisker    \\
Namibia         & $\sim$18 23 \ E       &           &                   &               \\
                & $\sim$1100            &           &                   &               \\
                
Hakos           & 23 14 11.0 \ S        & 0.28      & ?/?               & C. Boissel    \\
Namibia         & 16 21 41.5 \ E        & (IOC?)    &                   & A. Doressoundiram \\
                & 1825                  &           &                   &               \\

Les Makes       & 21 11 57.4.0 \ S      & 0.35      & clouded out       & B. Payet          \\
La R\'eunion Island & 55 24 34.5.0 \ E  & CCD/clear &                   &                   \\
France          & 972                   &           &                   &                   \\

Ma\"{\i}do          & 21 04 15.5 \ S    & 0.30      & wind shaking      & J. Fran\c{c}oise  \\
La R\'eunion Island & 55 23 14.2 \ E    & video/clear &                 & B. Mondon         \\
France              & 2200.0            &           &                   &                   \\

Piton Lacroix       & 21 12 52.0 \ S    & 0.28      & moisture          & A. Peyrot         \\
La R\'eunion Island & 55 38 35.0 \ E    & CCD/clear &                   & J.-P. Teng-Chuen-Yu \\
France              & 2350.0            &           &                   &                   \\
                
Springbok       & 29 39 40.3 \ S        & 0.30          & clouded out   & T. Widemann       \\
South Africa    & 17 52 58.9 \ E        & CCD/clear     &               &                   \\
                & 951                   &               &               &                   \\

Sutherland      & 32 22 43.8 \ S        & 1.0       & outside shadow    & G. Blanchard      \\
South Africa    & 20 48 42.0 \ E        & CCD/clear &                   & M. Castets        \\
                & 1760                  &           &                   & F. Colas          \\
                
Tivoli          & 23 27 40.2 \ S        & 0.27          & ?/?           & B. Thome          \\
Namibia         & 18 01 01.2 \ E        & CCD/clear     &               &                   \\
                & 1344                  &               &               &                   \\
\hline

\end{longtable}
}


\begin{appendix}
\section{Circumstances of observations}
See table \ref{tab_sites}.
\section{Retrieval of atmospheric structure}
\label{sec_appen_validity}

Here we discuss some limitations and caveats associated
with the approach mentioned in the main text.

\subsection{Upper and lower limits of probed atmosphere}

The density $n(r)$ is reliably retrieved up to the level where 
the flux standard deviation $\sigma_\Phi$ due to photometric noise is 
comparable to the drop of stellar flux caused by the occultation (DO15). 
This corresponds to a density of
\begin{equation}
n_{\rm upper} \sim \frac{\sigma_\Phi}{K} \sqrt{\frac{H^3}{2\pi r D^2}},
\label{eq_n}
\end{equation}
where 
$K$ is the molecular refractivity of N$_2$, 
$H= -n/(dn/dr)$ is the scale-height and 
$D$ is Triton's geocentric distance.

Using the values of $K$ and $D$ listed in Table~\ref{tab_param},
considering that $H \sim 30$~km around $r= 1500$~km, and taking
$\sigma_\Phi \sim 0.011$ for our best dataset (La Palma station), 
we find that reliable density values cannot be obtained above $r_{\rm upper} \sim 1540$~km, 
where $n \sim 4 \times 10^{11}$~cm$^{-3}$, corresponding to pressures of a few nanobars.

The relative error on $n(r)$ reduces as $\exp[-(r-r_{\rm up})/H]$ 
when deeper levels are probed.
In practice, only the inversion of the best light curves provide useful retrieved profiles. 
In our case, this concerns the light curves from La Palma and Helmos stations,
providing the profiles displayed in Figs.~\ref{fig_n_p_r}-\ref{fig_dTdr_r}.

The deepest layers probed when inverting an occultation light curve are 
those corresponding to the closest approach of the observing station to Triton's shadow centre.
In our case, La Palma's light curve provides data 
down to a radius of about $r=1375$~km 
(i.e. just above the 20~km altitude level).
However, the central flashes provide constraints that go further down, 
typically just under the 8 km altitude level (see Sect.~\ref{sec_lower_atmo}).

\subsection{Constructing the temperature profile template}
\label{sec_appen_phi0}

\begin{figure}[!t]
\centerline{\includegraphics[totalheight=7cm, angle=0]{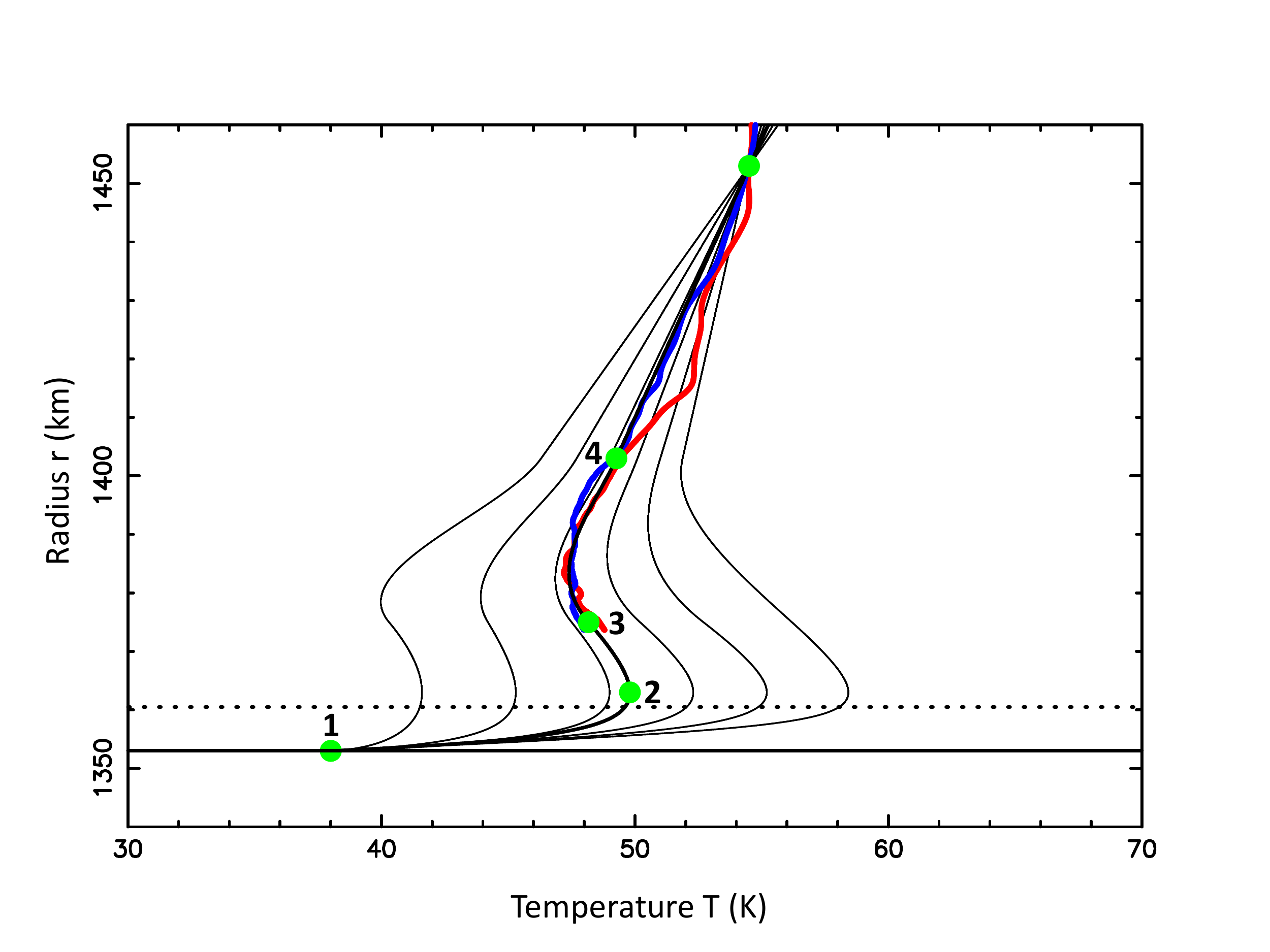}}
\centerline{\includegraphics[totalheight=7cm, angle=0]{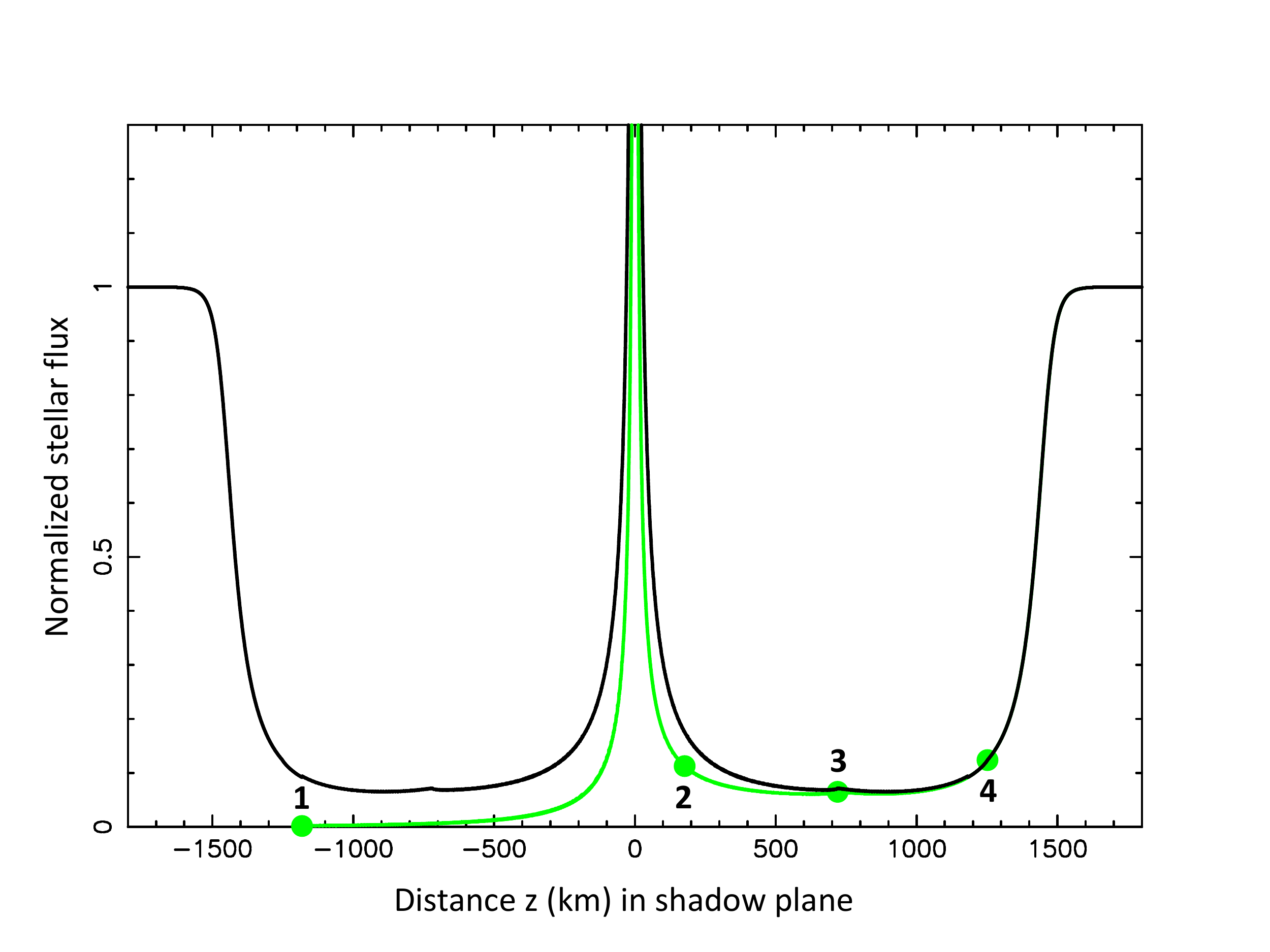}}
\caption{Template temperature profiles and result of the ray-tracing code.
\textit{Upper panel:}
Six template temperature profiles, indicated by the thin black lines, obtained taking values 
$\phi_0=$ 0.370, 0.365, 0.360, 0.355, 0.350, and 0.3445 (from left to right)
of Triton's contribution to the normalised occultation light curve at La Palma.
The rightmost profile (corresponding to $\phi_0= 0.3445$) 
is the profile obtained from the calibration images taken before the event.
Exploring $\phi_0$ with small incremental steps, 
we find a best fit to the central flash for 
$\phi_0=0.35885$ (see the thicker black template profile, where
the green dots labelled 1 to 4 are prescribed particular points in the \citealt{dia15} model,
as specified in Table~\ref{tab_T_template}).
We note that all the profiles go through the boundary condition $T=54.5$~K at $r=1453$~km
(100~km altitude, upper green dot).
The inverted profiles for La Palma corresponding to $\phi_0=0.35885$ 
are shown in red (ingress) and blue (egress).
The horizontal dotted line marks the altitude of the central flash layer.
\textit{Lower panel:}
Result of the ray-tracing code, using the best $T(r)$ profile of the upper panel
and the best fit value of $p_{1400}= 1.18$~$\mu$bar (Table~\ref{tab_pressure_time}).
The normalised stellar flux is plotted against the distance to Triton's shadow centre.
The green curve represents the flux if only one stellar image is present, and 
the green dots show the correspondence with the points labelled 1 to 4 in the upper panel.
The black curve is the sum of the fluxes from two stellar images, i.e. the sum of the green curve and its mirrored version with respect to the $z=0$ axis.
This black curve is then used to fit the observations.
}
\label{fig_T_r_various_phi0}
\end{figure}

A difficulty encountered during the inversion of the light curves is the assessment 
of Triton's contribution to the total flux. 
We define $\phi_0= F_T/(F_S +  F_T)$, 
where $F_T$ (resp.  $F_S$) is the flux coming from Triton (resp. the unocculted star).
Thus, $\phi_0$ corresponds to the zero stellar flux level 
in the normalised occultation light curves. 
Changing its value mainly changes the deepest parts of the retrieved profiles (see Fig.~\ref{fig_T_r_various_phi0}).

Measuring $\phi_0$ relies on images where Triton and the star are angularly separated. 
It is generally a difficult task, 
since a photometric accuracy of better that 1\% is necessary to bring useful constraints. 
To mitigate differential chromatic effects (as the star and Triton have different colours), 
images were taken at the same elevation as for the occultation, 
either during the same nights or during nights before or after the event.  
As a sanity check, it is also desirable to use another reference star with flux $F_R$ 
and see if the sum of the ratios $F_S/F_R$ and $F_T/F_R$ outside the event matches 
the ratio $(F_S +  F_T)/R_R$ during the event.
Mismatches may then reveal possible variabilities in any of the objects involved
(the target and reference stars, and/or Triton),
and serve as an estimator of systematic sources of errors.

Such calibration images were acquired at the La Palma and Helmos sites,
which provided the best datasets in terms of S/N. 
The focal lengths of those two telescopes are large enough 
to clearly resolve Triton from Neptune and from the occulted star and 
to avoid, in particular, flux contamination from the planet in the occultation light curve.

The calibration results are self-consistent for both instruments, 
with small internal error bars on $\phi_0$ (i.e.
$\phi_0=~0.3445~\pm~0.0003$ for La Palma and
$\phi_0=~0.360~\pm~0.013$ for Helmos).
We note that there is no reason why the $\phi_0$ should be the same for the two stations,
as different filters were used: I+z ($>$720~nm) at La Palma and V+R at Helmos.
Also, different observing conditions were prevailing at the two stations.
In particular, 
the different airmasses (1.3 and 2.4 at La Palma and Helmos, respectively) 
may also affect the flux ratios star/Triton.
However, the different values of $\phi_0$ at La Palma and Helmos
are mutually inconsistent (to within error bars) in the sense that 
they provide significantly different bottom parts for the $T(r)$ profiles.

This is particularly problematic for La Palma, where we obtain a temperature profile 
that peaks at $T \sim 58$~K just above the flash layer, using the value 
$\phi_0= 0.3445$ derived from the calibration (Fig.~\ref{fig_T_r_various_phi0}).
This imposes a strong inversion layer in order to connect the temperature profile 
38~K at the surface (see Fig.~\ref{fig_T_logp_bottom} and associated discussion).
When used in our ray-tracing code, this profile yields inconsistent 
synthetic central flashes when compared with observations
(Figs.~\ref{fig_fit_flash_1}-\ref{fig_fit_flash_2}). 
In particular, it is not possible to fit simultaneously the flashes 
lying north and south of the shadow centre.

We have no satisfactory explanations for the inconsistencies induced by 
the photometric calibration (i.e. by the retrieved values of $\phi_0$). 
It could stem from unaccounted light contamination from Neptune, 
although the biggest telescopes should be free of this problem, as mentioned above.
We tested the effect of digital coronagraphy to remove any such possible contamination.
However, no significant changes of $\phi_0$ are obtained with and without the 
use of coronagraphy.
Possible variations of Triton's flux due to rotational light curves of the satellite 
have also been considered. However, considering the low amplitude of those rotational
curves \citep{bur11}, and the fact that calibrations were made about 90 minutes before
the occultation at La Palma, Triton's flux variation should be well below the 1\% level,
and thus have a negligible effect on the retrieved value of $\phi_0$.

In these conditions, we have opted for another approach.
We have varied the value of $\phi_0$ for the best dataset (i.e. the La Palma light curve).
For each value, we have inverted the light curve to obtain $T(r)$,
and then derived a template temperature profile using the \cite{dia15} modelling,
except that the upper branch of $T(r)$ is not isothermal, 
but has a constant gradient $dT/dr \sim 0.1$~K~km$^{-1}$ 
to account for the general temperature increase (thermosphere) described in \cite{str17}.

Each particular value of $\phi_0$ provides a set of prescribed $T(r)$ 
boundary conditions (the green dots in Fig.~\ref{fig_T_r_various_phi0}). 
Then it is possible to interpolate those prescribed values for any $\phi_0$
using smooth polynomial functions, and finally get a one-parameter family
of $T(r)$ profiles depending only on $\phi_0$.

The main goal here is to find the value of $\phi_0$ that best matches all the
central flashes, thus constraining the deepest part of the $T(r)$ profile.
The best $T(r)$ profile (with $\phi_0= 0.35885$) is shown as a thicker line 
in the upper panel of Fig.~\ref{fig_T_r_various_phi0}, 
and is adopted throughout this paper every time we use our ray-tracing code.
An example of an output of our tracing code is displayed in the lower panel of
Fig.~\ref{fig_T_r_various_phi0}.

Using the notations of \cite{dia15}, the $T(r)$ profile is constructed 
by adopting the parameters of Table~\ref{tab_T_template}.
We note that although both the $T$ and $dT/dr$ profiles are continuous,
the $d^2T/dr^2$ profile is not. 
This is evident in Fig.~\ref{fig_dTdr_r}  at the inflection point labelled `3' 
in Fig.~\ref{fig_T_r_various_phi0}.
The discontinuity of $d^2T/dr^2$ creates a very small kink at the
corresponding point in the synthetic light curve (see the lower panel of
Fig.~\ref{fig_T_r_various_phi0}).
This kink is well below the noise level of all the observed light curves,
and thus, has a negligible effect on the fit to the data.

\begin{table}[!h]
\caption{%
Parameters of the temperature template profile.
\label{tab_T_template}
}%
\centering
\begin{tabular}{ll}
\hline \hline
$r_1$, $T_1$, $dT/dr(r_1)$      & 1353~km, 38~K, 5~K~km$^{-1}$ (surface)    \\
\hline
$r_2$, $T_2$                    & 1363~km, 49.8~K   (`elbow')              \\
\hline
$r_3$, $T_3$                    & 1375~km, 48.2~K   (inflection point)       \\
\hline
$r_4$, $T_4$, $dT/dr(r_4)$      & 1403~km, 49.3~K, 0.105~K~km$^{-1}$        \\
                                & (upper branch, thermosphere)              \\
\hline \hline
\end{tabular}
\end{table}

Admittedly, 
there is not a unique way to find a template $T(r)$ model that best fits the flashes.
However, our solution should capture the main properties of the real profile, 
with a temperature maximum reached just above the central flash layer, and 
a mesosphere with a mild negative temperature gradient above that temperature peak.
We note that the inversion layer connecting the profile to the surface at 38~K
(i.e. below the temperature peak) has a vanishing effect on the synthetic light curve
as the surface is approached, thus defining a `blind zone' 
as far as our data are concerned.

\subsection{The secondary stellar image issue}
\label{sec_appen_secondary_image}

The Abel inversion assumes that only one stellar image (the near-limb, or primary refracted image) contributes to the recorded flux. 
In reality, Triton's atmosphere also produces a far-limb (secondary) stellar image whose flux is added to the light 
curve\footnote{More images can be produced in the central flash region for
a non-spherical atmosphere.}. 
This is a source of error, especially in the central flash region, 
where the primary and secondary images have comparable fluxes (see for instance the lower panel of Fig.~\ref{fig_T_r_various_phi0}).

\begin{figure}[!t]
\centerline{
\includegraphics[totalheight=7cm, trim=0 0 0 0, angle=0]{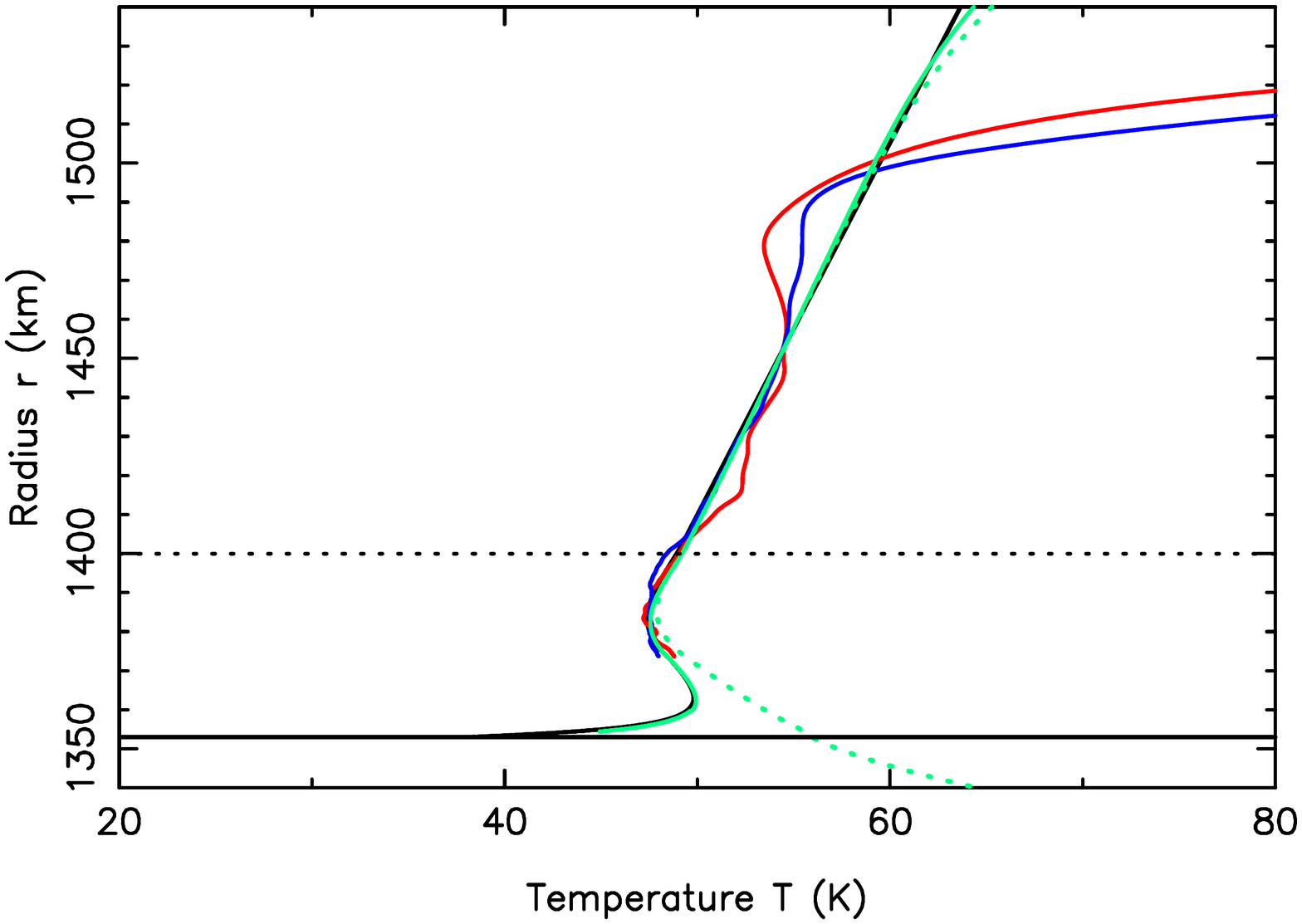}
}
\centerline{
\includegraphics[totalheight=7cm, trim=0 0 0 0, angle=0]{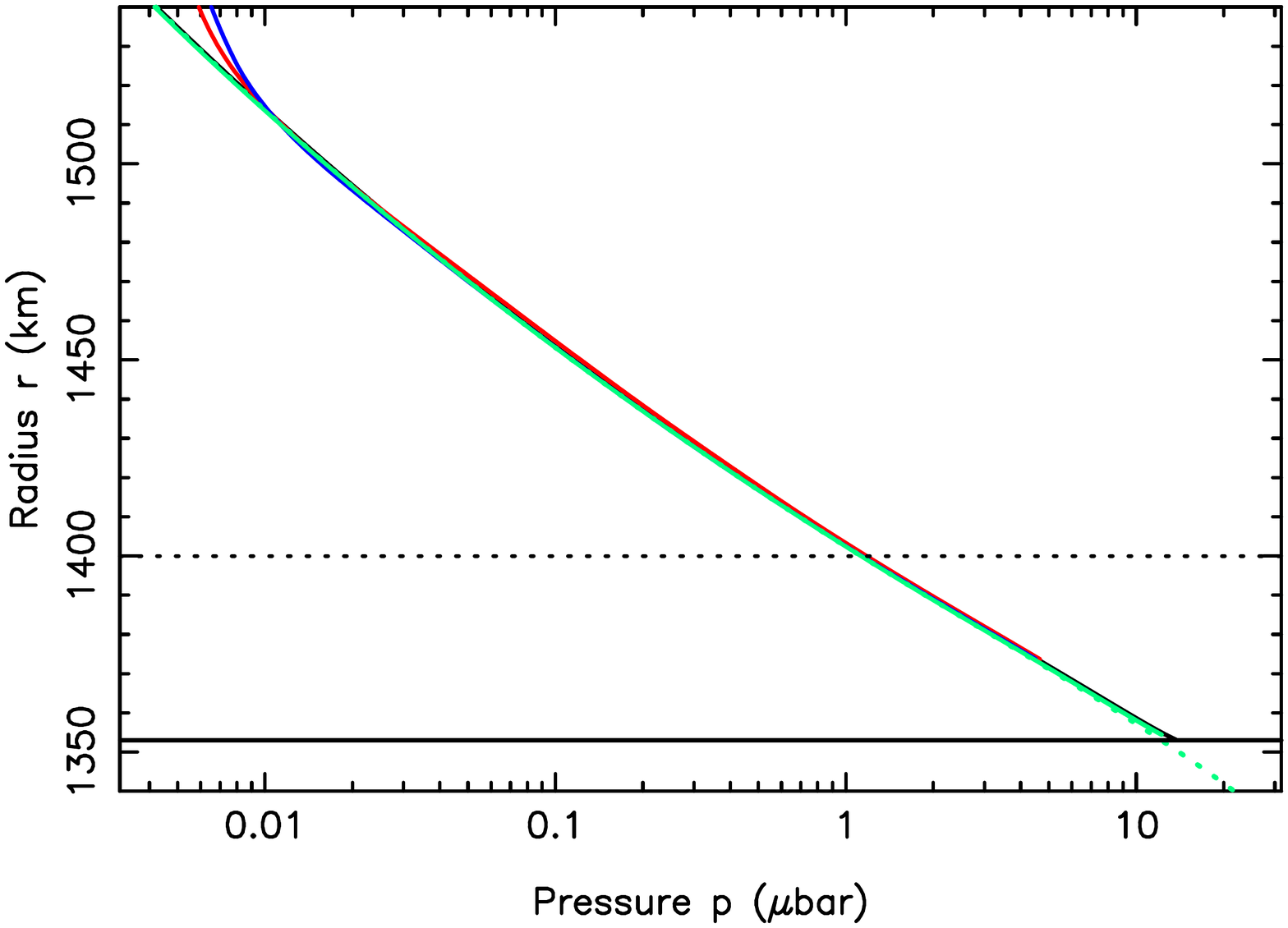}
}
\caption{Effect of the secondary stellar image on the inversion results.
\textit{Upper panel:}
Effect on the temperature profiles.
The smooth black line connects these profiles to the surface at 38~K, 
where the pressure is set to 14~$\mu$bar. The resulting density profile is 
used to generate synthetic light curves that feed the Abel inversion procedure.
The solid green line is the retrieved $T(r)$ profile with only the primary image accounted for. 
The green dotted line is the retrieved profile where the primary and secondary 
images have been added.
The profiles obtained from the inversion of La Palma's light curve
at ingress and egress (in red and blue, respectively) are shown for comparison (see also Figs.~\ref{fig_n_p_r}-\ref{fig_dTdr_r}).
Black horizontal solid and dotted lines show
Triton's surface and reference radius (1400~km).
\textit{Lower panel:}
Same, but with the pressure profiles.
}
\label{fig_effect_secondary_image}
\end{figure}

We have performed tests to assess this effect on La Palma's light curve.
At closest approach, the station was at about 685~km from the shadow centre, 
providing information down to the $\sim$20~km altitude level (Fig~\ref{fig_effect_secondary_image}).
A smooth $T(r)$ profile is used to generate synthetic
light curves at this station, 
one with the flux of the primary image only, 
and one with the sum of the fluxes of the primary and secondary images.

%
As expected, the inversion of the one-image light curve correctly retrieves the temperature at the 0.2~K accuracy level (Fig~\ref{fig_effect_secondary_image}),
and the density and pressure profiles at the 0.1\% accuracy level.
Conversely, the inversion of the two-image light curve correctly retrieves the upper parts of the profiles, but 
it fails in reproducing the lower parts. 
For instance, at the deepest point reached by La Palma's light curve,  
the primary flux is 0.061 (normalising the unocculted stellar flux to unity), 
while the secondary flux is 0.0072, about 8.5 times fainter than the primary flux.
At that point, the temperature is retrieved to within 0.9~K and the pressure at the 1\% level, 
a satisfactory result at our accuracy level.

However, this discrepancy rapidly increases as deeper levels are probed. For instance at $r= 1362$~km (9~km altitude level, where the temperature locally reaches a maximum; see the upper panel of Fig.~\ref{fig_effect_secondary_image}), 
the discrepancy between the original and retrieved temperature is about 3~K.
We also note that in this case the retrieved temperature profile has an
unrealistic behaviour as it extends below Triton's surface.
Although the pressure profile is satisfactorily retrieved even if the 
secondary image is present, it suffers nevertheless the same unrealistic
behaviour as $T(r)$, as it also extends below Triton's surface 
(lower panel of Fig.~\ref{fig_effect_secondary_image}).

In summary, the inversion procedure cannot provide reliable results 
below the 20 km altitude level. In particular the central flash region
cannot be used in the inversion procedure.
We note, however, that the direct (ray-tracing) approach does include the 
primary and secondary fluxes, and as such, it can be used to constrain
the atmospheric profiles in the central flash region.

%
%

\section{Fit to the data}
\label{sec_appen_fit_data}
\clearpage \newpage
\begin{figure*}[h]
\centerline{\includegraphics[width=135mm]{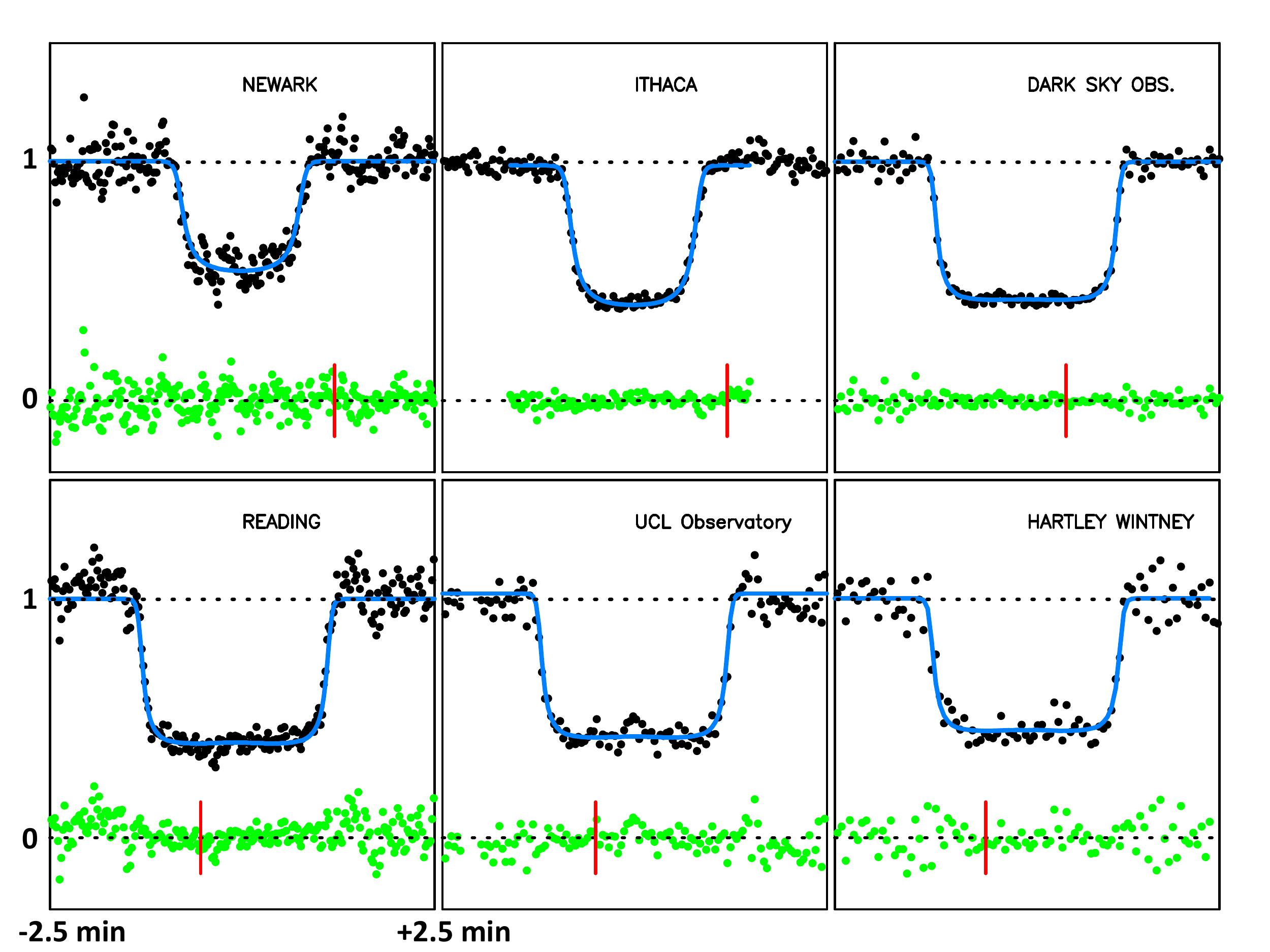}}
\centerline{\includegraphics[width=135mm]{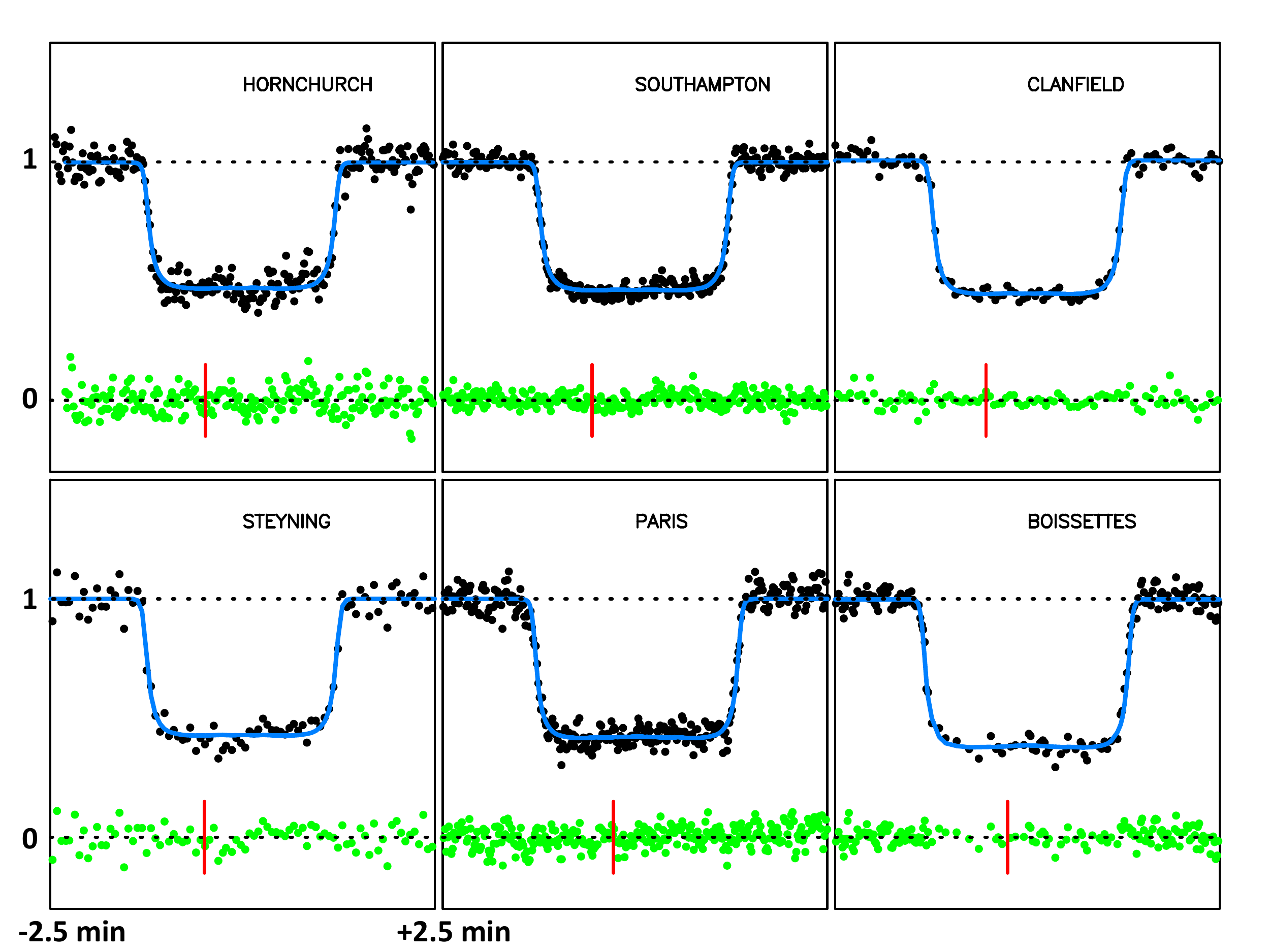}}
\caption{%
Data (black dots) fitted simultaneously with synthetic light curves 
(blue lines), based on the temperature profile displayed in Fig.~\ref{fig_T_r_logp} 
(black line) and the pressure boundary condition $p_{1400}= 1.18$~$\mu$bar
(Table~\ref{tab_pressure_time}).
The green dots are the residuals of the fits.
We note that the central flash regions have been excluded from the fit 
so that we obtain a global fit that is not influenced by the deepest atmospheric layers.
The stations with exposure times smaller than 1~s have been smoothed to have a sampling 
time close to 1~s, allowing a direct visual comparison of S/N of the various datasets.
The lower and upper horizontal dotted lines mark the zero-flux level and the 
total star plus Triton unocculted flux, respectively.
We note that the three central-most stations (Constância, Le Beausset, and Felsina observatories in Fig.~\ref{fig_fit_data_3})
are plotted at a different vertical scale to accommodate the presence of a strong central flash.
Each panel has a duration of five minutes and is centred around the time of closest approach 
(or mid-occultation time) of the station to Triton's shadow centre.
The stations are sorted from left to right and top to bottom from the northernmost track
(Newark) to the southernmost track (Athens; see Fig.~\ref{fig_fit_data_5}), 
projected on Triton in the sky plane (Fig.~\ref{fig_chords}).
For reference, the vertical red lines mark 23:48 UTC for the European and African stations, and 23:55 UTC for the US stations (Newark, Ithaca, and Dark Sky observatories).
}%
\label{fig_fit_data_1}
\end{figure*}

\begin{figure*}[h]
\centerline{\includegraphics[width=135mm]{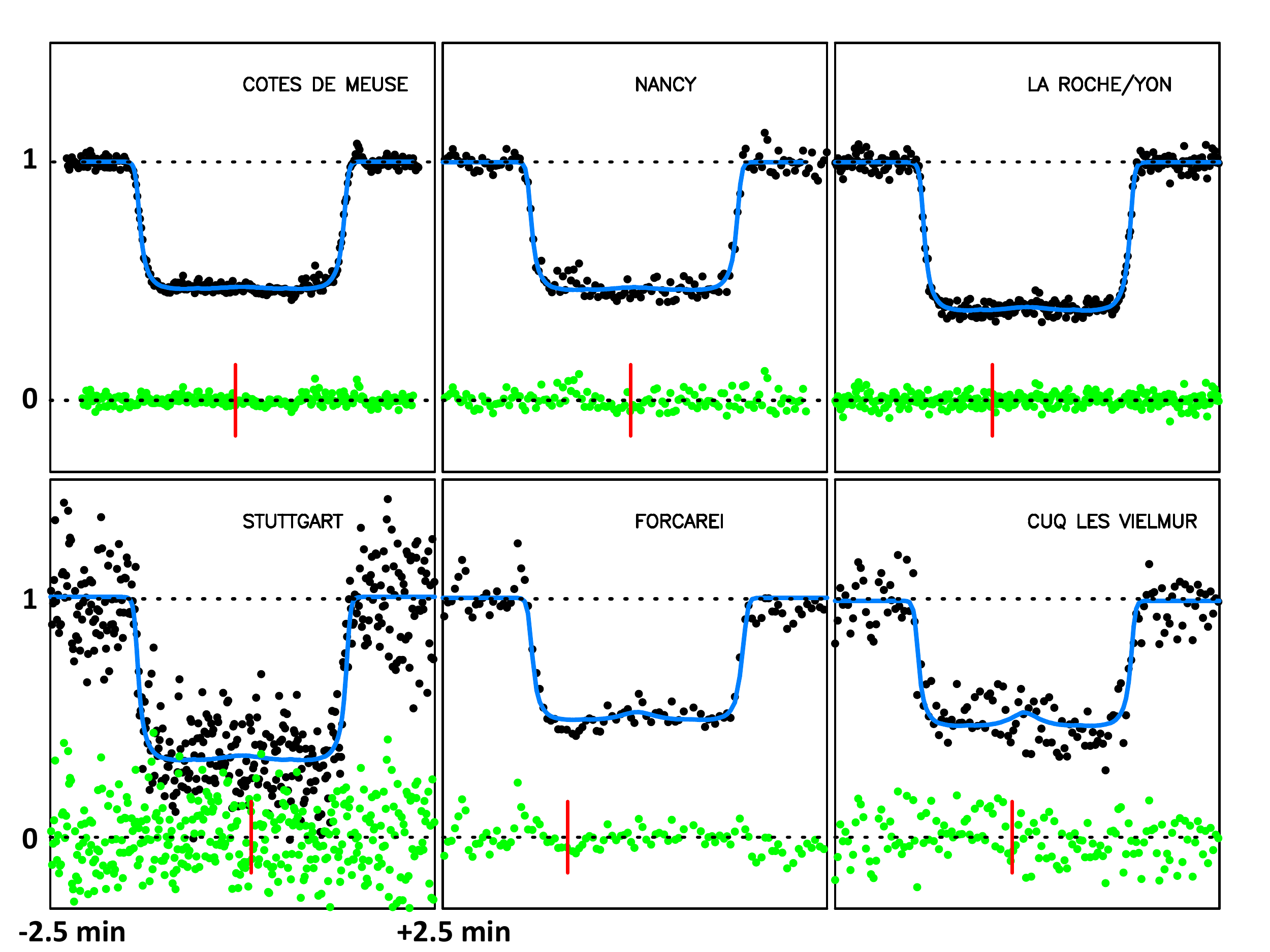}}
\centerline{\includegraphics[width=135mm]{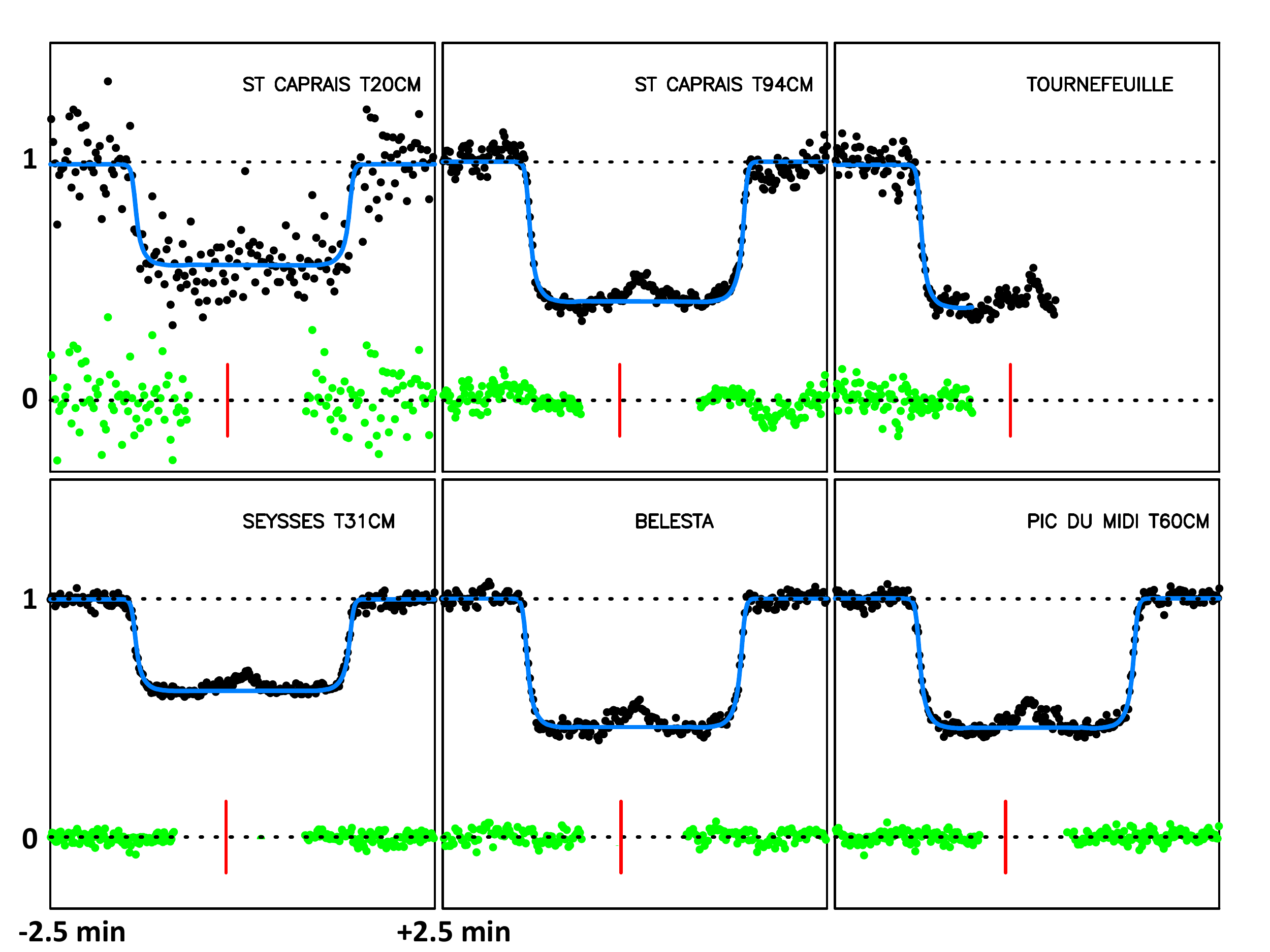}}
\caption{
Continuation of Fig.~\ref{fig_fit_data_1}.
}
\label{fig_fit_data_2}
\end{figure*}

\begin{figure*}[h]
\centerline{\includegraphics[width=135mm]{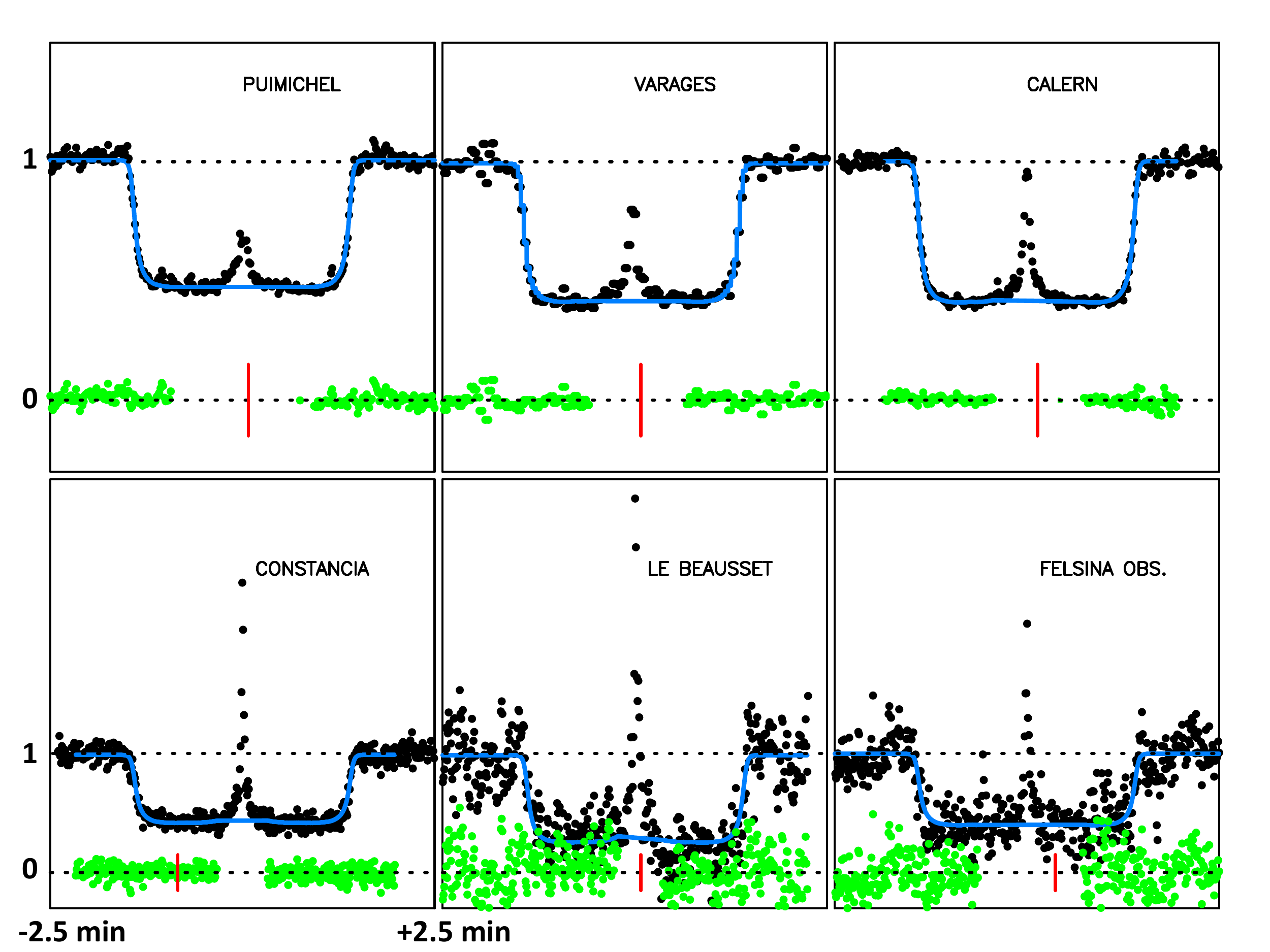}}
\centerline{\includegraphics[width=135mm]{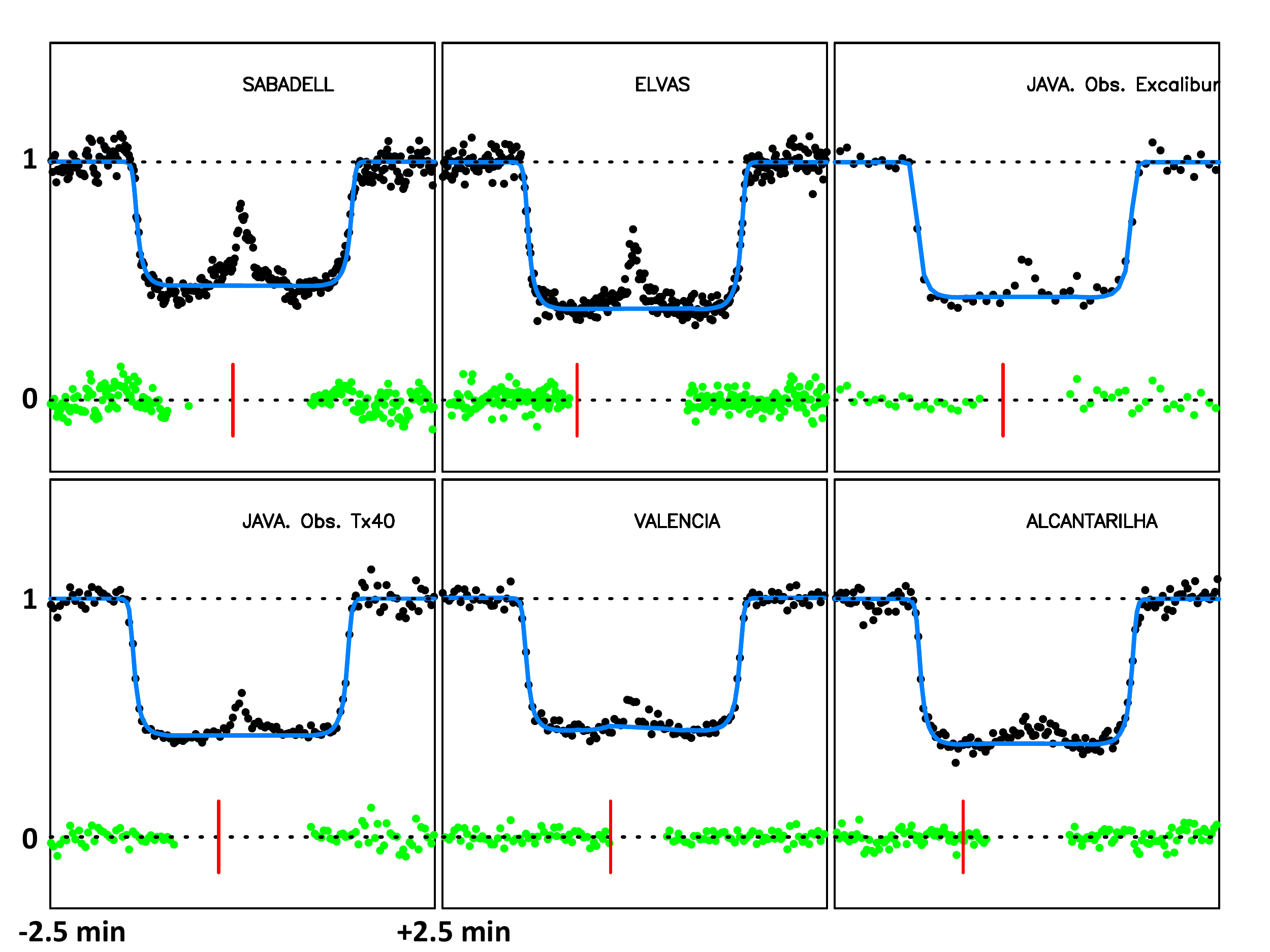}}
\caption{
Continuation of Fig.~\ref{fig_fit_data_2}.
NB. `JAVA.' is the abbreviation of Javalambre, 
used so that the name of the station fits into the plot.
}
\label{fig_fit_data_3}
\end{figure*}

\begin{figure*}[h]
\centerline{\includegraphics[width=135mm]{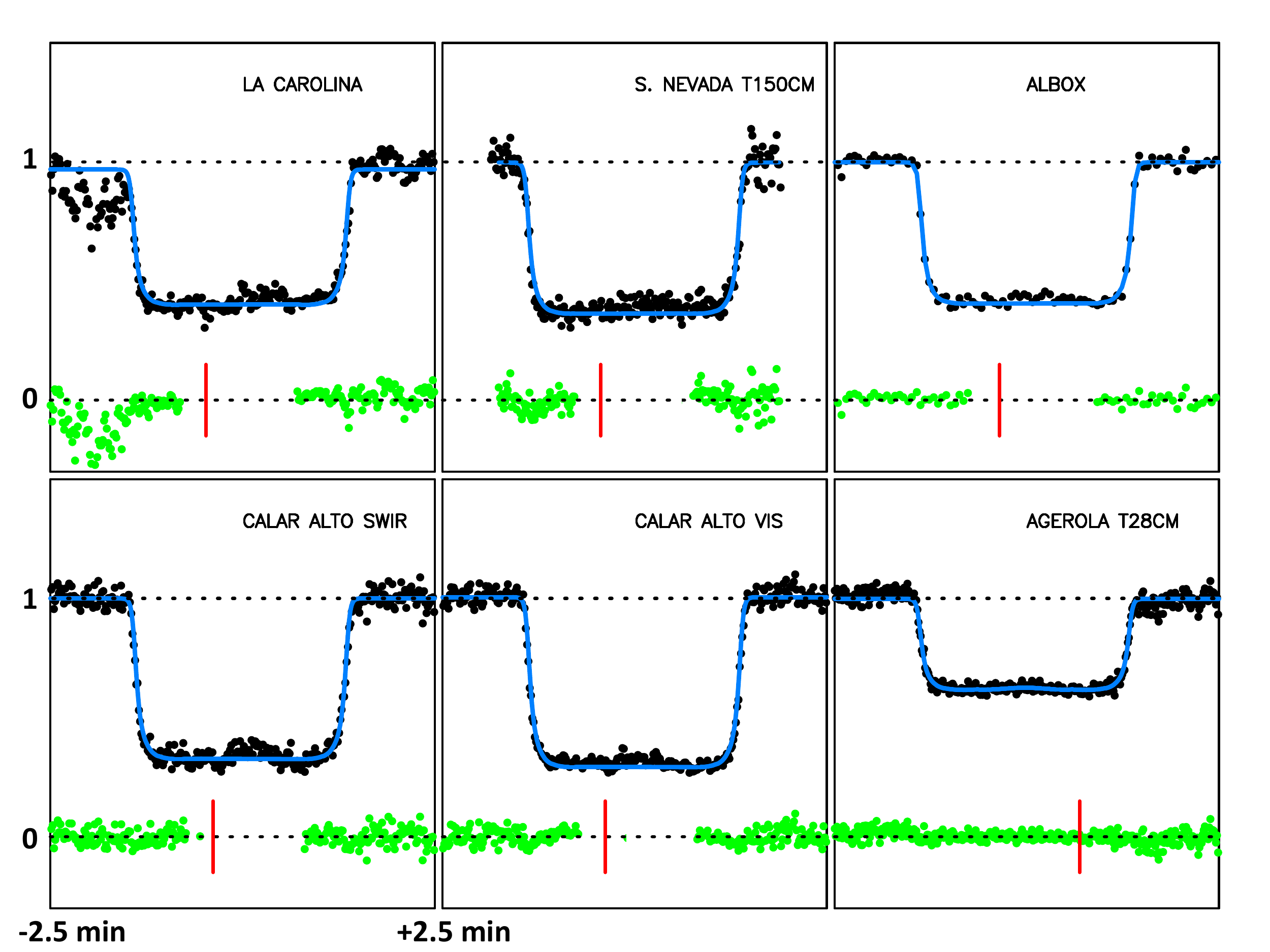}}
\centerline{\includegraphics[width=135mm]{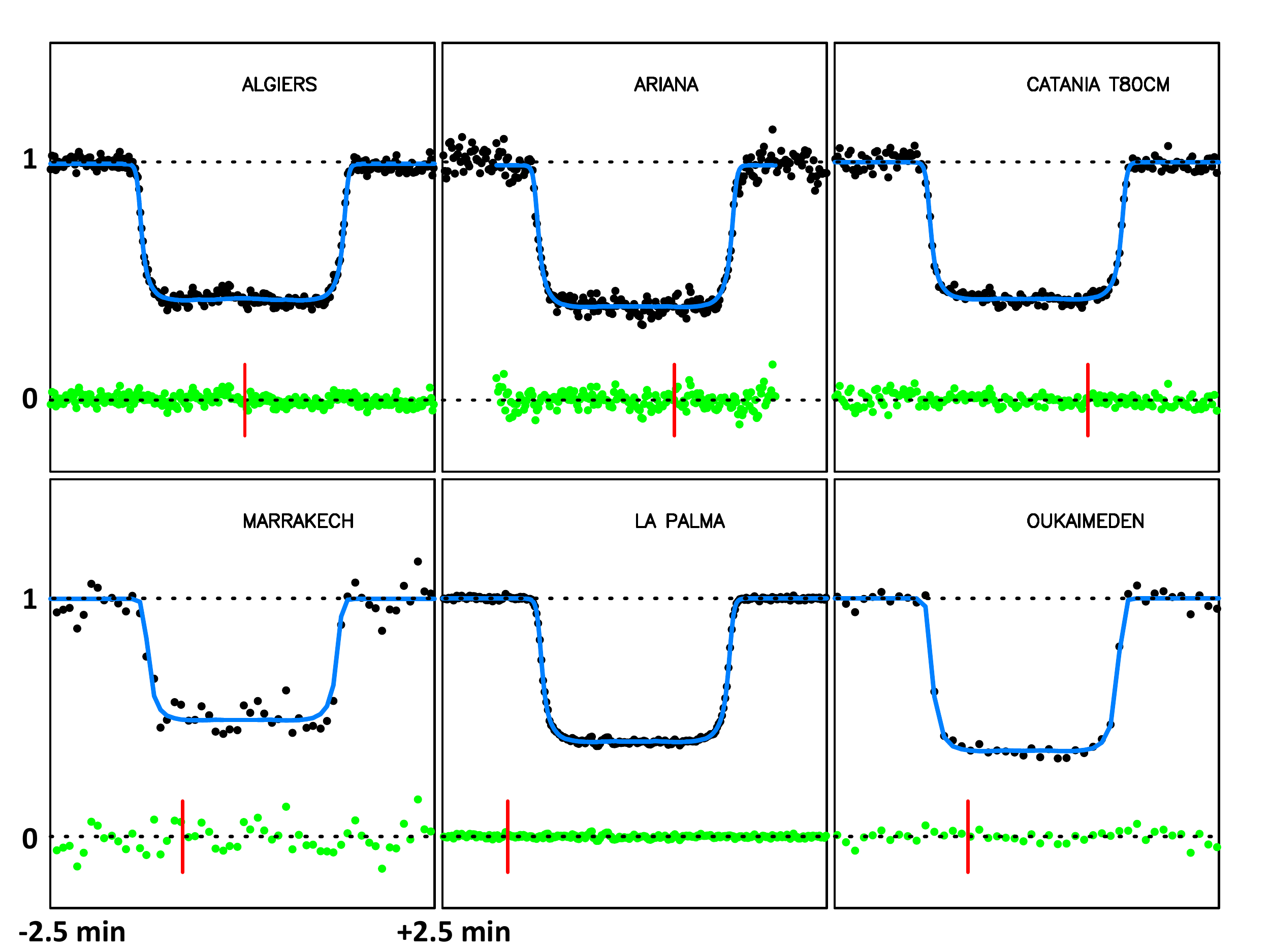}}
\caption{
Continuation of Fig.~\ref{fig_fit_data_3}.
}
\label{fig_fit_data_4}
\end{figure*}

\begin{figure*}[h]
\centerline{\includegraphics[width=135mm]{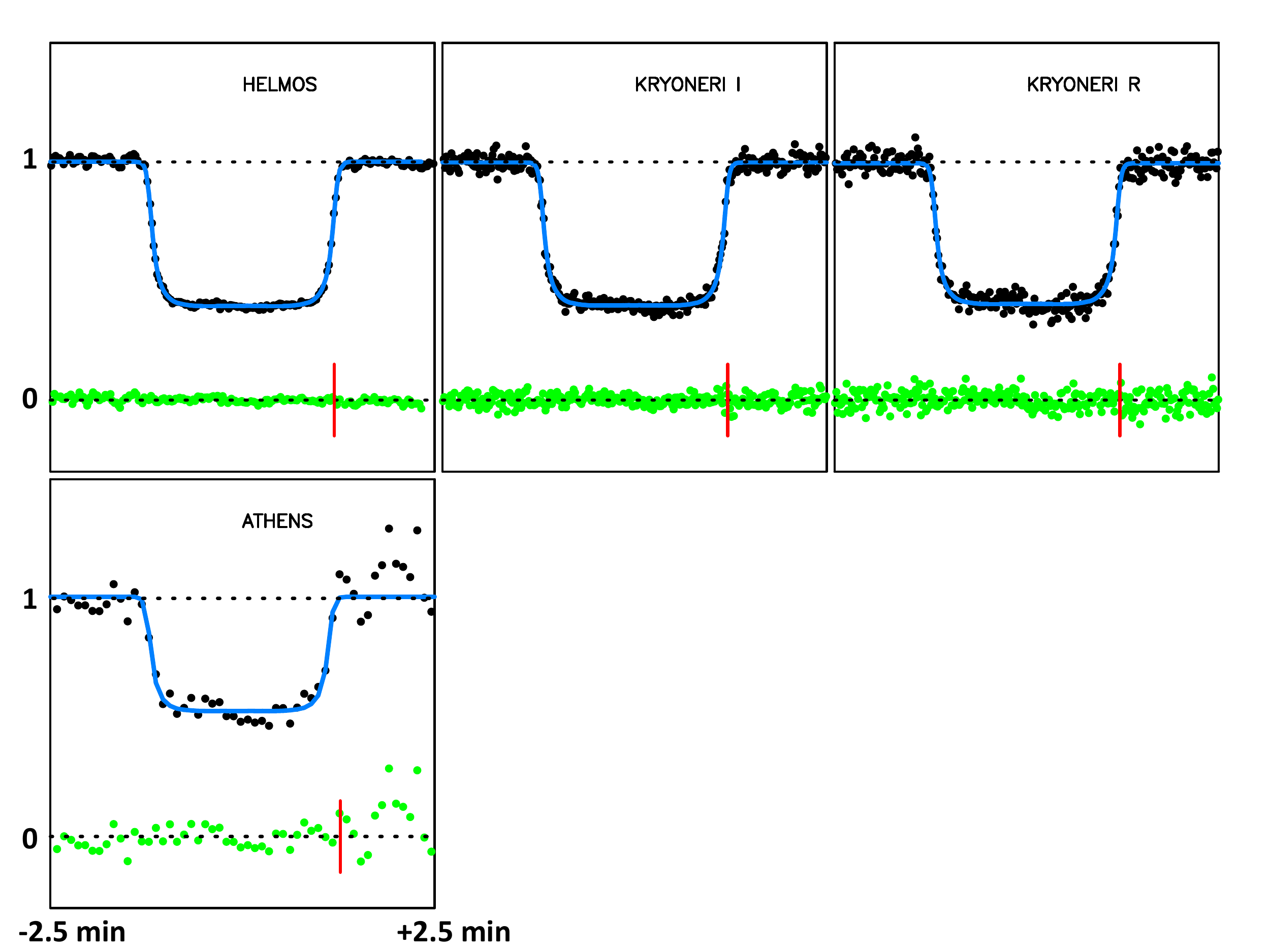}}
\caption{
Continuation of Fig.~\ref{fig_fit_data_4}.
}
\label{fig_fit_data_5}
\end{figure*}

\begin{figure*}[h]
\centerline{\includegraphics[width=135mm]{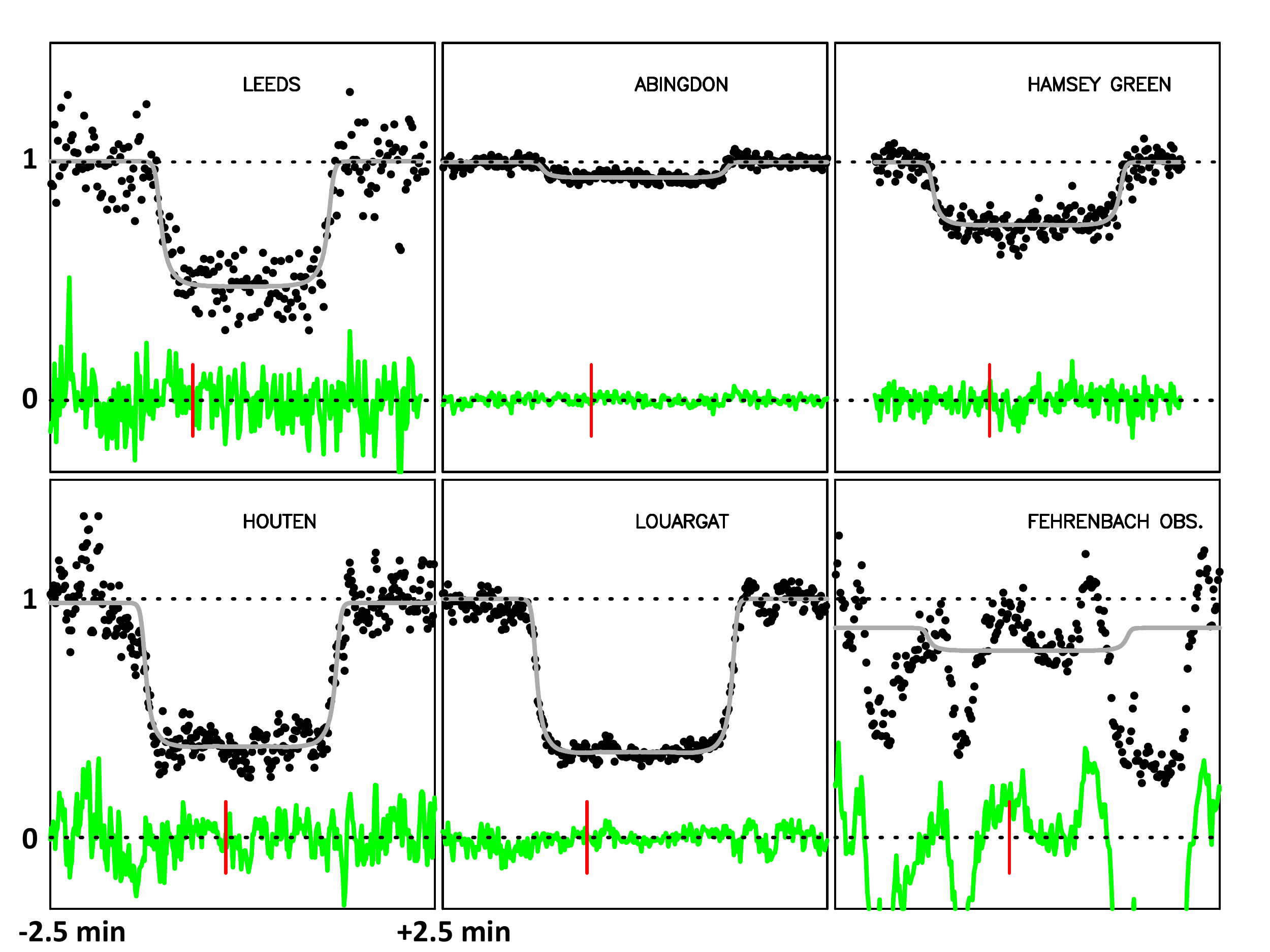}}
\centerline{\includegraphics[width=135mm]{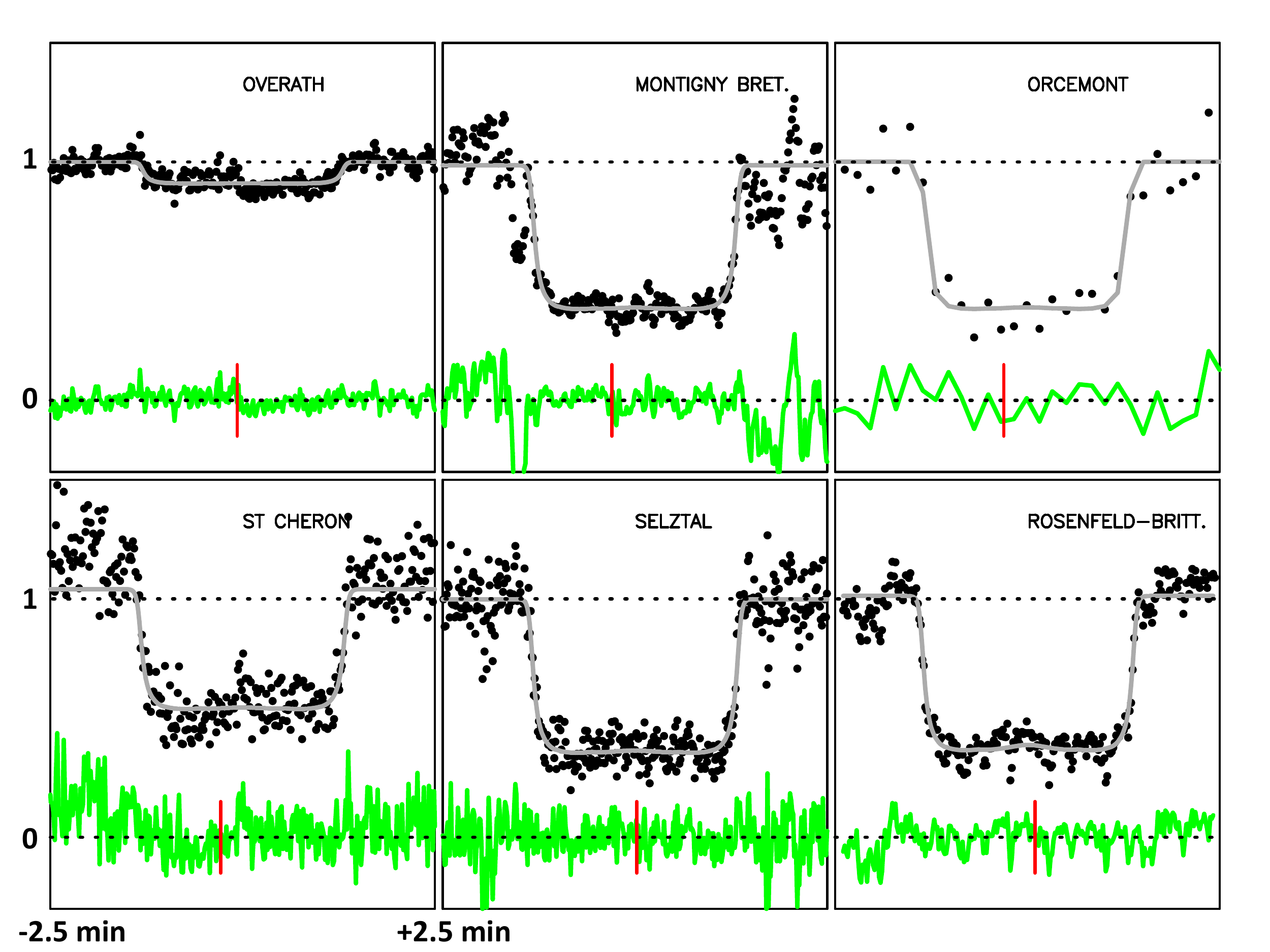}}
\caption{
The same as Figs.~\ref{fig_fit_data_1}-\ref{fig_fit_data_5}, but
for stations that were not used in the simultaneous fit.
}
\label{fig_no_fit_data_1}
\end{figure*}

\begin{figure*}[h]
\centerline{\includegraphics[width=135mm]{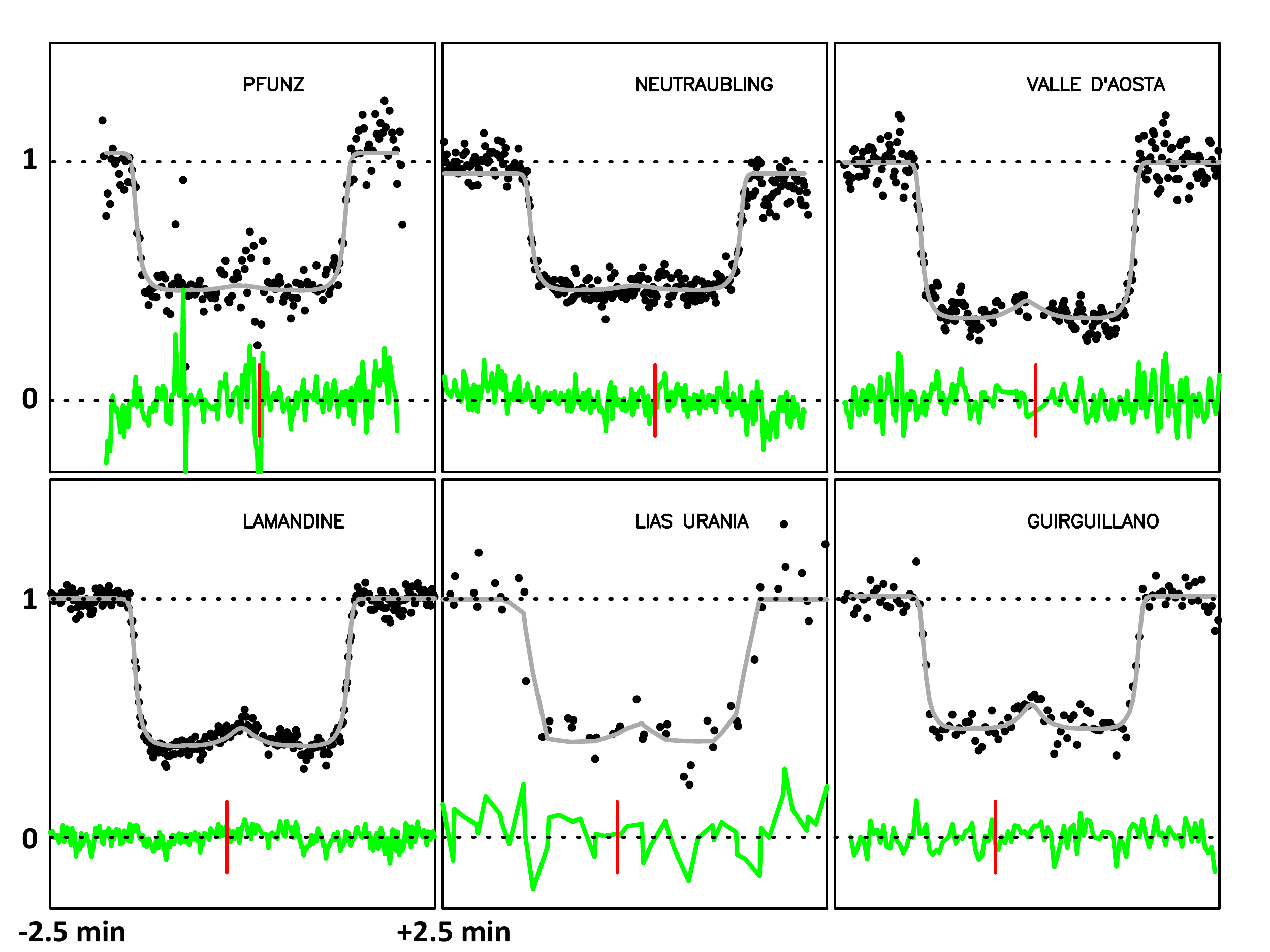}}
\centerline{\includegraphics[width=135mm]{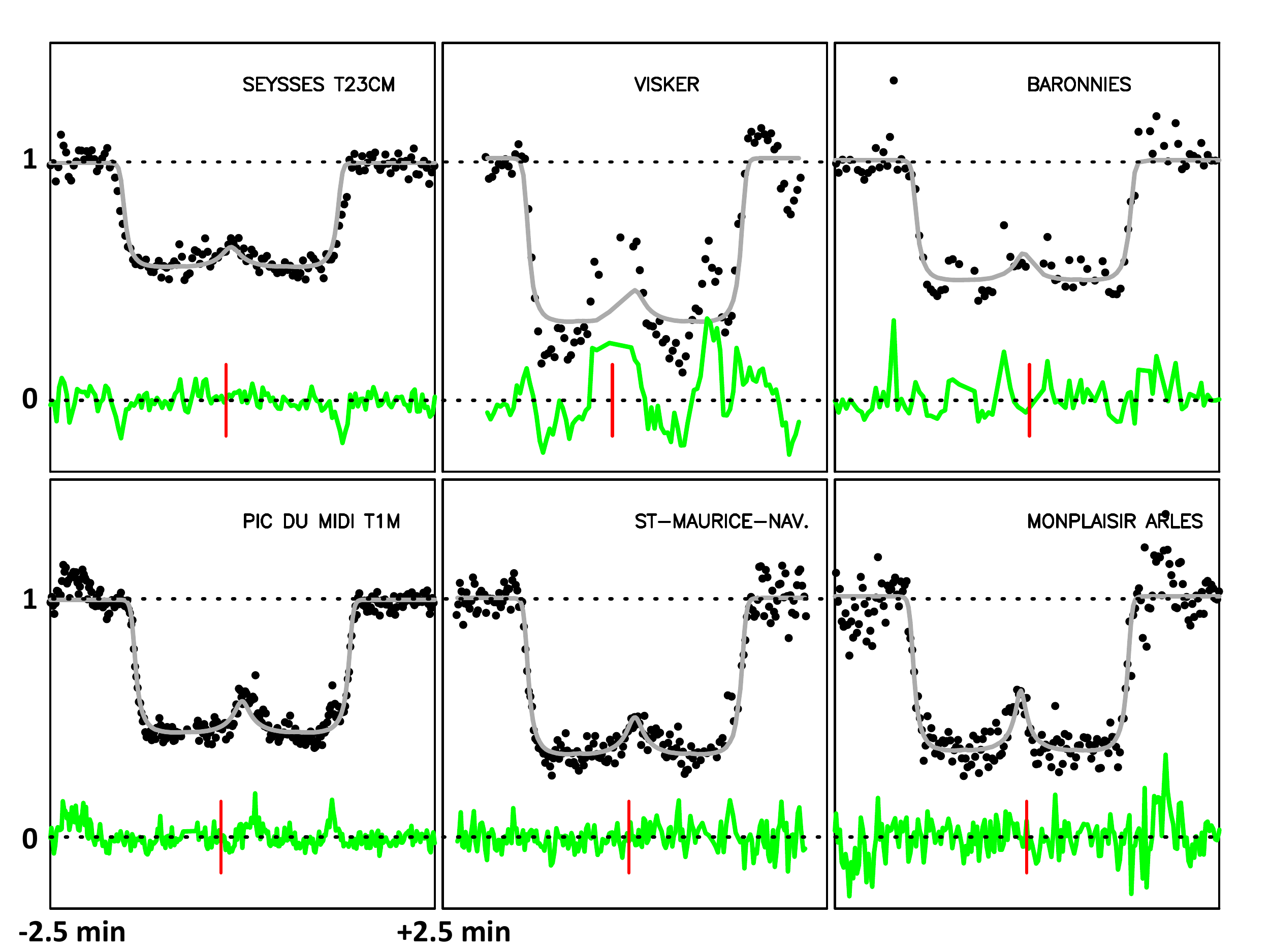}}
\caption{
Continuation of Fig.~\ref{fig_no_fit_data_1}.
}
\label{fig_no_fit_data_2}
\end{figure*}

\begin{figure*}[h]
\centerline{\includegraphics[width=135mm]{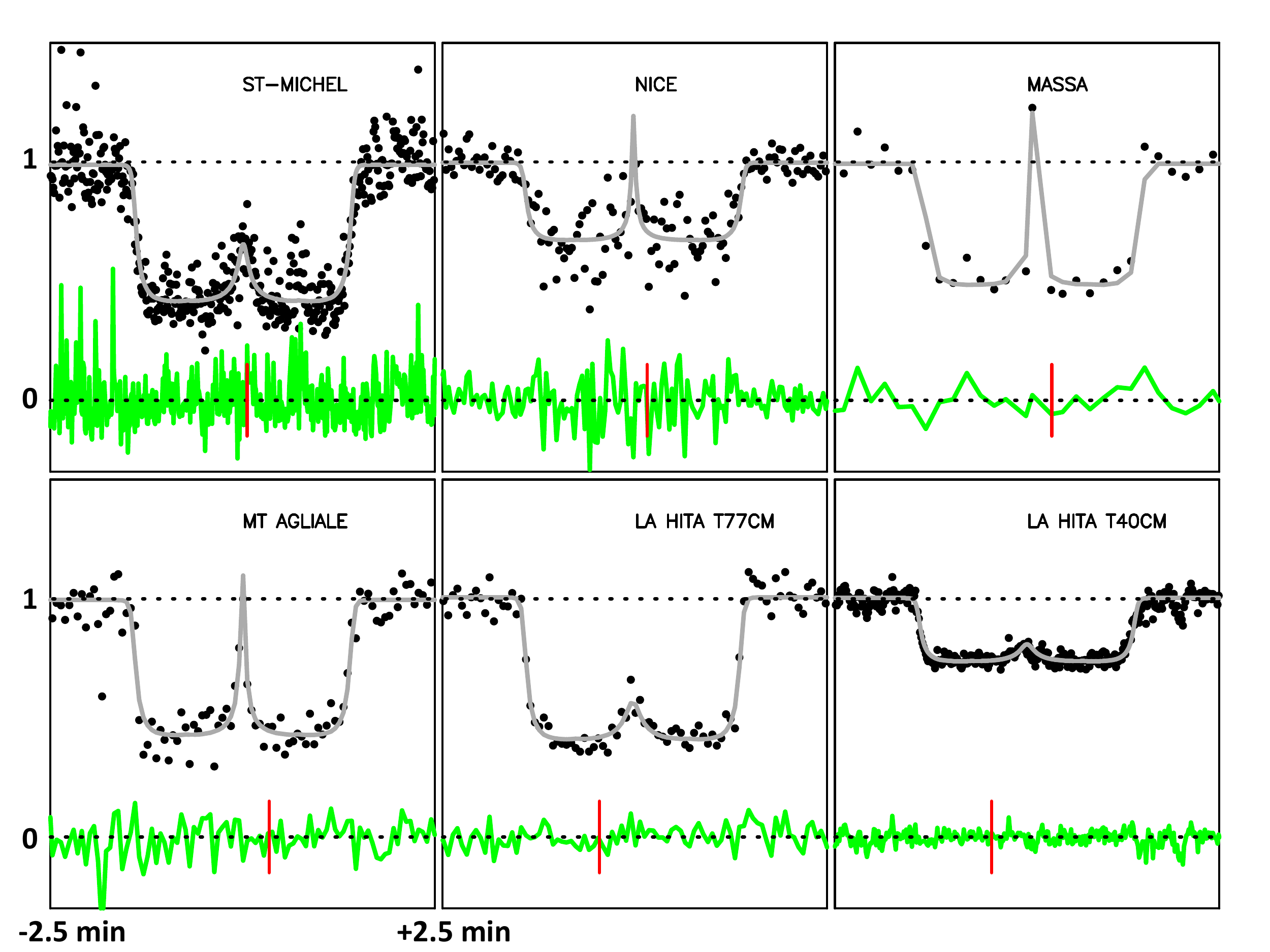}}
\centerline{\includegraphics[width=135mm]{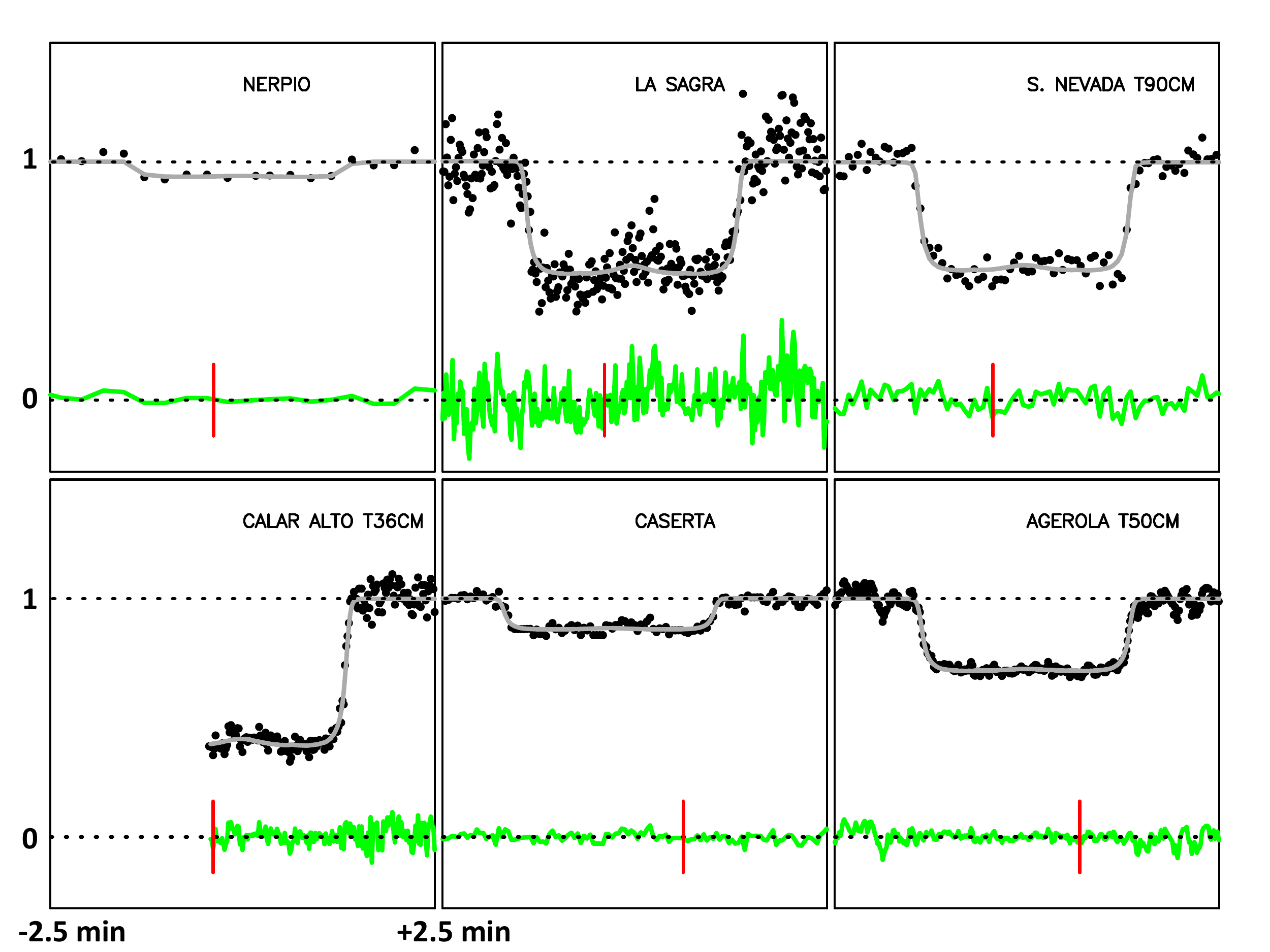}}
\caption{
Continuation of Fig.~\ref{fig_no_fit_data_2}.
}
\label{fig_no_fit_data_3}
\end{figure*}

\begin{figure*}[h]
\centerline{\includegraphics[width=135mm]{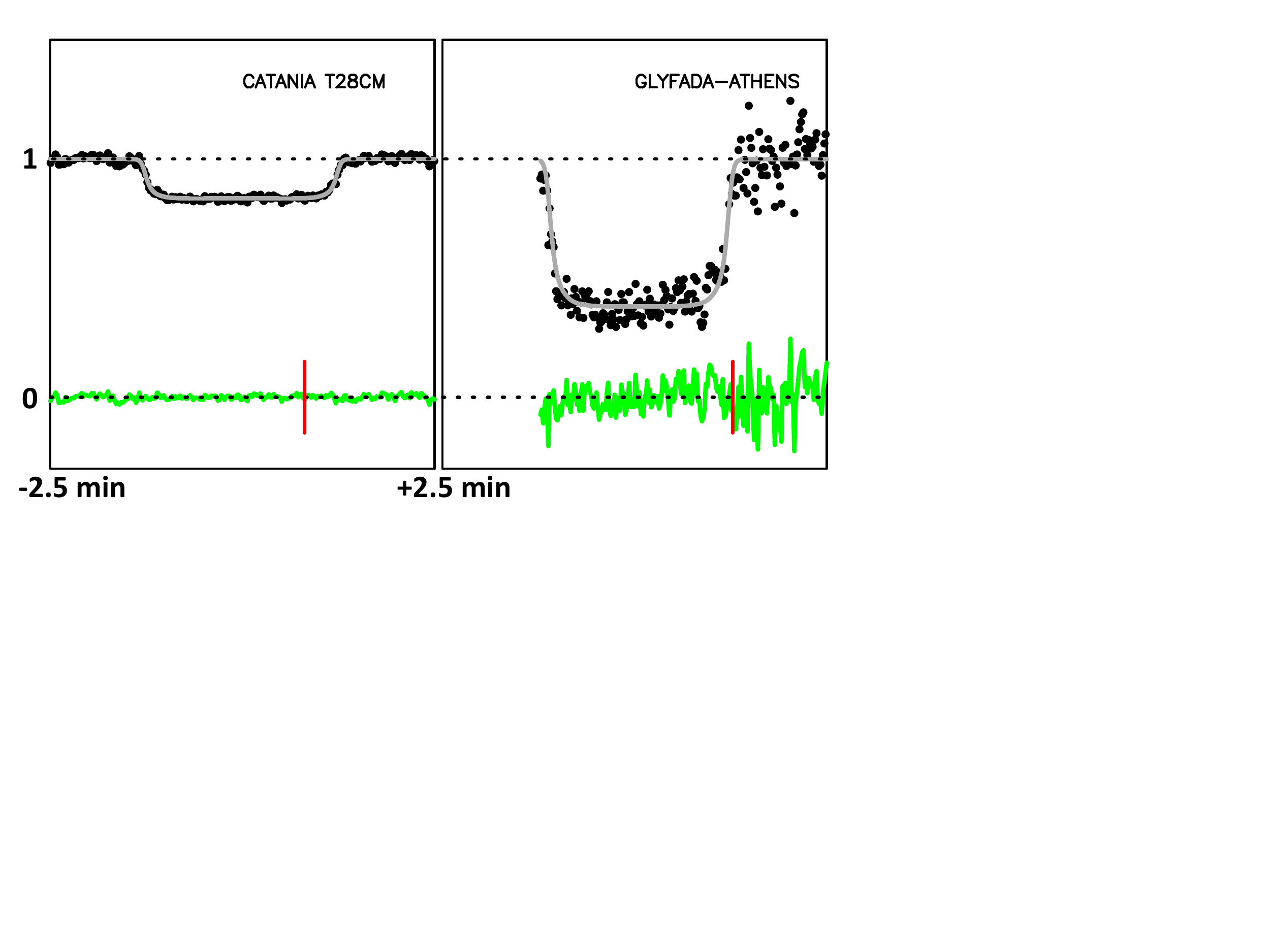}}
\caption{
Continuation of Fig.~\ref{fig_no_fit_data_3}.
}
\label{fig_no_fit_data_4}
\end{figure*}

\end{appendix} 

\end{document}